\documentclass[12pt]{article}
\usepackage{amsmath,amssymb,amsthm,color}
\usepackage{graphicx}
\usepackage{epsfig}
\usepackage{multirow}
\usepackage{dcolumn}
\usepackage{bm}
\usepackage{float}
\usepackage{chngcntr}
\counterwithout{figure}{section}
\renewcommand{\baselinestretch}{1.66}
\begin{document}

\title {Diverse quantization phenomena in layered materials}
%\small Chiun-Yan Lin, Thi-Nga Do, Jhao-Ying Wu, Po-Hsin Shih, Shih-Yang Lin, Ching-Hong Ho, Ming-Fa Lin $$\\
\author{
\small Chiun-Yan Lin,\textit{$^{a}$} Thi-Nga Do,\textit{$^{b,c,\ast}$} Jhao-Ying Wu,\textit{$^{d,\dag}$} Po-Hsin Shih,\textit{$^{a,\ddag}$}\\
\small Shih-Yang Lin,\textit{$^{e,\S}$} Ching-Hong Ho,\textit{$^{d,\P}$} Ming-Fa Lin\textit{$^{a}$} \\
\small $^{a}$ Department of Physics, National Cheng Kung University, Tainan 701, Taiwan. \\
\small $^{b}$ Laboratory of Magnetism and Magnetic Materials, Advanced Institute of Materials Science,\\
\small Ton Duc Thang University, Ho Chi Minh City, Vietnam. \\
\small $^{c}$ Faculty of Applied Sciences, Ton Duc Thang University, Ho Chi Minh City, Vietnam. \\
\small $^{d}$ Center of General Studies, National Kaohsiung University of Science and Technology,\\
\small Kaohsiung 811, Taiwan \\
\small $^{e}$ Department of physics, National Chung Cheng University, Chiayi 621, Taiwan. \\
}
\footnotetext{$\ast~$ E-mail: dothinga@tdtu.edu.vn}
\footnotetext{$\dag~$ E-mail: yarst5@gmail.com}
\footnotetext{$\ddag~$ E-mail: phshih@phys.ncku.edu.tw}
\footnotetext{$\S~$ E-mail: sylin.1985@gmail.com}
\footnotetext{$\P~$ E-mail: hohohosho@gmail.com}

\renewcommand{\baselinestretch}{1.66}
\maketitle

\renewcommand{\baselinestretch}{1.66}

\begin{abstract}

The diverse quantization phenomena in 2D condensed-matter systems, being due to a uniform perpendicular magnetic field and the geometry-created lattice symmetries, are the focuses of this book. They cover the diversified magneto-electronic properties, the various magneto-optical selection rules, the unusual quantum Hall conductivities, and the single- and many-particle magneto-Coulomb excitations. The rich and unique behaviors are clearly revealed in few-layer graphene systems with the distinct stacking configurations, the stacking-modulated structures, and the silicon-doped lattices, bilayer silicene/germanene systems with the bottom-top and bottom-bottom buckling structures, monolayer and bilayer phosphorene systems, and quantum topological insulators. The generalized tight-binding model, the static and dynamic Kubo formulas, and the random-phase approximation, are developed/modified to thoroughly explore the fundamental properties and propose the concise physical pictures. The different high-resolution experimental measurements are discussed in detail, and they are consistent with the theoretical predictions.
\end{abstract}

\vskip 1.0 truecm
\par\noindent

\newpage
\tableofcontents

\pagebreak
\renewcommand{\baselinestretch}{2}
\newpage

\newpage
\begin{center}
\section{Introduction}\label{ch1}
\author{ Shih-Yang Lin,\textit{$^{e,\S}$} Thi-Nga Do,\textit{$^{b,c,\ast}$} Chiun-Yan Lin,\textit{$^{a}$} Jhao-Ying Wu,\textit{$^{d,\dag}$}\\ Po-Hsin Shih,\textit{$^{a,\ddag}$} Ching-Hong Ho,\textit{$^{d,\P}$} Ming-Fa Lin\textit{$^{a,\sharp}$}}
\end{center}
\vskip 1.0 truecm

How to diversify the quantization phenomena is one of the main-stream issues in physics science. Up to now, they can be achieved by the intrinsic lattice symmetries\cite{1IOP;SC}, the distinct dimensions\cite{1CRCPress;CY,1IOPBook;CY}, the diminished scales\cite{1PCCP18;7573}, the various stacking configurations\cite{1IOP;SC,1CRCPress;CY}, the mechanical strains\cite{1JPCC116;8271}, the uniform and/or modulated magnetic fields\cite{1PRB83;195405,1IOP;SC,1CRCPress;CY}, and the electric fields\cite{1CRCPress;CY,1IOPBook;CY,1CPC184;1821}. The focuses of this book are the magnetically quantized behaviors of the emergent layered materials, in which they will be clearly revealed in the electronic properties, optical absorption spectra, quantum Hall transports, and magneto-Coulomb excitations. Such 2D condensed-matter systems are very suitable for exploring the rich and unique essential properties, such as, the twisted bilayer graphene systems\cite{1PRB91;155428,1PRL121;037702,1AdvMater22;3723}, the  stacking-modulated bilayer graphene ones\cite{1PRX3;021018,1SciRep9;859}, the sliding bilayer graphenes\cite{1IOPBook;CY,1SciRep4;7509}, the few-layer graphenes with the AAA\cite{1IOPBook;CY,1CRCPress;CY}, ABA\cite{1IOPBook;CY,1CRCPress;CY}, ABC\cite{1IOPBook;CY,1CRCPress;CY} and AAB stackings\cite{1IOPBook;CY}, the silicon-doped graphene systems\cite{1IOP;SC}, the AA- and AB-stacked  bilayer silicene/germanene systems with the bottom-top and bottom-bottom bucking structures\cite{1SciRep7;40600,1PRB97;125416}, the monolayer and bilayer phosphorene systems \cite{PRB95;115411,SRp8;13303}, and the quantum topological insulators\cite{1RMP82;3045}. To thoroughly present and comprehend the diverse magnetic quantization, the previous theoretical modes need to be further developed and modified. The significant magnetic Hamiltonians, which cover the single- or multi-orbital chemical bondings, the intralayer $\&$ interlayer hopping integrals, the important spin-orbital couplings, and the external field, will be efficiently solved by the generalized tight-binding model\cite{1CRCPress;CY,1IOP;SC}. Its direct combinations with the dynamic Kubo formula is reliable in calculating the magneto-optical excitation spectra and analyzing the selection rules under the gradient approximation\cite{1IOPBook;CY}. Another linking with the static one can investigate the unusual quantum conductivities\cite{1PCCP19;29525}. Moreover, the random-phase approximation (RPA), which is available for the electronic excitations in all the condensed-matter systems, requires the exact modifications being closely related to the layered structures\cite{1SciRep7;40600,1PRB74;085406,1PRB98;041408}. The detailed comparisons between the theoretical predictions and the high-resolution experimental measurements are also made in this work.\\

The magnetic quantization can be greatly diversified by modulating/changing the various geometric structures of condensed-matter systems and the external field forms. The diverse phenomena have been predicted/observed in the main-stream or emergent materials. For example, the hexagonal symmetry, stacking configuration play critical roles in graphene-related honeycomb lattices. Under a uniform perpendicular magnetic field [$B_z \hat{z}$], the low-lying Landau-level energies of monolayer graphene are proportional to the square root of $n B_z$ ($n$ quantum number)\cite{1PCCP17;26008}, the sliding bilayer graphene systems exhibit three kinds of Landau Levels [the well-behaved, perturbed and undefined Landau levels]\cite{1SciRep4;7509},  the twisted ones present the fractal energy spectra during the variation of the magnetic-field strength\cite{1NanoLett12;3833}, the AAA-stacked graphene systems only display the quasi-monolayer Landau levels with the different initial energies for the distinct groups (the vertical Dirac=cone structure), the trilayer ABA stacking shows the unusual superposition of monolayer- and bilayer-like Landau levels\cite{1PCCP17;26008}, the irregular Landau-level energy spectra and the frequent anti-crossing $\&$ crossings come to exist in the ABC-stacked graphene systems\cite{1PRB90;205434}, the strongly, significantly and weakly $k_z$-dependent Landau subbands, respectively, appear in the AA-, AB- and ABC-stacked 3D graphites\cite{1CRCPress;CY}. The low-energy quasi-Landau-levels survive in the 1D graphene nanotubes\cite{1PCCP18;7573}, and the dispersionless Landau levels are absent in carbon nanotubes/carbon tori\cite{1PRB70;075411}. Such diverse Landau levels are only created by the significant single-2p$_z$ orbital hybridizations of carbon atoms. Furthermore, the  multi-orbital chemical bondings and/or the non-negligible spin-orbital interactions, which can create new categories of Landau levels, are frequently  revealed in the emergent layered materials, such as, few-layer silicene \cite{PRB94;205427,PRB97;125416}, germanene \cite{PRB88;085434,PRL110;197402,PRB91;035423} stanene \cite{IOP2017,PRB94;045410}, bismuthene \cite{NJP20;062001}, antimonene \cite{PRB98;115117}, GaAs \cite{SRp9;2332}, and MoS$_2$-related systems \cite{RSCAdv5;20858,PRB89;55316}. The above-mentioned Landau levels are drastically changed by a uniform perpendicular electric field [the layer-dependent Coulomb potentials], e.g., the induced Landau-level splittings, crossings, and anti-crossings. When a opposite external field, a uniform perpendicular magnetic field accompanied with a spatially modulated magnetic/electric one, is present in 2D systems, the non-uniform magnetic quantization will lead to the abnormal energy dispersions without the very high state degeneracy and the irregular wave functions\cite{1PRB83;195405}. The previous theoretical studies are focused on monolayer graphene, and the similar methods could be generalized to 2D emergent materials. The up-to-date calculations on the main features of Landau levels and Landau subbands are covered in the following chapters. Some magnetic issues, being associated with the  novel geometries and the adatom chemisorptions, are worthy of the systematic studies, e.g., the Landau-level characteristics in the 1D folded\cite{1NatComm5;3189}, curved\cite{1PCCP18;7573,1JPSJ81;064719} and scrolled\cite{1NanoLett9;3766} graphene nanoribbons\cite{1PCCP18;7573}, amorphous and defect-enriched graphene systems\cite{1PRB75;125408,1PRL106;105505}, adatom-adsorbed graphene systems\cite{1CRCPress;SY}. \\

Various types of experimental equipment can accurately examine and then identify the diverse magnetic quantization phenomena in layered materials. Their measurements cover the magneto-electronic energy spectra $\&$ Landau level/Landau subband wave functions\cite{1CRCPress;CY,1IOPBook;CY,1IOP;SC}, magneto-optical absorption spectra\cite{1IOPBook;CY}, Hall conductivities\cite{1PCCP19;29525}, magnetic-field-dependent specific heat\cite{1PRB54;2896,1PhysicaE11;356}, and magneoplasmon modes\cite{1ACSNano5;1026}. First, the energy-related and energy-fixed measurements of scanning tunneling spectroscopy [STS; details in Chap. 2.1] are, respectively, available in exploring the main features of Landau-level energy spectra\cite{1PRL94;226403,1NatPhys3;623,1PRL94;226403} and probability distributions\cite{1NatPhys10;815}. Most of the experimental verifications are conducted on the Former, while the opposite is true for the latter, e.g., the various $B_z$ dependencies for few-layer graphene systems with different stacking configurations\cite{1CRCPress;CY,1IOPBook;CY,1IOP;SC}. The direction identifications of the number of zero points and distribution symmetries are worthy of the systematic studies under the great enhancement in STS measurement techniques\cite{1PRL94;226403,1NatPhys3;623,1PRL94;226403}. Second, the transmission\cite{1PRL100;136403}, absorption\cite{1SCIENCE322;1529}, reflection\cite{1PRB15;4077} and Rayleigh\cite{1NanoLett7;2711} optical spectroscopies, accompanied by an external magnetic field, are very useful in examining the special structures and selection rules of electronic excitation spectra [details in Chap. 3.2]. Up to now, the high-resolution measurements have confirmed and the symmetric absorption peaks and the specific magneto-selection rule of $\Delta n = \pm 1$ [$n$ quantum number] for the well-behaved Landau levels in monolayer-like graphene systems\cite{1PRL100;136403}. However, the different excitation categories and the extra selection rules, which are, respectively, due to the multi-group Landau levels and the abnormal spatial distributions, require the further experimental test. Third, the Hall transport measurements are frequently utilized to investigate the normal and unusual quantum conductivities arising from the static scatterings of Landau-level states. Specifically, the experimental and theoretical studies are consistent with each other for monolayer\cite{1Science315;5817}, AB\cite{1Nature438;201} and ABC\cite{1NatPhys7;953} stacking graphene systems. For example, magnetic transport measurements. have verified the unconventional half-integer Hall conductivity $\sigma_{xy}=(m + 1/2)4e^2/h$, in which $m$ is an integer and the factor of 4 stands for the spin- and sublattice-induced Landau-level state degeneracy. This novel quantization is attributed to the quantum anomaly of $n= 0$ Landau levels corresponding to the Dirac point. Fourth, a delicate calorimeter, which is sensitive to the temperature and magnetic-field strength, can accurately identify the very high Landau-level degeneracy, a weak but significant Zeeman splitting, and few Landau levels nearest to the Fermi level. Such measurements on graphene-related materials are absent up to date. However, they have been done a 3D electron gas in GaAs-GaALAs\cite{1PRL54;1820,1PRB33;2893}, showing the non-monotonous $T$- and $B_z$-dependences. Finally, in general, the electronic Coulomb excitations could be measured by the electron energy loss spectroscopy (EELS) and the hard-X ray inelastic scatterings. The EELS measurements are very difficult to the collective excitations in the presence of a magnetic field, since the incident electron beam is easily perturbed by such field. The light scattering experiments are very suitable for the full exploration of magnetoplasmon modes, e.g., the clear verifications on the magnetoplasmon modes due to electron gases in the doped semiconductor compounds\cite{1Winfried}. \\

Up to now, there exist the effective-mass approximation\cite{1RevModPhys81;109,1Rep76;056503,1PRB80;165409,1PRB77;115313} and the generalized tight-binding model\cite{1CRCPress;CY,1IOPBook;CY,1IOP;SC}, being frequently utilized in exploring the rich magnetic quantization of layered materials. It should be noticed that the first-principles method can solve the electronic properties\cite{1PRB77;235430,1PRB75;041401}, but not the magneto-electronic ones, as a result of the enlarged unit cell by the vector potential [details discussed later in Chap. 3.1]. Such calculations could provide the reliable energy bands at the high symmetry points [e.g., the K and $\Gamma$ point] and thus the reliable hopping integrals related to the single- or multi-orbital hybridizations in chemical bonds by using the Wanner functions\cite{1PRB86;165108}. As to the low-energy approximation, this method is suitable in dealing with the magneto-electronic properties, if the layered systems have the simple and monotonous valence and conduction bands near the Fermi level, e.g., monolayer graphene\cite{1Science315;5817}, silicene\cite{1PRL109;055502} and phosphorene\cite{1SciRep5;12295,1PRB93;075408}, and  the bilayer AA- and AB-stacked graphene systems\cite{1PRB80;165409,1PRB77;115313}. The zero-field and magnetic Hamiltonian matrices have the same dimension, so the main features of Landau levels could be solved very quickly. However, it might create the high barriers in calculating the other essential physical properties [detailed comments in Chap. 3.1], e.g., magneto-optical absorption spectra\cite{1IOPBook;CY}, unusual Hall conductivities\cite{1SciRep5;12295,1PRB93;075408}, and magnetoplasmons\cite{1PRB78;085401,1PRB77;125417}. On the other side, the generalized tight-binding model covers all the intrinsic interactions, the intralayer and interlayer hopping integrals due to the single- or multi-orbital hybridizations and the spin-orbital couplings; that is, such model does not ignore the significant atomic and spin interactions, e.g., the complicated non-vertical interlayer hopping integrals in the trilayer ABC-stacked graphene\cite{1PRB90;205434}. Such interactions and the magnetic field (the composite field) are included in the numerical calculations simultaneously. The concise physical pictures could be proposed to clarify how many groups of the different sublattice-dominated Landau levels can be classified\cite{1CRCPress;CY,1IOPBook;CY,1IOP;SC}. Furthermore, the magneto-electronic energy spectra and the Landau-level wave functions are very useful in the further understanding of the other physical properties. \\

It is well known that the geometric symmetries are one of the critical factors in determining the fundamental physical, chemical and material properties. The condensed-matter systems, which are purely made up of carbon atoms, include the significant ${sp^3}$\cite{1Science302;425,1PRL93;245502,1JPCC115;2705,1ChemMater16;1786}, ${sp^2}$ \cite{1Science302;425,1JPCC115;2705,1ChemMater16;1786}, and ${sp}$ chemical bondings\cite{1Science302;425,1APL91;131906}, being, respectively, revealed as the 3D bulk, 2D  surface, and 1D  line structures, such as, diamond, graphene and carbon chain. Apparently, the diversified orbital hybridizations in C-C bonds originate from the four half occupied orbitals of ${[2s, 2p_x, 2p_y, 2p_z] }$. Among carbon allotropes, the open/closed surface geometries might appear in the planar/nonplanar forms, covering  graphites\cite{1PRSLSA106;749}, layered graphenes\cite{1Nature438;197}, graphene nanoribbons\cite{1NatComm5;3189} and quantum dots\cite{1NanoLett12;844} [3D-0D]/carbon nanotubes\cite{1Nature354;56}, C$_{60}$-related fullerenes\cite{1Nature363;685} and carbon onions \cite{1CPL305;225} [1D-0D]. These unusual materials possess the unique honeycomb lattices with the hexagonal symmetries; therefore, the arrangements of normal hexagons are expected to play important roles in diversifying the various essential properties. A perfect 2D graphene crystal, with three nearest neighbors, is responsible for the isotropic Dirac-cone band structures\cite{1Nature438;197}. Very interesting, the achiral or chiral arrangement on a cylindrical nanotube surface/two open boundaries of nanoribbon has been clearly identified to dominate the metallic or semiconducting behaviors across the Fermi level [Refs\cite{1Nature391;59,1PRL79;2093}/Refs\cite{1PRL82;3520}]. The similar diverse phenomena are verified/predicted to appear in bilayer graphene systems through the twisting\cite{1PRL121;037702,1PRB91;155428} or sliding effects\cite{1SciRep4;7509}. The former behave as a Moire superlattice, with a lot of carbon atoms in an enlarged unit cell of bilayer graphene, while the latter only have four ones being identical to those in AA or AB stackings [details in Chap. 9.1]. The distinct geometric structures of twisted bilayer graphenes are accurately confirmed through the high-resolution STM measurements [details in Chap. 2.1], e.g., the twisted angles between two honeycomb lattices  of ${\theta\,=1.4^\circ}$, 3.5$^{\circ}$, 6.4$^{\circ}$, 9.6$^{\circ}$ and 1.79$^\circ$\cite{1PRL109;126801}. In addition, the spatially resolved Raman spectroscopy also provides the geometric information, a bilayer system with a mixture of AB stacking and twisted structure\cite{1NanoLett12;3162,1NatComm5;5309,1PRL97;187401}. Simultaneously, the STS examinations  clearly identify three kinds of van Hove singularities in density of states, covering the V-shape across the Fermi level, shoulders, and logarithmically divergent peaks at the negative and positive energies. The measured results are consistent with the theoretical calculations of the first-principles method\cite{1PRL109;196802,1NanoLett14;3353} and tight-binding model\cite{1PRB85;195458,1PRB81;161405}. Such special structures indicate the linear and parabolic energy dispersions, as verified from the angle-resolved photo-emission spectroscopy\cite{1NatMat12;887}. The main features of geometric and electronic properties are directly reflected in optical absorption spectra\cite{1PRB87;205404} and quantum Hall transports\cite{1PRL108;076601}. Very interesting, the specific bilayer system, with a magic angle of ${\theta\,\sim\,1.1^\circ}$, is thoroughly examined and verified to exhibit the superconducting characteristics [almost vanishing resistance and diamagnetism] at the critical temperature of ${T_c=1.7}$ K\cite{1Nature556;43}. However, the systematic theoretical investigations on the essential properties are absent up to date. For example, the magneto-electronic states, being closely related to the unusual optical selection rules and Hall conductivities, need to be well characterized through the magnetic wave functions with the normal or irregular spatial oscillations. This will be completely explored by the generalized tight-binding model in Chap. 4. \\

In addition to the variations among the well-behaved stacking configurations\cite{1CRCPress;CY,1IOP;SC,1IOPBook;CY}, a creation of the non-uniform sublattices will lead to the dramatic changes in the fundamental physical properties, especially for the thorough transformation from the 2D to quasi-1D behaviors. The modulated geometries in few-layer graphene systems have been successfully synthesized using the STM tips\cite{1NatNanotech13;204}, mechanical exfoliation\cite{1Nature520;650}, and chemical vapor deposition\cite{1Science312;1191}. From the theoretical point of view, the stacking symmetries of bilayer graphene systems can be drastically altered by the significant sliding\cite{1SciRep4;7509}, twisting\cite{1PRB81;161405}, and generation of domain walls\cite{1SciRep9;859}. When only the low-lying electronic states due to the C-2p$_z$ orbitals are taken into account, the first and second systems are predicted to exhibit two pairs and many pairs of valence and conduction bands along the high symmetry points in the hexagonal Brillouin zone, respectively. Energy bands have the regular dispersion relations with the 2D wave vectors. Furthermore, their band-edges states possess the 2D-specified structures in van Hove singularities of density of states \cite{1SciRep9;859} and the similar ones in absorption spectral structures\cite{1SciRep9;859}. The magnetic quantization in the first [second] system creates the highly degenerate Landau states with the well-behaved, perturbed, or undefined oscillation modes [the normal ones]. Apparently, the mandatory modulation of stacking configuration, which are due to the changes of bond lengths, have generate a large unit cell and the non-uniform intralayer and interlayer hopping integrals. Even in the zero-field case, the dependence of essential properties on the wave vector along the modulated direction is expected to be negligible. That is, the quasi-1D phenomena come to exist in certain special cases, e.g., plenty of 1D energy subbands with the unusual dispersion relations, and many significant band-edge states and absorption structures\cite{1SciRep9;859}. A uniform magnetic field strongly affects the commensurate periods and the vector-potential-induced Peierls phases on the neighboring atomic interactions. As a result, the non-uniform magnetic quantization can create a plenty of the low-degenerate Landau subbands with the sufficiently wide oscillating energy widths and the strong couplings\cite{1PRB83;195405}, being in sharp contrast with the high-degenerate Landau levels. It is worthy of making a systematic comparison for the main features of Landau subbands in stacking-modulated bilayer graphene systems\cite{1SciRep9;859}, graphene nanoribbons\cite{1PCCP18;7573}, and Bernal and rhombohedral graphites\cite{1CRCPress;CY}. \\

Obviously, the 3D silicon crystals, with the ${sp^3}$ chemical bondings, have shown the critical roles in the basic sciences\cite{1Nature408;440}, applied engineering\cite{1Nature460;974,1Science257;1906}, and high-technique semiconductor applications\cite{1Science291;851}. From the experimental and theoretical viewpoints, the reduction in dimension should be a very effective way in diversifying the various     physical phenomena. The researches on layered silicene systems are getting into one of the main-stream topics in 2D emergent materials. In general, the epitaxial growth is frequently utilized to synthesize monolayer and bilayer silicene systems on the different substrates, e.g., the former grown on Ag(111)\cite{1NanoLett12;3507,1PRL108;155501}, Ir(111)\cite{1NanoLett13;685} and ZrB$_2$(0001)\cite{1PRL108;245501}, with the ${4\times\,4}$, ${\sqrt 3\times\,\sqrt 3}$ and ${2\times\,2}$ unit cells, respectively. The STM [details in Chap. 2.1] and low-energy electron diffraction can identify a buckled single-layer honeycomb lattice, as well as an enlarged unit cell arising from the significant atomic interactions between the host and guest atoms. Such geometry clearly indicates the dominating ${sp^2}$ bonding and its competition with the weak, but important ${sp^3}$ one. The phenomenon of orbital hybridizations are further reflected in the low-lying valence and conduction bands. The former, which possesses a slightly separated Dirac-cone structure [approximately a zero-gap semiconductor; the monolayer-graphene-like behavior] is confirmed through the high-resolution angle-resolved photoemission spectroscopy [ARPES; \cite{1PRL108;155501}].   However, it is very difficult to successfully generate the pure bilayer silicene systems through the various experimental methods. The previous study   shows that there exist three types of bilayer silicenes after treating the calcium-intercalated monolayer silicene [CaSi$_2$] around a BF${_4^{-}}$-based ionic liquid. That is to say, the bilayer silicenes might consist of the four-, five- and six-membered silicon rings with three nearest neighbors\cite{1NanoLett12;3507}, where the third ones correspond to the AB-bottom-top stacking configuration [AB-bt]. In addition, the experimental evidences of AA stackings are absent up to now. Such images are accurately achieved by a high-angle annular dark field scanning transmission electron microscopy [HAADF-STEM]. Moreover, the measured absorption spectrum suggests an indirect optical gap of ${1.08}$ eV [or a threshold excitation frequency]. Energy gaps are quite different from in monolayer and bilayer material; therefore, the stacking configurations, the interlayer hopping integrals, and the spin-orbital couplings play a critical role in the fundamental physical properties.  Most important, the multi-metastable geometric structures come to exist simultaneously, clearly illustrating the very active chemical environments due to the four half-occupied orbitals of silicon atoms [${(3s, 3p_x, 3p_y, 3p_z)}$]. Few-layer silicene systems are expected to become the 2D materials in the next generation of electronic devices\cite{1NatNanotechnol10;227}, while their structural instabilities need to be overcome for the highly potential applications. \\

Also, a lot of theoretical predictions on 2D layered silicene materials present very interesting results, especially for the diversified physical phenomena. Generally speaking, their essential properties are thoroughly investigated by the first-principles calculations, ${\it ab}$ ${\it initio}$ molecular dynamics, the tight-binding model and the effective-mass approximation. The optimal geometric structures are accurately presented through the first and second methods. For example, monolayer silicene has a buckled honeycomb lattice, with a significant height difference between A and B sublattice\cite{1NanoLett12;3507}, and bilayer ones possesses the distinct members in silicon rings and the various stacking configurations\cite{1NanoLett12;3507}, such as, the metastable configurations of AA-bt, AA-bb, AB-bt and AB-bb in bilayer materials. Specifically, the Vienna  \emph{ab initio} simulation package is quite powerful in predicting electronic properties, covering band structures, spatial charge distributions, atom-, orbital- and spin-projected density of states, magnetic moments, and spin distribution configuration. For example, the chemisorption-diversified fundamental properties are clearly revealed in the hydrogenated and oxidized silicene systems\cite{1JPCC116;22916,1JPCC116;4163}. For pristine bilayer silicenes, their band properties across the Fermi level are predicted to be very sensitive to the stacking configurations, e.g., a semimetal with some free carrier densities\cite{1PRB88;245408}, and a finite indirect-gap/direct-gap semiconductor [${\sim,0.5-1.0}$ eV\cite{1NanoLett12;3507,1PRB97;125416}, being in sharp contrast with a very narrow gap  of monolayer system [${\sim,5}$ meV; \cite{1PRL108;155501}]. On the other hand, the third and fourth methods are quite efficient in studying the diverse physical phenomena after getting a set of reliable parameters for all the intrinsic interactions. It should be noticed that an outstanding fitting in non-monolayer silicene systems, which is done by a detailed comparison with the first-principles calculations, will become a giant barrier. The more complex interlayer hopping integrals and enhanced spin-orbital couplings, due to the rich bucklings and stackings, are very difficult to obtain even for bilayer silicenes. Up to date, the effective-mass can only deal with the fundamental properties for monolayer silicene under the external electric and magnetic fields, covering the in-orbital-induced slight separation of Dirac points, gate-voltage-created state splitting\cite{1PRB97;125416}, and the magnetic-field-generated degenerate Landau levels and specific selection rule\cite{1PRB97;125416}. There are no excellent parameters for two pairs of low-lying valence and conduction bands in AA¡V and AB-stacked systems. However, the first pair nearest to the Fermi level is well fitted for the non-monotonous energy dispersions and the irregular/normal valleys. Apparently, the diversified Coulomb excitations and magnetic quantizations are effectively solved by the generalized tight-binding model, e.g., the diverse [momentum, frequency]-phase diagrams in monolayer silicene under the composite effects of spin-orbital interactions, electric potentials and magnetic fields, the inter-Landau-level and electron-gas-like magnetoplasmon modes  with  the rather strong electron-hole dampings for the same system\cite{1PRB94;205427}, and the spin- and valley-split Landau levels of bilayer silicenes in the frequent non-crossings, crossings and anti-crossings\cite{1PRB97;125416}. The systematic investigations will be made on the fundamental properties of AA- and AB-stacked bilayer systems. \\

The chemical modifications, as revealed in the theoretical and experimental researches\cite{1PCCP17;992,1Catal2;54,1arXiv1806;0529}, are one of the most efficient methods in dramatically changing the essential properties of condensed-matter systems. They are very effective for the layered materials, since such systems can provide the outstanding chemical environments. For example, there are a lot of dangling C-2p$_z$ orbitals on the planar structures of few-layer graphene systems. Recently, the adatom chemisorptions\cite{1PCCP17;992,1Catal2;54} and the guest-adatom substitutions\cite{1arXiv1806;0529} in the emergent 2D materials are easily achieved during the experimental growths. Apparently, how to generate the uniform adatom/guest-atom distributions in experimental laboratories is the studying focus/issue. Up to now, many graphene-related binary and ternary compounds have been successfully synthesized using various methods. Among them, the silicon carbide compounds, with the sp$^2$ or sp$^3$ multi-orbital chemical bondings, frequently appear in the distinct dimensions, e.g., the 3D  bulk compounds\cite{1JPCC110;10645,1CST67;2390,1ChemMater22;2790}, the 2D systems\cite{1Adv16;561}, and the 1D nanotube ones\cite{1APL85;2932,1APL89;013105}. Similarly, there also exist the dimension-dependent $B_x C_y N_z$ compounds. The 3D silicon carbide systems possess the sp$^3$ multi-orbital hybridizations, mainly owing to the four half-occupied orbitals of $(3s, 3p_x, 3p_y, 3p_z)$ and $\&$ $(2s, 2p_x, 2p_y, 2p_z)$ orbitals, Such chemical bondings are similar to those in diamond\cite{1NanoLett7;570}, so the SiC materials have the outstanding mechanical properties, as observed in the consistence between the calculated results \cite{1PRB81;174301,1PRB53;4458} and experimental measurements\cite{1ComStruc67;115}. Such SiC-related materials have presented the high potentials in applications, such as, the energy storage\cite{1NanoLett6;1581}, and the semi-conducting optical devices\cite{1APL82;3107}. Recently, the planar silicon-carbon compounds have been synthesized in the experimental laboratories\cite{1NatMater6;479}. The theoretical predictions under the first-principles calculations clearly show that these 2D compounds display the sp$^2$ and $\pi$ chemical bondings \cite{1PRB82;035431}. The latter will be responsible for the low-energy essential physical properties. Specifically, the $\pi$ bondings, which mainly originate from the 2p$_z$-2p$_z$ hybridization in C-C bond and the 2p$_z$-3p$_z$ mixing in C-Si one, is expected to be significantly distorted by the Si-substitutions. These are very sensitive to the concentrations and distributions of the Si-guest atoms, and so do the other physical, material and chemical phenomena \cite{1PRB82;035431}. It is very interesting to fully comprehend the close relations between the substitution-enriched chemical bondings and the magnetic quantization behaviors. The non-uniform ionization potentials, C-C $\&$ Si-C bond lengths and hooping integrals are deduced to be the critical factors in determining how many kinds of Landau levels and magneto-optical selection rules can be created during the variation of the Si-substituted geometric structures\cite{1CRCPress;CY,1IOP;SC,1IOPBook;CY}. \\

For the quantum transport properties, the 2D graphene-related systems are very suitable for fully exploring the diverse magnetic quantization phenomena. Apparently, such layered materials sharply contrast with the doped semiconductors/2D electron gases revealing the integral and fractional Hall effects\cite{1RMP67;357}. The previous investigations clearly show that honeycomb lattices\cite{1Science315;5817}, few layers\cite{1Nature438;201,1NatPhys7;953}, stacking configurations\cite{1NatPhys7;953}, planar/rippled/curved/folded structures\cite{1PRB85;195458}, negligible spin-orbital interactions \cite{1PRL111;136804}, dimensions\cite{1PRL106;126803}, nanodefects\cite{1PRB70;041303}, chemical absorptions\cite{1JPCC117;6049}, and guest-atom substitutions\cite{1arXiv1806;0529} are responsible for the rich band structures and thus the unusual magneto-electronic and transport properties. The close and complex relations between the critical factors and the diversified Hall conductivities are the studying focus of Chap. 9. Up to date, the experimental measurements and theoretical predictions [the tight-binding model and the effective-mass approximation have delicately identified the rich characteristics of quantum Hall conductivities in monolayer\cite{1Science315;5817}, bilayer\cite{1Nature438;201} $\&$ trilayer AB stackings\cite{1NatPhys7;953}, and trilayer ABC configuration\cite{1NatPhys7;953} under a low Fermi level and magnetic-field strength. For example, there are one, two $\&$ six, and three Landau levels at/across the Fermi energy, leading to the anomalous structures there in terms of height, width and sequence of quantum plateaus\cite{1Science315;5817,1Nature438;201,1NatPhys7;953}. This clearly illustrates the significant roles due to the well-defined modes, state degeneracies, and normal energy spectra. Obviously, the more complicated transport phenomena, which might be created by the non-well-behaved magneto-electronic states and the frequently crossing $\&$ anti-crossing behaviors, deserve a systematic investigation. The sliding bilayer graphenes and AAB trilayer stacking will be chosen for a model study, in which the former have the structural transformation between two high-symmetry configurations\cite{1SciRep4;7509,1Carbon94;619}, and the low-symmetry latter presents the very complex interlayer hopping integrals\cite{1SciRep4;7509,1Carbon94;619}. These systems are predicted to exhibit the diversified energy bands and magneto-electronic properties, such as, three kinds of Landau levels [the well-behaved, perturbed and undefined ones] with the complex $B_z$-dependent energy spectra. To propose the completely physical pictures, the quantum Hall effects are conducted on the trilayer AAA, ABA, ABC and AAB stacking. On the theoretical side, the linear static Kubo formula is directly combined with the generalized tight-binding model [discussed later in Chap. 3.3.1], based on the model consistence between the gradient approximation in evaluating the electric dipole moment and the sublattice-defined subenvelope functions. Such direct association is expected to be very efficient in solving the open transport issues due to the abnormal Landau-level wave functions and energy spectra\cite{1SciRep4;7509,1Carbon94;619}. That is to say, the developed theoretical framework is suitable and reliable for any kinds of Landau levels, e.g., the quantum transports arising from  the well-behaved, perturbed, and undefined oscillation modes. In addition, the effective-mass perturbation method might be quite difficult in exploring the irregular magneto-electronic states and even the higher-energy/strong-field transport properties of the well-behaved Landau levels, mainly owing to the complicated effects coupled by the intrinsic interactions and the magnetic field. \\

The magneto-electronic specific heats of conventional quantum Hall systems are very suitable for fully understanding the diverse magnetic quantization phenomena. They have been extensively investigated both theoretically\cite{1PRL73;2744} and experimentally\cite{1PRL59;1341,1NatNanotechnol6;418,1PRL54;1820}. For example, a lot of researches on GaAs¡VGaAlAs\cite{1PRL54;1820,1PRL59;1341} show that the thermal properties are very sensitive to the temperature and magnetic-field strength. The temperature-dependent specific heat reveals an obvious peak structure at a low critical temperature\cite{1PRL54;1820,1PRL59;1341}. Furthermore, the magnetic-field dependence exhibits an oscillation behavior\cite{1PRL54;1820,1PRL59;1341}. These two
kinds of thermal behaviors directly reflect the main features of Landau levels, such as, the filling factor and energy spacing. Also, the high-resolution calorimeter measurements are conducted on the graphene-related materials, such as 1D carbon nanotubes\cite{1Polymer45;4227} and 3D graphite systems\cite{1ACSAppl1;2256}. The former clearly display a linear temperature-dependence in ${C(T)}$, and so do for the latter. These results are deduced to be closely related to the 1D and 3D metals. This book will illustrate the rich thermal excitation phenomena on monolayer graphene, e.g., their unique dependences on the Zeeman splitting energy, doping level of  conduction electrons/valence holes, temperature, and magnetic field. \\

The 2D emergent materials are ideal condensed-matter systems in thoroughly exploring the many-particle Coulomb excitations under the magnetic quantization. The strong competition in magneto-excitation spectra, that due to the longitudinal electron-electron interactions and the transverse cyclotron forces, is a very interesting research focus. Recently, the various experimental methods are delicately developed to successfully synthesize a lot of layered materials, e.g., few-layer germanene, silicene and tinene being, respectively, grown on [Pt(111), Au(111) $\&$ Al(111)]\cite{1ACSNano11;3553,1NJP16;095002,1NanoLett15;2510}, [Ag(111), Ir(111) $\&$ ZrB$_2$]\cite{1NanoLett12;3507,1NanoLett13;685,1PRL108;155501,1PRL108;245501} and Bi$_2$Te$_3$ surfaces. Apparently, the obvious buckled structures, the important orbital hybridizations, the significant spin-orbital interactions, and the external electric and magnetic fields would play critical role in the essential physical properties, especially for their strong effects in creating the diverse magneto-electronic excitation. The previous theoretical studies, which are focused on monolayer graphene and bilayer AA and AB stackings\cite{1IOP;SC,1CRCPress;CY,1IOPBook;CY} clearly show the rich and unique phenomena: a lot of inter-Landau-level single-particle and collective excitations, the unusual magnetoplasmon modes with the  non-monotonic momentum dispersion relations, the very strong dependences on the temperature, doping level and interlater atomic interactions, the stacking-enriched [momentum, frequency]-phase diagrams, and the 2D electron-gas behavior only under a sufficiently high free carrier density. On the experimental side [details in Chap. 2.4], the high-resolution electron energy loss spectroscopy is very powerful in the full examinations of the carbon-orbital-\cite{1PRB83;161403}, dimension-\cite{1Carbon50;183}, layer-\cite{1PRB72;125117},  stacking- \cite{1PRB72;125117}, doping-\cite{1APL99;082110}, and temperature-diversified plasmon modes\cite{1NatNanoTech6;630} . Up to now, most of the measurements are consistent with the calculated results\cite{1PRB74;085406}. A thorough investigation on monolayer germanene is expected to present the distinct magneto-electronic excitation phenomena, since its band structure might be quite differ from those of graphene and silicene systems\cite{1SciRep7;40600}. The composite effects arising from the significant spin-orbital coupling, free carrier doping, static Coulomb on-site energies, and magnetic field would play critical roles in diversifying magnetoplasmons. \\

As the final topic, a topological concern is addressed in view of the extensively and intensively studied topological phases of condensed matter\cite{volovik03,chiu16,wen17}. In this realm, the phases with topological order cannot be characterized by symmetry alone. Moreover, the relevant phase transitions do occur without spontaneous symmetry breaking, beyond the scope of Landau's theory\cite{landau58}. The first realization of such phases is the discovery of the integer quantum Hall effect (QHE)\cite{klitzing80}, which was followed soon by a topological interpretation\cite{thouless82,avron83,simon83}. Later on, a distinct, half-integer QHE was also found from graphene, which has almost spin degeneracy described by $SU(2)$ symmetry\cite{novoselov05,zhang05}. The previous theoretical predictions\cite{zheng02,gusynin05} were realized in this finding. It has been well understood that the anomaly of this half-integer QHE originates from the presence of $2$D massless Dirac fermions around the zero energy with respect to the original Dirac points (DPs). As described in Chap. 3.3, the Hall conductivity and the associated Landau levels (LLs) are of the relativistic type in graphene. The very characteristic lies in that there exists a topologically robust zero-mode LL that is a constant function of the perpendicular magnetic field. More deeply, this zero-mode LL is protected by the local chiral symmetry (CS), against CS preserving perturbations provided that intervalley scattering between the double DPs is inhibited\cite{ostrovsky08,kawarabayashi09,konig12}, where the CS arises from the global sublattice symmetry in spinless graphene. Since massless Dirac particles are broadly present in condensed matter with various symmetries, not to mention Dirac bosonic systems, it is of interest to see how about the situations in other systems with $2$D massless Dirac fermions\cite{kopelevich03,kempa06,xu14,yoshimi15}. In practice, one can use such a zero-mode LL to signify the existence of $2$D massless Dirac fermions. This topic is central to Chap. 11 for the subject material of this book, that is, $3$D layered systems. \\

In addition to the interesting topics in this work, the open issuers, being clearly illustrated through a final chapter [details in Chap. 13], are proposed to promote the understanding of magnetic quantizations and explore the other diverse phenomena in 2D emergent materials. Certain main-stream researches might be very difficult to thoroughly investigate their main features according to the current methods/the present calculations, mainly owing to the complicated geometric structures with the non-uniform/quasi-random chemical environments, or a lot of important intrinsic interactions [such as, the significant couplings of multi-orbital hybridizations and spin-orbital interactions]. For example, a single-layer graphene could present the vacancy and Stone-Wales defects\cite{1PRB80;033407}; that is, a defective and amorphous system possesses hexagons, pentagons and heptagons. It is deduced that non-homogenous ${sp^2}$-bonding network can create the great modifications on the basic magnetic properties, e.g., the strong dependences of energy spectra and wave functions on the concentrations and distribution configurations of various defects, and the close relationships of the defect-created sublattices and the vector-potential-dependent Peierls phases. Whether the highly degenerate Landau levels appear in the seriously deformed/curved surfaces is one of the main-stream topics. The folded graphene nanoribbons\cite{1NatComm5;3189}, curved ones\cite{1ACSNano4;1362}, carbon nanoscrolls\cite{1NanoLett9;3766}, carbon toroids\cite{1PRB70;075411}, and carbon onions\cite{1CPL305;225} are very suitable for the model studies on the curvature effects [the misorientation of carbon-${2p_z}$ orbitals and hybridization of $\pi$ and $\sigma$ electrons], critical roles of the position-dependent hopping integrals, the  strong competition of the quantum-confinement-induced standing wave functions with the magnetically  localized ones, and the diamagnetic, paramagnetic, ferromagnetic and  anti-ferromagnetic behaviors\cite{1NatComm5;3189,1ACSNano4;1362,1NanoLett9;3766,1PRB70;075411,1CPL305;225}. The more complex magnetic quantization phenomena will frequently come to exist in the adatom-chemisorption/substitution layered materials [the chemically modified systems]. The previous first-principles calculations show that the halogenated\cite{1JMCC3;3087}, hydrogenated and oxidized graphenes/silicenes \cite{1JPCC116;22916,1JPCC116;4163} exhibit the main characteristics of Moire superlattices in the presence of multi-orbital chemical bondings. The superposition of guest and host sublattices should be the critical mechanism responsible for the localized magneto-electronic states; therefore, plenty of non-equivalent sublattices can take part in the highly irregular magnetic quantizations. Apparently, the above-mentioned geometry- and chemical-modification-dominated novel materials could serve as the typical models in fully exploring the diversified optical properties\cite{1IOPBook;CY}, quantum Hall transports \cite{1PCCP19;29525}, and Coulomb excitations\cite{1PRB74;085406} under any external fields. The studying focuses might cover the different magneto-optical selection rules of Landau levels or subbands due to the combined effects of dimensions, edge structure, curvatures and external fields, the integer or      non-integer Hall conductivities with the regular/abnormal dependences on the magnetic-field strength and the Fermi energy\cite{1Science315;5817}, and the [momentum. frequency]-phase diagrams of the inter-Landau-level magnetoplasmon modes and electron-hole excitations\cite{1PRB78;085401}. Also noticed that the theoretical frameworks of the different physical/chemical/material properties, belonging to the single- and many-particle models, might need to  make the great modifications for fully understanding their diverse phenomena. Among them, the suitable Hamiltonians with any number of layers, which clearly show the dramatic transformation in multi-orbital hybridizations from  few-layer  ${sp^2}$-bonding silicenes to 3D diamond-like silicon, are one of the challenges.  \\

This book covers Chaps. 1-10 for fully exploring the diverse magnetic quantization phenomena. Chapter 2 discusses the various high-resolution experimental measurements, including scanning tunneling microscopy [STM], scanning tunneling spectroscopy [STS], optical absorption reflection/transmission spectroscopies, transport/thermal instruments, and light inelastic scattering spectroscopy. The theoretical frameworks, the generalized tight-binding model, its combinations with the static and dynamic Kubo formulas, and the modified random-phase approximation, are systematically developed in Chap. 3. They are, respectively, responsible for the magnetically quantized electronic states, the unusual Hall conductivities, the magneto-optical selection rules, and the magneto-electronic excitations. Chapter 4 will thoroughly investigate the diversified magneto-electronic states and optical properties of bilayer graphene systems due to the very strong cooperations/competitions between the twist-induced Moir\'{e} superlattice and the enlarged unit cell by the vector potential. The modulation of stacking configuration, which is created by the domain walls in one graphene layer, is clearly illustrated for the AB-stacked bilayer graphene in Chap. 5. Its significant effects on the electronic and optical properties are studied in detail, and furthermore, the detailed comparisons with the sliding and twisted bilayer graphene systems, graphene nanoribbons, and Bernal and rhombohedral graphites are also made. Chap. 6 fully explores the rich and unique phenomena in an AA-bt bilayer silicene with the very strong interlayer  hopping integrals and the negligible  spin-orbital couplings; furthermore, the significant differences two- and single-layer systems
are made in detail. The composite effects, due to the intrinsic atomic interactions and the magnetic $\&$ electric fields, on the essential properties are the studying focuses through the proposed viewpoint: the valley structures in electronic energy spectra will play critical roles in the fundamental physical properties, such as the diverse magnetic quantizations due to the different valleys in monolayer and bilayer silicene systems. The layer-dependent spin-orbital interactions, as reported in Chap. 7, will come to exist in AB-stacked silicene systems. Apparently, the more complicated effects arising from stacking configuration and buckling structure are expected to diversify the magnetic quantization. Chapter 8 clearly illustrate the dramatically chemical modifications of the fundamental properties by using the silicon guest-atom substitution in monolayer graphene. Four kinds of Landau levels and magneto-optical selection rules will be classified according to the main features of magneto-electronic properties and absorption transition channels. Few-layer graphenes in Chap. 9, with the distinct layer numbers and stacking configurations, can present the rich and unique quantum transport properties, the unusual Hall conductivities and magneto-heat capacity. The available scattering events are clarified by delicately analyzing the dipole matrix elements associated with the initial and final Landau-level-state spatial distributions. The magneto-electronic specific heat of monolayer graphene provides another magnetic quantization phenomenon, especially for its rich dependences on the Dirac cone/the unusual Landau-level energy spectrum, Zeeman splitting effect,  temperature and magnetic-field strength. The superposition of the transverse magnetic force and the longitudinal Coulomb one leads to the abnormal electronic excitations, where germanene, silicene and graphene will exhibit the diverse magnetoplasmon modes induced by the significant/negligible spin-orbital couplings, as shown in Chap. 10. Chap. 11 covers the topological characterization for the Landau levels in $3$D layered systems of topological matter. The focus is put, in a topological viewpoint, on the existence and stability of $2$D massless Dirac fermions when an external perpendicular magnetic field is applied. If this is the case, the half-integer QHE can be achieved and dictated by the topologically robust zero-mode LL in the relativistic LL spectrum, as having been understood. This is distinct from the nontrivial topological phase of the normal integer quantum Hall effect (QHE) occurring at any $2$D plane of a topological trivial system. Henceforth, Two systems are considered after introducing the relevant topological background. Chap. 11.1 aims at certain $3$D layered spinless nodal-line topological semimetals (TSMs), exemplified by rhombohedral graphite (RG), a $3$D stack of graphene layers in ABC configuration ($SU(2)$ symmetry assumed), where $2$D massless Dirac fermions are hosted along the nodal lines in the bulk. This topological phase is shown here to be protected by the local chiral symmetry, which arises from the global sublattice symmetry. A comparison is made to graphene, including the $3$D half-integer QHE in RG v.s. the $2$D half-integer QHE in graphene. In Chap. 11.2, time reversal invariant spinful strong topological insulators (TIs), such as Bi$_{2}$Se$_{3}$ and Bi$_{2}$Te$_{3}$, are covered (spinful with $SU(2)$ symmetry broken). There are gapless surface states corresponding to the nontrivial insulating phase of the bulk. The surface state on each surface layer is manifested as a single Dirac cone, where $2$D massless Dirac fermions are hosted owing to the presence of the time reversal symmetry (TRS). One could by intuition think of gapping the surface DCs with some means for violating the TRS. Nevertheless, as known from existing experiments, this thought is true in the case of magnetic doping or a proximate magnetic material while being false in the case of applying an external perpendicular magnetic field. Thus, the zero-mode LLs still signify $2$D massless Dirac fermions in these strong TIs. This odd fact motivates the content of Chap. 11.2. Finally, Chaps. 12 and 13, respectively, present the concluding remarks, and future perspectives $\&$ open issues. \\

\newpage
\begin{center}
\section{Experimental measurements on magnetic quantization}\label{ch2}
\author{ Shih-Yang Lin,\textit{$^{e,\S}$} Thi-Nga Do,\textit{$^{b,c,\ast}$} Chiun-Yan Lin,\textit{$^{a}$} Jhao-Ying Wu,\textit{$^{d,\dag}$}\\ Po-Hsin Shih,\textit{$^{a,\ddag}$} Ching-Hong Ho,\textit{$^{d,\P}$} Ming-Fa Lin\textit{$^{a,\sharp}$}}
\end{center}
\vskip 1.0 truecm

High-resolution experimental equipment  has now been developed to fully explore the novel physical phenomena in the emergent layered materials. In general, they can measure the fundamental quantities very accurately and efficiently, covering the geometric, electronic, optical, transport, thermal and Coulomb-excitation properties. Specifically, the observed phenomena will become complex and diverse in the presence of a uniform perpendicular magnetic field. The greatly diversified magnetic quantization is clearly revealed in the distinct physical properties, as thoroughly discussed in the following sections associated with the various experimental measurements.\\

\subsection{Scanning tunneling spectroscopy}\label{ch2.1}

Scanning tunneling microscopy has become one of the most important experimental techniques in resolving the diversified nanostructures since the first discovery by Binnig and Rohrer in 1982\cite{2IBM30;355}. STM is capable of revealing the surface topographies in the real space with both lateral and vertical atomic resolutions, e.g., the nanoscaled bond lengths, crystal symmetries, planar/nonplanar geometries, step edges, local vacancies, amorphous dislocations, adsorbed adatoms $\&$ molecules, and nanoclusters $\&$ nanoislands\cite{2IBM30;355}. The STM instrument is composed of a conducting (semi-conducting) solid surface and a sharp metal tip, where the distance of several angstroms is modulated by piezoelectric feedback devices. A very weak current will be generated from the quantum tunneling effect in the presence of a bias voltage between surface and tip. The tunneling current presents a strong dependence on their distance; furthermore, the relation roughly agrees with the exponential decay form. It flows from the occupied electronic states of tip into the unoccupied ones of the surface under the positive bias voltage $V>0$, and  vice versa. This current is required to serve as a feedback signal, so that the most commonly used mode, a constant tunneling one (current and voltage), will be operated to resolve the surface structure very delicately by using a piezoelectric device. By combining the tip of a metal probe and a precise scanning device, the spectroscopic information of surface morphology is thoroughly revealed at the preselected positions with the well-defined conditions. The structural response hardly depends on the background effects, being attributed to an ultra-high vacuum environment. The up-to-date spatial resolution of STM measurements is $\sim 0.1$ $\AA$. The rather accurate measurements on the graphene-related systems have clearly illustrated the complicated relations among the honeycomb lattice, the finite-size quantum confinement, the flexible feature, and the active chemical bondings of carbon atoms. For example, the experimental identifications on the rich and unique geometric structures cover the achiral (armchair and zigzag) and chiral graphene nanoribbons with the nanoscale-width planar honeycomb lattices\cite{2APL110;051601,2JACS137;6097}, the curved\cite{2ACSNano4;1362}, folded \cite{2NatComm5;3189}, scrolled\cite{2NanoLett9;3766} and stacked graphene nanoribbons \cite{2NanoLett9;3766}, the achiral and chiral arrangements of the hexagons on cylindrical surfaces of carbon nanotubes\cite{2NANOTECHNOLOGY27;055602}, the AB\cite{2APL107;263101}, ABC\cite{2ACSNano9;5432,2APL107;263101,2PRB91;035410} and AAB stackings\cite{2PRB75;235449,2PRB48;17427} in few-layer graphene systems, the twisted bilayer graphenes\cite{2PRB91;155428,2PRL109;126801}, the corrugated substrate and buffer graphene layer\cite{2ACSNANO9;8997}, the rippled structures of graphene islands\cite{2PRL100;056807,2PRB77;235412}, the adatom distributions on graphene surfaces\cite{2PRL110;136804}, and  the 2D networks of local defects on graphite surface\cite{2Carbon50;4633,2ChemPhys348;233}. \\

Scanning tunneling spectroscopy (STS), an extension of scanning tunneling microscopy in Fig. 2.1, creates the tunneling current through the tip-surface junction under a constant height mode [details in Ref \cite{2IBM30;355}]. The electronic properties are characterized by both $I$-$V$ and $dI/dV$ curves sweeping over the bias voltage $V$. In the STS measurements, the normalized differential conductance, being created by a small AC modulation of $dV$, could be measured by adding a lock-in amplifier. Furthermore, the noise involved in the observed conductance is largely reduced. The up-to-now experiments present the highest resolution of $\sim 10$ pA. STS is a very efficient method for examining the electronic energy spectra of condensed-matter systems. The tunneling differential conductance is approximately proportional to the DOS and directly presents the main features, being useful in identifying the semiconducting, semi-metallic or metallic behaviors. For example, a zero-gap semiconducting monolayer graphene presents a V-shape DOS with a vanishing value at the Fermi level (the Dirac point) \cite{2PRL102;176804}. The semi-metallic ABC-stacked graphene systems, a pronounced peak at the Fermi level is revealed in tri-layer and penta-layer cases as a result of the surface-localized states\cite{2PRB91;035410,2ACSNano9;5432}. The high-resolution STS could serve as a powerful experimental method for investigating the magnetically quantized energy spectra of layered graphenes. The measured tunneling differential conductance directly reflects the structure, energy, number and degeneracy of the Landau-level delta-function-like peaks. Part of the theoretical predictions on the Landau-level energy spectra have been confirmed by the STS measurements, such as the $\sqrt{B_z}$-dependent Landau-level energy in monolayer graphene\cite{2Science324;924,2PRB91;115405}, the linear $B_z$-dependence in AB-stacked bilayer graphene\cite{2PRB91;115405}, the coexistent square-root and linear $B_z$-dependences in trilayer ABA stacking\cite{2PRB91;115405}, and the 2D and 3D characteristics of the Landau subbands in Bernal (AB-stacked) graphite\cite{2PRL94;226403,2NatPhys3;623}. Specifically, STS is suitable for identifying the spin-split density of states, when the magnetic tips are utilized in the experimental measurements. The spin-polarized STS (SPSTS) is very powerful in fully comprehending the magneto-electronic properties, e.g., the spin-split Landau levels/Landau subbands due to the significant spin-orbital couplings and the gate voltage (perpendicular electric field, e.g., those in silicene and germanene systems [detailed in Chaps. 6 and 9]. \\

The high-resolution STS measurements under the specific energies can directly map the spatial probability distributions of wave functions in the absence and presence of a uniform perpendicular magnetic field. Up to now, they have confirmed the rich and unique electronic states in graphene-related systems. For example, the normal standing waves along the tubular axis, with the finite zero points, are clearly identified to survive in finite-length carbon nanotubes\cite{2Nature391;59}. The topological edge states are created by the AB-BA domain wall of bilayer graphenes\cite{2PRL121;037702}. Moreover, the well-behaved Landau-level wave functions, being similar to those of an oscillator, come to exist in few-layer graphene systems wit the distinct stacking configurations, such as, the $n^{c,v}=0$, and $\pm 1$ Landau levels for the monolayer and AB-stacked bilayer graphenes. The higher-$n$\cite{2CRCPress;CY}, undefined\cite{2SciRep4;7509}, and anti-crossing Landau levels\cite{5PRB90;205434} are worthy of a series of experimental examinations. The similar STS measurements are available in testing the theoretical predictions on the non-uniform magnetic quantization, establishing the direct relations between the band-edge state energies and the distribution range of the regular or irregular wave functions quantization, e.g., the main features of  Landau subbands in Chap. 5]. \\

\subsection{Magneto-optical spectroscopies}\label{ch2.2}

Optical spectroscopies are very power in characterizing the absorption\cite{2PRB81;155413}, transmission\cite{2PRL106;046401} and reflectance\cite{2APPOPT48;5713} of any condensed-matter systems, especially for the diverse optical properties of emergent layered materials. They can provide the rather sufficient information on the single-particle and many-body vertical optical excitations, directly reflecting the main features of energy bands and strongly coupled excitons [electron-hole pairs attracted by the longitudinal Coulomb interactions; \cite{2PRL106;046401,2PRB81;155413}], e.g., the experimental examinations on the predicted energy gap, the threshold excitation frequency, the band-edge state energies, the special absorption structures, the electric-field effects, and the magneto-optical selection rules. Absorption spectroscopy, which is based on the analytical method of measuring the fraction of incident radiation absorbed by a sample, is one of the most versatile and widely used techniques in the sciences of physics, chemistry and materials. This well-developed tool is available within a wide frequency range. The measured spectral functions are frequency-dependent functions for characterizing the fundamental electronic properties of materials. The experimental setup related to the light source, sample arrangement and detection technique is sensitive to  the frequency range and the experimental purpose. \\

The most common optical setup, as clearly shown in Fig. 2.2, is to emit a radiation beam on the specific sample and then delicately measure the transmitted intensity that is responsible for the significant frequency-dependent absorption spectrum. In general, optical experiments are performed by a broadband light source, where the intensity and frequency could be adjusted within a broad range. According to the frequency range of optical operations, the most utilized light sources are classified into three kinds: (${\bf I}$) the xenon-mercury arc lamp under the high pressure in the far-infrared region\cite{2ANALCHEM48;528,2RNPFEB;33}, (${\bf II}$) the black-body source of  the heated SiC element in the mid- to near-infrared spectral range\cite{2JMATSCI36;137,2CPL22;1806,2JRNBS59;405} and (${\bf III}$) the tungsten halogen lamp in a continuous spectrum from the visible to near-ultraviolet\cite{2APPOPT54;4447,2APPOPT54;2289}. These optical sources are commonly adopted for the analytical characterizations of optical properties in emergent materials over a broad spectrum, since they are successfully operated even at a high temperature [$>3000$ K] under the inert-halogen mixture atmosphere.\\

Generally speaking, the optical absorption spectra are delicately measured using the Fourier transform spectrometer\cite{2SCIENCE322;1529,2FARDIS150;71}, linear photodiode array spectrometer\cite{2SPECTROSCOPY15;16,2JChromSci14;195}, or charge coupled device spectrometer [CCD; \cite{2REVSCIINS83;093108,2APPOPT48;5713}]. Fourier transform spectrometer is based on the coherent observations of electromagnetic radiations, in either the time- or space-domain ones, e.g., Bruker IFS125 with the high-resolution linewidths narrower 0.001 cm\cite{2SCIENCE322;1529,2FARDIS150;71}. On the other side, the photodiode array spectrophotometer, which is built from hundreds of linear high-speed detectors integrated on a single chip, simultaneously measures the dispersive light over a wide frequency range. CCD spectrometer is also a multichannel detector; it can simultaneously detect as many frequency channels as the number of the individual resolution pixels. According to the linear dynamic Kubo formula\cite{2REVSCIINS83;093108,2APPOPT48;5713}, the optical absorption structures are closely related to electronic properties [energy spectra and wave functions], in which they have finite widths and specific forms arising from the initial and final band-edge states [the critical points in energy-wave-vector space]. \\

Up to now, the high-resolution optical observations\cite{2PRB78;235408,2NatPhys7;944,2PRL104;176404,2PR138;A197,2PRL106;046401} on layered graphene systems are consistent with part of theoretical calculations\cite{2CRCPress;CY} in the absence./presence of a uniform perpendicular electric field [$E_z \hat{z}$]. For example, the AB-stacked bilayer graphene, the $\sim 0.3$-eV shoulder structure under a zero field\cite{2PRB78;235408,2PRB79;115441}, the $E_z$-created semimetal-semiconductor transition and two low-frequency asymmetric peaks, and two very prominent $\pi$-electronic absorption peaks at middle frequency\cite{2PRB78;235408,2PRB79;115441}. The similar examinations performed for the trilayer ABA-stacked graphene clearly show one shoulder at $\sim 0.5 eV$\cite{2NatPhys7;944}, the gapless behavior unchanged by the gate voltages, the $E_z$-induced low-frequency multi-peak structures\cite{2NatPhys7;944}, and several strong $\pi$-electronic absorption peaks at higher frequencies\cite{2NatPhys7;944}. The identified features of absorption spectra in the trilayer ABC stacking cover two low-frequency characteristic peaks\cite{2PRL104;176404,2NatPhys7;944} and gap opening under the electric field\cite{2NatPhys7;944}. The above-mentioned strong $\pi$ peaks, which are due to all the valence states of carbon-2p$_z$ orbitals, have the most high frequency in Bernal graphite because of the infinite graphene layers\cite{2PR138;A197,2PRL106;046401}. On the other hand, certain theoretical predictions  deserve further experimental verifications, e.g., the rich and unique optical properties of the trilayer AAB and AAA stackings \cite{2CRCPress;CY} and sliding bilayer graphene systems\cite{2CRCPress;CY,2SciRep4;7509}. \\

Magneto-optical spectroscopies are available for a full exploration of the various magnetic quantization phenomena in the dimension-dependent condensed-matter systems, e.g., the 1D-3D graphene-related sp$^2$-bonding systems. A uniform magnetic field is generated by using the superconducting magnet, in which a 25-T cryogen-free superconducting magnet are successfully developed at the High Field Laboratories\cite{2SUPSCITECH30;065001,2IEEETAS27;4603805}. Furthermore, the super high magnetic fields can be built under a semidestructive single-turn coil technique that provides a pulsed magnetic field larger than 100 T for pulse lengths of tens microseconds\cite{2JLTP159;297}. Up to now, the high-resolution magneto-optical measurements have confirmed the well-known Aharnov-Bohm effect due to the periodical boundary condition in cylindrical carbon nanotubes\cite{2Science304;1129}. The infrared transmission spectra clearly identify the $\sqrt{B_z}$-dependent absorption frequencies of the interband Landau-level transitions in mono- and few-layer graphene systems\cite{2PRL100;136403,2PRB15;4077}. Specifically, the magneto-Raman spectroscopy is utilized to observe the low-frequency Landau-level excitation spectra for the AB-stacked graphenes up to five layers\cite{2NanoLett14;4548}. Only the well-behaved selection rule of $\Delta n=1$ is revealed in the above-mentioned inter-Landau-level excitation experiments. Apparently, the extra magneto-optical selection rules in few-layer graphene systems are worthy of a series of detailed examinations, e.g., those in trilayer ABC\cite{2CRCPress;CY} and AAB \cite{2CRCPress;CY} stackings, and the electric-field-applied AB-stacked systems\cite{2CRCPress;CY}. As for the magneto-optical reflection/absorption/transmission/Raman experiments on the Bernal [AB-stacked] graphite\cite{2PR138;A197,2PRB80;161410,2PRL102;166401,2PRB86;155409,2JAP117;112803}, the vertical optical excitations, which mainly originate from a lot of 1D Landau subbands with the significant $k_z$-dispersions, are verified to present the monolayer- and bilayer-like graphene behaviors. Such phenomena respectively, correspond to the magnetic quantizations initiated from K and H points [$k_z=0$ and $\pi$]. \\

\subsection{Quantum transport apparatus, and differential thermal calorimeter/laser flash analysis}\label{ch2.3}

In general, the Hall-bar method is available in fully exploring quantum transport properties since its first discovery by Edwin Herbert Hall in 1879 \cite{Hall;287}. The measurements of electrical resistivities for the rectangular thin films are carried out by a five-probe technique, as clearly displayed in Fig. 2.3. There exist the typical dimensions of 2 mm in length, 1 mm in width, and 100 nm in thickness \cite{Nitta;1335}. This equipment is very efficient and reliable in simultaneously measuring the longitudinal resistivity and the transverse Hall voltage [the positive or negative charges of free carriers and their density]. Five gold-pads are evaporated onto the contact areas to ensure the excellent electrical contacts for the transverse Hall and longitudinal resistivity measurements using the standard direct current methods. The Hall voltages are delicately analyzed through reversing the field direction at a fixed temperature to eliminate the offset voltage associated with the unbalanced Hall terminals. The Hall coefficients are taken in a magnetic-field range of ${\sim\,1-10}$ T, and a typical dc current density of ${\sim\,10^3-10^4}$ A/cm$^2$ is applied along the longitudinal direction of the sample. In addition, another method, the four-probe measurement of van der Pauw, is frequently utilized for the irregular finite-width materials. \\

Recently, there are a lot of magneto-transport measurements on quantum Hall conductivities of few-layer graphene systems with the various layer numbers and stacking configurations under the low temperatures. Monolayer graphene has been clearly identified to present the unconventional half-integer transverse conductivity $\sigma_{xy} = (m + 1/2)4e^2/h$ \cite{Novoselov;177}, where $m$ is an integer and the 4 factor represents the spin and the equivalence of A and  B sublattices. This unique quantization mainly comes from the quantum anomaly of $n^{c,v} = 0$ Landau levels corresponding to the Dirac point, as discussed later in Chap. 3.3.1. As to bilayer AB stacking \cite{Sanchez;076601}, the quantum Hall conductivity is verified to be $\sigma_{xy}= 4m' e^2/h$ [$m'$ a non-zero integer]. Furthermore, an unusual integer quantum Hall conductivity, a double step height of $\sigma_{xy} = 8e^2/h$, appears at zero energy under a sufficiently low magnetic field. This mainly comes from $n = 0$ and $n=1$ Landau levels of the first group.  The low-lying quantum Hall plateaus in trilayer ABA graphene are revealed as $2e^2/h$, $4e^/h$, $6e^2/h$, and $8e^2/h$ \cite{Henriksen;011004}, with a step height of $2e^2/h$, especially for the energy range of $\sim \pm 20$ meV. Such observation is consistent with the calculated Landau-level energy spectra \cite{Yuan;125455}. The neighboring- and next-neighboring-layer  hopping integrals, respectively, create the separated Dirac cone and parabolic bands, and the valley splitting of the latter. At low energy, these further lead to six quantized Landau levels with double spin degeneracy and thus the quantum Hall step of $2e^2/h$. Specifically, the higher-energy quantum transport properties in trilayer ABA stacking are roughly regarded as the superposition of monolayer and AB bilayer graphenes \cite{Kumar;126806}. Moreover, the ABC-stacked trilayer graphene, shows the significant differences in the main features of quantum Hall conductivities, compared with the ABA trilayer system. Its quantum Hall conductivity \cite{Kumar;126806} behaves as a sequence of $\sigma_{xy} = 4(\pm | m' | \pm 1/2)e^2/h$ in the absence of the $\sigma_{xy} = \pm 2e^2/h$, $\pm 4e^2/h$, and $\pm 8e^2/h$ plateaus. Specifically, a $\sigma_{xy} = 12e^2/h$ step comes to exist near zero energy, being created the $n = 0$, 1 and 2 Landau levels the first group due to the quantization of surface-localized flat bands \cite{Yuan;125455}. There also exists the four-fold spin and valley degeneracy in ABC trilayer stacking, resulting in the quantum conductivity step height of $4e^2/h$. The above-mentioned interesting quantum  transports open the door for fully exploring the configuration-modulated Hall conductivities in other layered graphenes, such as the trilayer AAB stacking, sliding bilayer systems, and twisted bilayer ones. \\

Up to now, two kinds of high-resolution instruments, differential scanning calorimeter [DSC; \cite{Schuller;142,Hohne;dsc2013,Gill;931}] and laser flash analysis [LFA; \cite{Zhou;105,Kover;151}], are frequently utilized to measure the thermal properties, especially for the specific heats of condensed-matter systems. As to the former [Fig. 2.4(a)], the sample material is subjected to a linear temperature algorithm. The heat flow rate into the sample, which corresponds to the instantaneous specific heat, is under the continuous and delicate measurements \cite{Hohne;dsc2013}. Specifically, two sample holders are symmetrically mounted inside an enclosure being normally held at room temperature. A primary temperature-control system dominates the average temperature of two sample holders, in which it covers platinum resistance thermometers and heating elements embedded in the sample holders. Furthermore, a secondary temperature-control system measures the temperature difference between the two sample holders, and generates the vanishing one by controlling a differential component of the total heating power. This differential power is accurately measured and recorded. Very interesting, an extension of DSC, a modulated differential scanning calorimeter \cite{Gill;931,Schuller;142}, is further developed to provide the more reliable results through measuring the heat capacity of the sample and the heat flow at the same time. MDSC utilizes a sinusoidal temperature oscillation instead of the traditional linear ramp. How to examine the magnetic quantization phenomena in thermal properties is a high-technique measurement. The close association of magnetic-field apparatus and DSC is the so-called PPMS calorimeter [a Quantum Design Physical Property Measurement System for magnetization]; furthermore, this combined instrument can measure the $T$- and $B_z$-dependent and specific heat and magnetic susceptibility simultaneously. It should be noticed that the stationary and pulsed magnetic fields can achieve the strengths of 10 T and 50 T in standard laboratories, respectively. \\

LFA \cite{Schuller;142}, as clearly shown in Fig. 2.4(b), represents a recent technical progress in identifying the crucial thermal phenomena.  The sample is positioned on a robot being surrounded by a furnace. For the accurate measurements, the furnace remains at a predetermined temperature and a programmable energy pulse irradiates the back side of the specimen. These create a homogeneous temperature rise at the sample surface. The resulting temperature rise of the surface of the sample is measured by a very sensitive high-speed infrared detector. Both properties, the specific heat and thermal diffusivity, are simultaneously determined from the temperature versus time data. The thermal conductivity could also be evaluated from the estimated carrier density. In short, LFA covers the experimental operation processes: uniform heating by a homogenized laser beam through an optical fiber with a mode mixer, measuring the transient temperature of a sample with a calibrated radiation thermometer, analyzing a transient temperature curve with a curve fitting method, and achieving the differential laser flash calorimetry. Recently, LFA is commonly used for the measurements of thermal properties, mainly owing to the excellent performances in the high accuracy, good repeatability, and wide temperature-range \cite{Schuller;142,Hohne;dsc2013,Gill;931}. \\

The high-resolution experimental measurements on specific heats of condensed-matter systems are very useful in examining the theoretical predictions about the electronic and phonon energy spectra, e.g., those measured for the various carbon-related materials. For example,  The thermal examinations on the AAA- and ABC-stacked graphites  [simple hexagonal and rhombohedral configurations] are lacking so far, while they have been conducted for the AB-stacked graphite [Bernal type], including electronic specific heat \cite{Van;1318,Bowman;367} and lattice specific heat \cite{DeSorbo;1660;Krumhansl;1663}. The previous experiments clearly show that the former presents a linear T-dependence at very low temperature [${T < 1.2}$ K; \cite{Van;1318}], and the latter follows the $T^3$ law at low temperature and approximately $T^2$ law at higher temperature \cite{Krumhansl;1663}. The applications of graphite systems promise an economical way to a new class of efficient thermal management materials, such as, the relatively high thermal conductivities very suitable for thermal interface applications \cite{Yu;7565,Lin;295}. Specifically, the magneto-electronic specific heats of conventional quantum Hall systems have been successfully verified for many years. In the research on GaAs¡VGaAlAs \cite{Gornik;1820}, the thermal properties are accurately measured under the influences of temperature and magnetic field. For the temperature-dependent specific heat \cite{Bayot;4584}, the specific heat reveals a peak structure at a low critical temperature. Furthermore,  the magnetic-field dependence heat obviously presents an oscillation behavior \cite{Gornik;1820}. These two kinds of behaviors directly reflect the main characteristics of the Landau levels, such as the filling factors, the energy spacings and the Zeeman effects. The similar examinations could be generalized to the emergent layered material in fully understanding the diverse thermal phenomena under the various magnetic quantizations. \\

The magnet-electronic specific heats could present another magnetic quantization phenomenon. The Landau levels in monolayer graphene are very suitable for a model investigation. It is well known that the hexagonal crystal structure accounts for the unique low-energy electronic properties, the isotropic and linear valence and conduction bands  intersecting at the Dirac point and the square-dependent Landau-level energy spectrum of ${E^{c,v}\propto\,\sqrt {n^{c,v}B_z}}$. The Zeeman effect further splits the magneto-electronic states into the spin-up and spin-down ones. In general, the level spacings between two neighboring quantum numbers are much wider than the Zeeman splitting energy. This clearly indicates that the former is insensitive to temperature. Furthermore, the latter might be comparable to the thermal energy at low temperature. The Zeeman effect is expected to plays an important role in the thermal properties. On the other hand, monolayer graphene could be rolled up to become a hollow cylinder, [a single-walled carbon nanotube]. Apparently, the periodic boundary condition is responsible for the creation of one-dimensional energy bands with  linear and parabolic energy dispersions. However, it is almost impossible to induce the dispersionless Landau levels in nanotube surfaces under various high magnetic fields \cite{Lin;7592}. Dimensionality and magnetic quantization have a great influence on the van Have singularities of the density of states (DOS), so the magneto-electronic specific heats would be very different between these two systems. \\

\subsection{Electron energy loss spectroscopy and light inelastic scatterings}\label{ch2.4}

Electron energy loss spectroscopy [EELS; \cite{2PRB49;2888,2CPL305;225,2ZPhys243;229,2NanoRes10;234,2PRL89;076402,2PRB31;4773,
2Carbon37;733,2PRB88;075433,2Carbon114;70,2PRB91;045418,2ACSNano12;1837,2Nature468;1088} and inelastic x-ray scatterings [IXS; \cite{2Winfried,2PRB76;035439,2PRB89;014206,2PCM19;046207,2JPSJ84;084701,
2PRB71;060504,2PRL101;266406,2Kittel,2Ultramicroscopy96;367,2PRB55;13961,2PRB86;195424,
2PRB38;2112,2PRB86;245430,2RSI82;113108,RSI77;053102,RMP73;203,RMP79;175,RMP83;705}] are the only two available tools in delicately examining excitation spectra of condensed-matter systems. They have successfully identified the reliable Coulomb excitations and phonon dispersion spectra in any dimensional materials, being supported by the theoretical predictions\cite{2PRL89;076402,2PRB31;4773,
2Carbon37;733,2PRB88;075433,2PRB91;045418,2Nature468;1088}. In general, EELS is achieved by the scattered reflection \cite{2SurSci602;2069,2APL89;213106,2SurSci601;L109} and transmission\cite{2Ultramicroscopy107;575,2HIbach,2Ultramicroscopy96;367,2Micron34;235,2JMicrosc194;203,2Ultramicroscopy106;1091} of a narrow electron beam, in which the former is very suitable for thoroughly exploring the low-energy excitation properties lower than 1 eV\cite{2SurSci602;2069}. RELLS is chosen for a clear illustration. This outstanding instrument extracts the bulk or surface energy loss functions from the backward electrons scattered from a specific sample\cite{2Ultramicroscopy96;367}. If the incident electron beam has a kinetic energy of several hundreds eVs, the very efficient technique can provide the screened response functions with an energy resolution of a few meVs, being sufficient in resolving atomic and electronic excitation modes\cite{2Ultramicroscopy96;367,2Micron34;235}. Up to date, REELS is widely utilized to explore the rich physical, chemistry and material properties of condensed-matter surfaces \cite{2HIbach}. It frequently operates under a 25-meV energy resolution within the energy range between 15 and 70 eV\cite{2HIbach}, while controlling the momentum resolution down to 0.013 $\AA^{-1}$, about one percent of a typical Brillouin zone\cite{2HIbach}. Moreover, the energy resolution could be greatly enhanced to $\sim 1$ 1 meV under the specific condition, in which much weaker electron beams are adopted using a high-resolution monochromator at an ultrahigh vacuum base pressure [$\sim 2 \times 10^{-10}$ Torr]. The momentum-dependent dispersion relations of collective excitation modes can be accurately detected by an angle-resolved EELS, which is performed at the low kinetic energy and utilizes an analyzer to measure the scattered electrons. The delicate analyzer consists of a magnetic-prism system, as clearly displayed in Fig. 2.5, where the commercially available Gatan spectrometer is installed beneath the camera and the basic interface and ray paths are shown as well. The surface of this prism is curved to largely reduce the spherical and chromatic aberrations. The scattered electrons in the drift tube are deflected by a uniform magnetic field into a variable entrance aperture (typical variation from 1 to 5 mm in diameter). The electrons that lose more energies, deflect further away from the zero-energy-loss electrons according to the Lorenz force law. Furthermore,  all the electrons in any directions are focused on the dispersion plane of the spectrometer. The magnetic prism projects the electron energy-loss spectrum  onto a CCD camera, which is straightforward to capture the whole energy distribution simultaneously. It is possible to modulate the resolution of the transferred momentum by varying the half-angle of the incident beam in TEM and the scattering one in the spectrometer. For example, based on the angle variation of a few milliradians, the momentum resolution lies in the order of $\sim 0.01$ $\AA^{-1}$\cite{2Ultramicroscopy107;575,2HIbach,2Ultramicroscopy96;367}. \\

The high-resolution EELS measurements are very powerful in thoroughly exploring electronic Coulomb excitation modes in various condensed-matter systems. They have clearly identified the significant single-particle and collective excitations in the sp$^2$-bonding graphene-related materials, such as, 3D layered graphites\cite{2ZPhys243;229,2NanoRes10;234,2PRL89;076402}, graphite intercalation compounds\cite{2PRB31;4773}, 2D monolayer and few-layer graphene systems\cite{2PRB88;075433,2Carbon114;70,2PRB91;045418,2ACSNano12;1837}, 1D single- and multi-walled carbon nanotubes\cite{2Carbon37;733}, graphene nanoribbons\cite{2Nature468;1088}, 0D C$_{60}$-associated fullerenes\cite{2PRB49;2888}, and carbon onions \cite{2CPL305;225}. In general, the distinct dimensions, geometric symmetries, stacking configurations, interlayer atomic interactions, temperature, and chemical dopings might induce the high/some/few free conduction-electron [valence-hole] density, leading to the low-frequency acoustic or optical plasmon modes with frequencies lower than 1.5 eV. For room temperature, the AB-stacked graphite presents the low-frequency optical plasmons at $\sim 45-50$ meV and 128 meV under the long wavelength limit, corresponding to the electric polarizations parallel and perpendicular to the $(x,y)$- plane, respectively\cite{2ZPhys243;229}. Furthermore, the detailed analyses on temperature-dependent energy loss spectra verify the strong temperature effects in the former\cite{2ZPhys243;229}. The high-density free electrons and holes, which are, respectively, doped in the donor- and acceptor type graphite intercalation compounds, lead to the $\sim 1$-eV optical plasmons\cite{2PRB31;4773}. This represents their coherent collective oscillations in the  periodical  graphitic layers, strongly depending on the transferred momenta\cite{2PRB31;4773}. Concerning the layered graphene systems, with the metal-adatom chemisorptions\cite{2PRB88;075433,2Carbon114;70,2PRB91;045418,2ACSNano12;1837}, the low-frequency plasmon modes become 2D acoustic ones, i.e., they have the $\sqrt{\bf q}$ dependence at small transferred momenta\cite{2PRB88;075433,2Carbon114;70,2PRB91;045418,2ACSNano12;1837}. The above-mentioned important results are consistent with the theoretical illustrations\cite{\cite{2PRL89;076402,2PRB31;4773,
2Carbon37;733,2PRB88;075433,2PRB91;045418,2Nature468;1088}}. \\

The inelastic X-ray scatterings [IXS] are capable of directly detecting the behaviors of dynamic charge screenings in crystal/non-crystal systems. They are successfully conducted on a wide range of physical phenomena, e.g., phonon energy dispersions in solids\cite{2Winfried,2PRB76;035439}, carrier dynamics of disordered materials\cite{2PRB89;014206,2PCM19;046207,2JPSJ84;084701,2PRB71;060504} $\&$ biological structures\cite{2Winfried}, and electronic excitations in condensed-matter systems\cite{2PRL101;266406,
2Kittel,2Ultramicroscopy96;367,2PRB55;13961,2PRB86;195424,2PRB38;2112,2PRB86;245430}.
The transferred energy and momentum are available parameters and can cover the whole spectra of the dielectric responses. The uniform x-ray beam line is designed to provide a super-high photon flux within the typical wave-vector range of the first Brillouin zone. Furthermore, photon energy is distributed from 4.9 keV to 15 keV, being accompanied by energy and momentum resolutions of $\sim 70$ meV and $0.02-0.03$ $\AA ^{-1}$, respectively. Specifically, the IXS instrument in the Swiss Light Source possesses an extremely good energy resolution of $\sim 30$ meV \cite{Strocov;631}. The super high resolution, $\sim 10$ meV, is expected to be achieved in the further development of new synchrotron sources. IXS is available in fully exploring all kinds of electronic excitations because the electronic charges strongly interact with the high-energy photon beam under the various transferred momenta and energies. Using hard X-ray synchrotron sources, this spectroscopy becomes a very powerful technique to observe the intrinsic properties of condensed-matter systems, and it is suitable for the external electric and magnetic fields \cite{Qiao;033106}. The analyzer, which is built on the basis of Bragg optics, as depicted in Fig. 2.5, efficiently collects and analyzes the energy and momentum distributions of the scattered photons with a small solid angle [$d \Omega$]. This instrument can provide full information on the screened response function. To maximize the light intensity, a spherically bent analyzer ($\sim 10$ cm in diameter) is frequently utilized to capture all the scattered radiation of the momentum-transferred photons in $d \Omega$. The transferred energy is projected onto a CCD detector, and then the full energy loss spectrum is scanned by varying the Bragg angle of the specific crystal. When IXS is operated in the Rowland circle geometry, the specific measurement of the double differential scattering cross-section corresponds to the screened electronic excitations in the characteristic energy-loss regime by using the dissipation-fluctuation theorem. \\

There exist certain important differences between EELS and IXS in terms of the physical environments. The incident electron and photon beams can be, respectively, focused on the spatial range of $\sim 10$ $\AA$ \cite{Egerton;2011} and 100 $\AA$ \cite{Egerton;575}. The former possesses more outstanding resolutions for the transferred momenta and energies during the many-particle interactions. Much more inelastic scattering events are conducted by using the EELS within a short time, i.e., the EELS measurements on the full excitation spectra are performed more quickly and accurately. EELS is very reliable for the low-dimensional materials and nano-scale structures, mainly owing to the simultaneous identifications on their positions. On the other hand, IXS, which arises from the continuous synchrotron radiation sources, presents a rather strong intensity with the tunable energies and momenta. Most importantly, the extreme surrounding environments, induced by the applications of magnetic and electric fields as well various temperatures and pressures, can be overcome under the inelastic light scattering. The external fields have strong effects on the incident charges and the sample chamber is too narrow; therefore, EELS cannot work under such environments. As a result, the high-resolution IXS measurements are very useful in thoroughly understanding the magneto-electronic single-particle and collective excitations in any condensed-matter systems, especially for those in graphene-related materials. \\

\newpage
\begin{center}
\section{Theoretical models}\label{ch3}
\author{ Jhao-Ying Wu,\textit{$^{d,\dag}$} Chiun-Yan Lin,\textit{$^{a}$} Thi-Nga Do,\textit{$^{b,c,\ast}$} Po-Hsin Shih,\textit{$^{a,\ddag}$}\\ Shih-Yang Lin,\textit{$^{e,\S}$} Ching-Hong Ho,\textit{$^{d,\P}$} Ming-Fa Lin\textit{$^{a,\sharp}$}}
\end{center}
\vskip 1.0 truecm

The further modifications on the theoretical models are necessary for the emergent layer materials. The modified $\&$ new theoretical framework is available in fully exploring the fundamental physical properties and thus creating the rich and unique phenomena, being clearly illustrated in the following sections.\\

\subsection{The generalized tight-binding models for typical layered systems}\label{ch3.1}

To fully comprehend the essential electronic properties of 2D condensed-matter matter with the significant geometric symmetries, the intrinsic atomic interactions, and any external fields, we propose the generalized tight-binding model to diagonalize the various Hamiltonians efficiently. Furthermore, this model is accompanied with the concise physical pictures for the thorough understanding of the diverse physical phenomena. The main-stream typical systems, few-layer graphene, silicene, germanene, and phosphorene ones, are very suitable for a model study. The planar/buckled/puckered and layered structures, layer numbers, stacking configurations, and geometric modulations, are taken into account. The lattice- and atom-induced important interactions, the distinct site energies, the single- or multi-orbital hybridizations for the low-lying energy bands, the strong/weak spin-orbital couplings, and the intralayer and interlayer hopping integrals, are included in the various Hamiltonians. The field-induced independent Hamiltonian matrix elements will be derived in the analytic form, especially for those under a uniform perpendicular magnetic field. Finally, a giant magnetic Hermitian matrix can be accurately solved by the exact diagonalization method. \\

The bilayer AB-bt silicene system is very suitable for a full understanding of the magneto-electronic theoretical model, mainly owing to the observable buckling structure, the comparable intralayer and interlayer hopping integrals, and the significant layer-dependent spin-orbital interactions. The sp$^2$ chemical bondings still dominate the non-planar honeycomb lattice; therefore, the $\pi$ bondings due to the 3p$_z$ orbitals of silicon atoms are responsible for the low-energy fundamental properties. The generalized tight-binding model will be developed to investigate the feature-rich electronic properties of AB-bt bilayer silicene system. The external magnetic and electric fields are included in the calculations simultaneously. This honeycomb structure, as shown in Figs. 3.1(a) and 3.1(b) [the top and side views], possesses four silicon atoms in a unit cell, in which the two primitive unit vectors, $\bf{a_1}$ and $\bf{a_2}$, have a lattice constant of $a =3.86$ $\AA$ \cite{Zhao;24}. Apparently, the well-stacked bilayer silicene consists of four sublattices of (A$^1$, B$^1$) and (A$^2$, B$^2$). For each layer, the two sublattices lie on two different buckling planes with a separation of $l_z =0.46$ $\AA$. The B$^1$ and B$^2$ sublattices are situated at the higher and lower planes, respectively, where the inter-layer distance is 2.54 $\AA$. The buckled angle due to the intra-layer Si-Si bond and the $z$ axis is $\theta =$ 78.3 $^{\circ}$. \\

The low-energy Hamiltonian, which is closely related to the silicon-3p$_z$ orbitals, built from the tight-binding model covering the intra- and inter-layer atomic interactions, and two kinds of layer-dependent spin-orbital interactions. Among all the intrinsic interactions, the intralayer hopping integral and two kinds of spin-orbital couplings are similar to those in monolayer silicene\cite{3PRB97;125416}. On the other side, such a bilayer system exhibits the layer-environment-induced spin-orbital couplings, a vertical and two non-vertical interlayer atomic interactions, being absent in monolayer silicene. The very complex Hamiltonian is expressed as
\begin{align}
%\begin{array}{lll}
H  = &\sum_{I,l}(\epsilon_I^l+U_I^l)c_{I \alpha}^{\dagger l}c_{I \alpha}^{l}
+\sum_{ I,J  , \alpha, l, l'} \gamma_{IJ}^{ll'} c_{I \alpha}^{\dagger l}
c_{J \alpha}^{l'} \nonumber \\
& + \frac {i} {3\sqrt{3}} \sum_{ \langle \langle I,J \rangle \rangle, \alpha, \beta, l}
\lambda^{SOC}_{l} \gamma_l v_{IJ} c_{I\alpha}^{\dagger l} \sigma_{\alpha\beta}^{z} c_{J\beta}^{l} \nonumber \\
& - \frac{2i}{3} \sum_{\langle\langle I,J \rangle \rangle, \alpha, \beta, l} \lambda^{R}_{l} \gamma_l u_{IJ}
c_{I\alpha}^{\dagger l} (\vec{\sigma} \times \hat{d}_{IJ})_{\alpha\beta}^{z} c_{J\beta}^{l}\ .  \tag{3.1}
%\end{array}
\end{align}
$\epsilon_I^l (A^l,B^l)$ represents the sublattice-dependent site energy associated with the chemical environment difference [e.g.,$\epsilon_I^l (A^l) = 0$; $\epsilon_I^l (B^l) = -0.12$ eV]. $U_I^{l} (A^l,B^l)$ is the height-created Coulomb potential energy due to a uniform perpendicular electric field. The $c_{I\alpha}^{l}$ and $c_{I\alpha}^{\dagger l}$ operators, respectively, correspond to the annilation and creation of an electronic state with the spin polarization of $\alpha$ at the $I$-th site of the $l$-th layer. The atomic interactions in the second term include the nearest-neighbor intra-layer hopping integral ($\gamma_0=1.13$ eV and three inter-layer hopping integrals due to (A$^1$, A$^2$), (B$^1$, A$^2$) or (A$^1$, B$^2$) and (B$^1$, B$^2$) [$\gamma_1=-2.2$ eV, $\gamma_2=0.1$ eV, and $\gamma_3=0.54$ eV in Fig. 3.1(b)]. Quite importantly, the largest inter-layer vertical hopping integral of $t_1$, which originates from the on-line hybridization of Si-3p$_z$ orbitals (the parallel interaction characterized by the orbital  distribution and center-of-mass), induces very strong orbital hybridizations in bilayer silicene. Apparently, $t_1$ sharply contrasts with $t_0$ in the single-orbital bonding; that is. the latter belongs to the perpendicular atomic interactions. The traditional spin-orbital coupling (the third term) and the Bychkov-Rashba spin-orbital coupling (the fourth term) are only taken for the next-nearest-neighbor pairs $\langle \langle I,J \rangle \rangle$. $\vec{\sigma}$ is the Pauli spin matrix and $\hat{d}_{IJ} = \vec {d}_{IJ} / |d_{IJ}|$ denotes a unit vector linking the $I$- and $J$-th lattice sites. $v_{IJ} = 1$/-1 if the next-nearest-neighbor hopping is anticlockwise/clockwise with respect to the positive $z$ axis. $u_{IJ KL} = 1$ and -1, respectively, corresponds to the A and B sublattices. $\gamma_{l} = \pm 1$ presents the layer-dependent spin-orbital interactions due to the opposite buckled ordering of AB-bt bilayer silicene. Two kinds of spin-orbital couplings come to exist in the diagonal elements of the Hermitian matrix. They are fitted as: $\lambda_1^{SOC}=0.06$ eV, $\lambda_2^{SOC}=0.046$ eV, $\lambda_1^{R}=-0.054$ eV and $\lambda_2^{R}=-0.043$ eV so that the calculated low-lying energy bands are closed to those from the first-principles method\cite{3APL104;131904,3JPCC119;3818}. \\

A uniform perpendicular magnetic field creates an extra Peierls phase in the tight-binding function through the vector potential $\vec{A}$. The ${\bf B}$-induced spatial period is chosen to be the integral times of honeycomb lattice, as clearly illustrated by a yellow-green rectangle in Fig. 3.2. The well-known phase is defined by $G_R =\frac{2\pi}{\phi_0}\int_{R}^{r} \vec{A}\cdot d\vec{l}$, where $\phi_0 =hc/e$ is the magnetic flux quantum and $\phi = B_z \sqrt{3}a^2/2$ is the magnetic flux through a hexagon. There are totally ${8R_B}$ ($R_B = \phi_0 /\phi$) Si atoms in an enlarged unit cell. As a result, the magnetic Hamiltonian, being based on the tight-binding functions of the periodical 3p$_z$ orbitals, is a 16$R_B$ $\times$ 16$R_B$ Hermitian matrix. In general, its dimension is giant under the typical magnetic-field strength in laboratory. For example, the Hamiltonian is a $50000\times 50000$ matrix at $B_z=10$ T. It should be noticed that the Hamiltonian matrix elements might be real or imaginary numbers, being sensitive to the intrinsic geometric symmetries and orbital hybridizations. The eigenvalues and eigenfunctions are efficiently numerically solved by the exact diagonalization method: the band-like method [the rearrangement of the tight-binding functions in Refs.] and the spatial localizations of the magnetic wavefunctions [as discussed in Chaps. 6.1 $\&$ 4.1]. \\

Through the delicate calculations, the independent magnetic Hamiltonian matrix elements are expressed as: \\

\begin{equation}
\nonumber
\langle A_m^{1\uparrow} |H|A_m^{1 \uparrow} \rangle = -i\lambda_1^{SOC} (Q_1 P_4 - Q_1^* P_4^*) +E_z l_1,
\tag{3.1.1}
\end{equation}
\begin{equation}
\nonumber
\langle A_m^{1\uparrow} |H|A_m^{1 \downarrow} \rangle = i\lambda_1^{R} (Q_1 P_4 - Q_1^* P_4^*) ,
\tag{3.1.2}
\end{equation}
\begin{equation}
\nonumber
\langle A_m^{1\uparrow} |H|B_m^{1 \uparrow} \rangle = \langle A_m^{1\downarrow} |H|B_m^{1 \downarrow} \rangle = \gamma_0 (Q_3 P_2 + Q_3^* P_3) ,
\tag{3.1.3}
\end{equation}
\begin{equation}
\nonumber
\langle A_m^{1\downarrow} |H|A_m^{1 \downarrow} \rangle = i\lambda_1^{SOC} (Q_1 P_4 - Q_1^* P_4^*) +E_z l_1,
\tag{3.1.4}
\end{equation}
\begin{equation}
\nonumber
\langle B_m^{1\uparrow} |H|B_m^{1 \uparrow} \rangle =i\lambda_2^{SOC} (Q_2 P_4 + Q_2^* P_4^*) +E_z l_2 +\gamma_4,
\tag{3.1.5}
\end{equation}
\begin{equation}
\nonumber
\langle B_m^{1\uparrow} |H|B_m^{1 \downarrow} \rangle = i\lambda_2^{R} (Q_2 P_4-Q_2^* P_4^*),
\tag{3.1.6}
\end{equation}
\begin{equation}
\nonumber
\langle B_m^{1\downarrow} |H|B_m^{1 \downarrow} \rangle =-i\lambda_2^{SOC} (Q_2 P_4 -Q_2^* P_4^*) +E_z l_2 +\gamma_4,
\tag{3.1.7}
\end{equation}
%%%%%%%%%%%%%%%%%%%%%%%%%
\begin{equation}
\nonumber
\langle A_m^{2\uparrow} |H|A_m^{2 \uparrow} \rangle = i\lambda_1^{SOC} (Q_1 P_4 - Q_1^* P_4^*) -E_z l_1,
\tag{3.1.8}
\end{equation}
\begin{equation}
\nonumber
\langle A_m^{2\uparrow} |H|A_m^{2 \downarrow} \rangle = -i\lambda_1^{R} (Q_1 P_4 - Q_1^* P_4^*) ,
\tag{3.1.9}
\end{equation}
\begin{equation}
\nonumber
\langle A_m^{2\uparrow} |H|B_m^{2 \uparrow} \rangle = \langle A_m^{2\downarrow} |H|B_m^{2 \downarrow} \rangle = \gamma_0 P_1 ,
\tag{3.1.10}
\end{equation}
\begin{equation}
\nonumber
\langle A_m^{2\uparrow} |H|B_m^{2 \uparrow} \rangle = \langle A_m^{2\downarrow} |H|B_m^{2 \downarrow} \rangle = \gamma_0 P_1,
\tag{3.1.11}
\end{equation}
\begin{equation}
\nonumber
\langle A_m^{2\downarrow} |H|A_m^{2 \downarrow} \rangle = -i\lambda_1^{SOC} (Q_1 P_4 - Q_1^* P_4^*) -E_z l_1,
\tag{3.1.12}
\end{equation}
\begin{equation}
\nonumber
\langle B_m^{2\uparrow} |H|B_m^{2 \uparrow} \rangle = i\lambda_2^{SOC} (Q_5 P_4 - Q_5^* P_4^*) -E_z l_2 + \gamma_4,
\tag{3.1.13}
\end{equation}
\begin{equation}
\nonumber
\langle B_m^{2\uparrow} |H|B_m^{2 \downarrow} \rangle = i\lambda_2^{R} (Q_5 P_4 + Q_5^* P_4^*),
\tag{3.1.14}
\end{equation}
\begin{equation}
\nonumber
\langle B_m^{2\downarrow} |H|B_m^{2 \downarrow} \rangle = i\lambda_2^{SOC} (Q_5 P_4 - Q_5^* P_4^*) -E_z l_2 + \gamma_4,
\tag{3.1.15}
\end{equation}
%%%%%%%%%%%%%%%%%%%%%%%%%%%%%%%%%%%%%%%
\begin{equation}
\nonumber
\langle A_m^{1\uparrow} |H|A_m^{2 \uparrow} \rangle = \gamma_1,
\tag{3.1.16}
\end{equation}
\begin{equation}
\nonumber
\langle A_m^{1\uparrow} |H|B_m^{2 \uparrow} \rangle = \gamma_2 P_1,
\tag{3.1.17}
\end{equation}
\begin{equation}
\nonumber
\langle A_m^{1\downarrow} |H|A_m^{2 \downarrow} \rangle = \gamma_1,
\tag{3.1.18}
\end{equation}
\begin{equation}
\nonumber
\langle A_m^{1\downarrow} |H|B_m^{2 \downarrow} \rangle = \gamma_2 P_1,
\tag{3.1.19}
\end{equation}
\begin{equation}
\nonumber
\langle B_m^{1\uparrow} |H|A_m^{2 \uparrow} \rangle = \gamma_2 (Q_3 P_2^* - Q_3^* P_3^*),
\tag{3.1.20}
\end{equation}
\begin{equation}
\nonumber
\langle B_m^{1\uparrow} |H|B_m^{2 \uparrow} \rangle = \gamma_3 (Q_6 P_2 + Q_6^* P_3),
\tag{3.1.21}
\end{equation}
\begin{equation}
\nonumber
\langle B_m^{1\downarrow} |H|A_m^{2 \downarrow} \rangle = \gamma_2 (Q_3 P_2^* - Q_3^* P_3^*),
\tag{3.1.22}
\end{equation}
\begin{equation}
\nonumber
\langle B_m^{1\downarrow} |H|B_m^{2 \downarrow} \rangle = \gamma_3 (Q_6 P_2 + Q_6^* P_3),
\tag{3.1.23}
\end{equation}
%%%%%%%%%%%%%%%%%%%%%%%%%%%%%%%%%%%%%%%
\begin{equation}
\nonumber
\langle A_m^{1\uparrow} |H|A_{m+1}^{1 \uparrow} \rangle = i \lambda_1^{SOC}(P_6 Q_4^* + P_5 Q_4),
\tag{3.1.24}
\end{equation}
\begin{equation}
\nonumber
\langle A_m^{1\uparrow} |H|A_{m+1}^{1 \downarrow} \rangle = i \lambda_1^{R}(P_5 Q_4 - P_6 Q_4^*),
\tag{3.1.25}
\end{equation}
\begin{equation}
\nonumber
\langle A_m^{1\downarrow} |H|A_{m+1}^{1 \uparrow} \rangle = i \lambda_1^{R}(P_5 Q_4 e^{i\pi/6} + P_6 Q_4^* e^{-i\pi/6}),
\tag{3.1.26}
\end{equation}
\begin{equation}
\nonumber
\langle A_m^{1\downarrow} |H|A_{m+1}^{1 \downarrow} \rangle = -i \lambda_1^{SOC}(P_6 Q_4^* + P_5 Q_4),
\tag{3.1.27}
\end{equation}
\begin{equation}
\nonumber
\langle B_m^{1\uparrow} |H|A_{m+1}^{1 \uparrow} \rangle = \gamma_2 P_1,
\tag{3.1.28}
\end{equation}
\begin{equation}
\nonumber
\langle B_m^{1\uparrow} |H|B_{m+1}^{1 \uparrow} \rangle = i \lambda_2^{SOC}(P_5 Q_6 - P_6 Q_6^*),
\tag{3.1.29}
\end{equation}
\begin{equation}
\nonumber
\langle B_m^{1\uparrow} |H|B_{m+1}^{1 \downarrow} \rangle = i \lambda_2^{R}(P_5 Q_6 e^{-i\pi/6} + P_6 Q_6^* e^{i\pi/6}),
\tag{3.1.30}
\end{equation}
\begin{equation}
\nonumber
\langle B_m^{1\uparrow} |H|A_{m+1}^{2 \uparrow} \rangle = \gamma_2 P_1,
\tag{3.1.31}
\end{equation}
\begin{equation}
\nonumber
\langle B_m^{1\downarrow} |H|A_{m+1}^{1 \downarrow} \rangle = \gamma_2 P_1,
\tag{3.1.32}
\end{equation}
\begin{equation}
\nonumber
\langle B_m^{1\downarrow} |H|B_{m+1}^{1 \uparrow} \rangle = i \lambda_2^{R}(P_5 Q_6 e^{i\pi/6} - P_6 Q_6^* e^{-i\pi/6}),
\tag{3.1.33}
\end{equation}
\begin{equation}
\nonumber
\langle B_m^{1\downarrow} |H|B_{m+1}^{1 \downarrow} \rangle = i \lambda_2^{SOC}(P_5 Q_6 + P_6 Q_6^*),
\tag{3.1.34}
\end{equation}
\begin{equation}
\nonumber
\langle B_m^{1\downarrow} |H|A_{m+1}^{2 \downarrow} \rangle = \gamma_2 P_1,
\tag{3.1.35}
\end{equation}
%%%%%%%%%%%%%%%%%%%%%%%%%
\begin{equation}
\nonumber
\langle A_m^{2\uparrow} |H|A_{m+1}^{2 \uparrow} \rangle = i \lambda_1^{SOC} (P_5 Q_4 - P_6 Q_4^*),
\tag{3.1.36}
\end{equation}
\begin{equation}
\nonumber
\langle A_m^{2\uparrow} |H|A_{m+1}^{2 \downarrow} \rangle = i \lambda_1^{R} (P_6 Q_4^* - P_5 Q_4),
\tag{3.1.37}
\end{equation}
\begin{equation}
\nonumber
\langle A_m^{2\downarrow} |H|A_{m+1}^{2 \uparrow} \rangle = i \lambda_1^{R} (P_6 Q_4^* e^{-i\pi/6} - P_5 Q_4 e^{i\pi/6}),
\tag{3.1.38}
\end{equation}
\begin{equation}
\nonumber
\langle A_m^{2\downarrow} |H|A_{m+1}^{2 \downarrow} \rangle = -i \lambda_1^{SOC} (P_5 Q_4 + P_6 Q_4^*),
\tag{3.1.39}
\end{equation}
\begin{equation}
\nonumber
\langle B_m^{2\uparrow} |H|A_{m+1}^{2 \uparrow} \rangle = \gamma_0 (P_2 Q_6 + P_3 Q_6^*),
\tag{3.1.40}
\end{equation}
\begin{equation}
\nonumber
\langle B_m^{2\uparrow} |H|B_{m+1}^{2 \uparrow} \rangle = i \lambda_2^{SOC} (P_5 Q_7 + P_6 Q_7^*),
\tag{3.1.41}
\end{equation}
\begin{equation}
\nonumber
\langle B_m^{2\uparrow} |H|B_{m+1}^{2 \downarrow} \rangle = i \lambda_2^{R} (P_6 Q_7^* e^{i\pi/6} - P_5 Q_7 e^{-i\pi/6}),
\tag{3.1.42}
\end{equation}
\begin{equation}
\nonumber
\langle B_m^{2\downarrow} |H|A_{m+1}^{2 \downarrow} \rangle = \gamma_0 (P_2 Q_6 + P_3 Q_6^*),
\tag{3.1.43}
\end{equation}
\begin{equation}
\nonumber
\langle B_m^{2\downarrow} |H|B_{m+1}^{2 \uparrow} \rangle = i \lambda_2^{R} (P_6 Q_7^* e^{-i\pi/6} + P_5 Q_7 e^{i\pi/6}),
\tag{3.1.44}
\end{equation}
\begin{equation}
\nonumber
\langle B_m^{2\downarrow} |H|B_{m+1}^{2 \downarrow} \rangle = -i \lambda_2^{SOC} (P_5 Q_7 - P_6 Q_7^*),
\tag{3.1.45}
\end{equation}
%%%%%%%%%%%%%%%%%%%%%%%%%
\begin{equation}
\nonumber
\langle B_m^{2\uparrow} |H|A_{m+1}^{1 \uparrow} \rangle = \gamma_2 (P_2 Q_6 - P_3 Q_6^*),
\tag{3.1.46}
\end{equation}
\begin{equation}
\nonumber
\langle B_m^{2\uparrow} |H|B_{m+1}^{1 \uparrow} \rangle = \gamma_3 P_1,
\tag{3.1.47}
\end{equation}
\begin{equation}
\nonumber
\langle B_m^{2\downarrow} |H|A_{m+1}^{1 \downarrow} \rangle = \gamma_2 (P_2 Q_6 - P_3 Q_6^*),
\tag{3.1.48}
\end{equation}
\begin{equation}
\nonumber
\langle B_m^{2\downarrow} |H|B_{m+1}^{1 \downarrow} \rangle = \gamma_3 P_1,
\tag{3.1.49}
\end{equation}

Moreover, the wave-vector-dependent phase terms are given by \\
\begin{eqnarray}
\nonumber
P_1 = exp [ik_x b],
\end{eqnarray}
\begin{eqnarray}
\nonumber
P_2 = exp [i(k_x b/2 +k_y a/2)],
\end{eqnarray}
\begin{eqnarray}
\nonumber
P_3 = exp [i(k_x b/2 - k_y a/2)],
\end{eqnarray}
\begin{eqnarray}
\nonumber
P_4 = exp [ik_y a],
\end{eqnarray}
\begin{eqnarray}
\nonumber
P_5 = exp [i(k_x 3b/2 + k_y a/2)],
\end{eqnarray}
\begin{eqnarray}
\nonumber
P_6 = exp [i(k_x 3b/2 - k_y a/2)],
\end{eqnarray}
\begin{eqnarray}
\nonumber
Q_1 = exp [i2\pi \psi (j-1)]
\end{eqnarray}
\begin{eqnarray}
\nonumber
Q_2 = exp [i2\pi \psi (j-1+2/6)],
\end{eqnarray}
\begin{eqnarray}
\nonumber
Q_3 = exp [i\pi \psi (j-1+1/6)],
\end{eqnarray}
\begin{eqnarray}
\nonumber
Q_4 = exp [i\pi \psi (j-1+3/6)],
\end{eqnarray}
\begin{eqnarray}
\nonumber
Q_5 = exp [i\pi \psi (j-1+4/6)],
\end{eqnarray}
\begin{eqnarray}
\nonumber
Q_6 = exp [i\pi \psi (j-1+5/6)],
\end{eqnarray}
\begin{eqnarray}
\nonumber
Q_7 = exp [i\pi \psi (j-1+7/6)],
\end{eqnarray}

in which $j$ (an integer) indicates the lattice site of the atoms. \\

After the diagonalization of bilayer magnetic Hamiltonian, the Landau level wavefunction, with a specific quantum number $n$, is given by
\begin{align}
\Psi (n,\bf{k})=\sum_{{\it l}=1,2}\sum_{{\it I}=1}^{R_B}\sum_{\alpha} [ {\it A}_{\alpha}^{\it l,I}({\it n},\bf{k}) |\psi_{\alpha}^{\it l,I}({\it A})\rangle + {\it B}_{\alpha}^{\it l,I}({\it n},\bf{k}) |\psi_{\alpha}^{\it l,I} ({\it B})\rangle]. \tag{3.2}
\end{align}
$\psi_{\alpha}^{l,I}$ is the 3p$_z$-orbital tight-binding function situated at the $I$-th site of the $l$-th layer with the spin $\alpha$ configuration; furthermore, $A_{\alpha}^{l,I} (n,\bf{k})$ [$B_{\alpha}^{l,I} (n,\bf{k})$] is the corresponding amplitude on the sublattice-dependent lattice site. Most importantly, all the amplitudes in an enlarged unit cell could be regarded as the continuous spatial distributions of the sub-envelope functions on the distinct sublattice, since the magnetic length [$l_B=\sqrt{hc/eB_z}$] is much smaller/longer than the length of the enlarged unit cell [Fig. 3.2]/the original lattice constant. Such subenvelope functions can provide much information for fully understanding the unusual Landau-level features, such as the sublattice-dominated Landau-level quantum number, the different localization centers, and the frequent crossing $\&$ anti-crossing phenomena. For bilayer silicene, the buckled honeycomb structure, the complicated intra- and inter-layer atomic interactions and the significant spin-orbital interactions need to be thoroughly taken into account in the theoretical model calculations. Such a system is expected to exhibit diverse physical properties under various external fields. \\

In general, there are two kinds of theoretical models in studying the magnetic quantization phenomena, namely, the low-energy elective-mass approximation and the tight-binding model. Concerning the low-energy perturbation method\cite{3PRB83;165443,3PRL86;1062,3PRB72;165304}, the zero-field Hamiltonian matrix elements are expanded about the high-symmetry points (e.g., the K point in graphene). And then, the magnetic quantization is further done from an approximate Hamiltonian matrix, in which the magnetic Bloch wave function is assumed to be the superposition of the well-behaved and  sublattice-dependent oscillator states\cite{3PRB83;165443}. The zero-field and magnetic Hamiltonian matrices present the same dimension. It should be noticed that certain significant interlayer hopping integrals in 2D layered materials will create much difficulty in the investigation of magnetic quantization; therefore, they are usually ignored under the effective-mass approximation. Consequently, the unique and diverse magnetic quantization phenomena are frequently lost within this method, e.g., the frequent Landau-level anticrossing phenomenon in ABC-stacked trilayer graphene, and the extra magneto-optical selection rules\cite{3PRB90;205434,3PCCP17;15921}. In general, this perturbation method cannot solve the low-symmetry systems with the non-monotonous energy dispersions and the multi-pair band structures. For AB-bt bilayer silicene, it should be impossible to deal with the low-lying energy bands from the perturbation approximation, mainly owing to very complicated intrinsic interactions and energy bands. That is to say, the effective-mass model is not suitable for expanding the low-energy electronic states from the K and T points simultaneously [discussed later in Fig. 7.1]. This model becomes too cumbersome  to generate the further magnetic quantization. Apparently, it is very difficult to comprehend the unusual Landau levels, being attributed to the unique Hamiltonian in bilayer silicene. It is in sharp contrast with the monolayer silicene case. \\

In the previous theoretical studies, the tight-binding model is developed using the $\vec{\bf{k}}$-scheme, but not the $\vec{\bf{r}}$-scheme. The magneto-electronic states are directly built from the original electronic states in the first Brillouin zone (the hexagonal Brillouin zone in silicene/graphene). However, it is not suitable to present the main features of Landau-level wavefunctions (oscillatory distributions in real space with the distinct localization centers; discussed later). Explicitly, the subenvelope functions could not be identified as the Landau-level wavefunctions since they are randomly distributed.. This scheme is very difficult to explore the fundamental physical properties under the spatially modulated/non-uniform magnetic field, the modulated electric field, and the composite magnetic and electric fields, e.g., the magneto-optical properties and magneto-Coulomb excitations. For the generalized tight-binding model developed in this book, the calculations are based on the layer-dependent sublattices in an enlarged unit cell, in which the magnetic-electronic energy spectra and wave functions are closely linked to the well-known Kubo formula and the modified random-phase approximation. \\

\subsection{Magneto-optical excitation theory}\label{ch3.2}

The main features of magneto-electronic  energy spectra and Landau-level wave functions will create the diverse optical excitation properties. When an electromagnetic wave exists in a condensed-matter systems, the occupied electronic states are excited to the unoccupied ones according to the conservation of momentum $\&$ energy and the Pauli principle. The optical excitations belong to the vertical channels, since the momenta of photon in the experimental frequency range are negligible, compared with the typical wave vectors of electrons. From the dynamic Kubo formula [the Fermi golden rule], the spectral functions in the presence and absence of external fields are
\begin{align}
A(\omega) \propto \sum_{h,h',m,m'} \int_{1stBZ} \frac {d\bf{k}}{(2\pi)^2}
\Big| \Big\langle \Psi^{h'} (\bf{k},m')
\Big| \frac{   \hat{\bf{E}}\cdot \bf{P}   } {m_e}
\Big| \Psi^{h}(\bf{k},m)    \Big\rangle \Big|^2 \nonumber \\
\times \mbox{Im} \Big[ \frac{f(E^{h'} (\bf{k},m')) - f(E^h (\bf{k},m))}
{E^{h'} (\bf{k},m')-E^h (\bf{k},m)-\omega - i\Gamma}  \Big]. \tag{3.3}
\end{align}
$h$ represents the valence or conduction band, $\textbf{P}$ is the momentum operator, $f(E^{h} (\bf{k},m)$ the Fermi-Dirac distribution function; $\Gamma$ the broadening parameter due to the various deexcitation mechanisms. The absorption spectrum is associated with the velocity matrix elements [the dipole perturbation; the first term] and the joint density of states [the second term]. The former will determine whether the inter-Landau-level transitions are available during the optical excitations. \\

The velocity matrix elements, as successfully done for graphene-related materials\cite{3IOPBook;CY} are evaluated under the gradient approximation in the form of
\begin{align}
\Big\langle \Psi^{h'} (\bf{k},m') \Big| \frac{\hat{\bf{E}}\cdot \bf{P}} {m_e}
\Big| \Psi^{h}(\bf{k},m) \Big\rangle & \cong  \frac{ \mathrm{\partial}}{\mathrm{\partial}k_y}
\Big\langle \Psi^{h'} (\bf{k},m') \Big| H \Big| \Psi^{h}(\bf{k},m)\Big\rangle \nonumber \\
& =  \sum_{\alpha, l,l'} \sum_{m,m' = 1}^{2R_B} \bigg (c_{A_{\alpha,\bf{k}}^{l,m}}^*  c_{A_{\alpha,\bf{k}}^{l',m'}}
\frac{ \mathrm{\partial}}{\mathrm{\partial}k_y}
\Big\langle A_{\alpha,\bf{k}}^{l,m} \Big| H \Big|  A_{\alpha,\bf{k'}}^{l',m'}\Big\rangle \nonumber \\
& \hspace{5mm} + c_{A_{\alpha,\bf{k}}^{l,m}}^*  c_{B_{\alpha,\bf{k'}}^{l',m'}}
\frac{ \mathrm{\partial}}{\mathrm{\partial}k_y  }
\Big\langle A_{\alpha,\bf{k}}^{l,m} \Big| H \Big|  B_{\alpha,\bf{k'}}^{l',m'}\Big\rangle \nonumber \\
& \hspace{5mm} + c_{B_{\alpha,\bf{k}}^{l,m}}^*  c_{A_{\alpha,\bf{k'}}^{l',m'}}
\frac{ \mathrm{\partial}}{\mathrm{\partial}k_y  }
\Big\langle B_{\alpha,\bf{k}}^{l,m} \Big| H \Big|  A_{\alpha,\bf{k'}}^{l',m'}\Big\rangle \nonumber \\
& \hspace{5mm} + c_{B_{\alpha,\bf{k}}^{l,m}}^*  c_{B_{\alpha,\bf{k'}}^{l',m'}}
\frac{ \mathrm{\partial}}{\mathrm{\partial}k_y  }
\Big\langle B_{\alpha,\bf{k}}^{l,m} \Big| H \Big|  B_{\alpha,\bf{k'}}^{l',m'}\Big\rangle
\bigg ). \tag{3.4}
\end{align}
Under this approximation, we do not need to really do the inner product of the left side in Eq. (4); that is, the wave functions of the 3$p_z$ orbitals are not included in the calculations, but only amplitudes [sub-envelope functions] are sufficient in the right-hand side of Eq. (4).
It clearly indicates that, the subenvelope functions can be used to investigate the magneto-absorption spectra. Additionally, the similar theoretical framework is very useful in understanding the quantum Hall conductivities. In general, one can fully explore the critical factors purely due to the characteristics of Landau levels, e.g., many symmetric delta-function-like absorption peaks with a uniform intensity in monolayer graphene\cite{3IOPBook;CY}. \\

The model calculations of magneto-absorption spectra are worthy of a closer examination, since they account for the existence of diverse selection rules. After solving the magneto-energy spectra and Landau-level wave functions, the matrices associated with them are directly utilized in evaluating the magneto-optical properties, e.g., the $1\times 16R_B$ and $16R_B\times 16R_B$ matrices for bilayer AB-bt silicene system. Apparently, the accurate magneto-optical data consume very much computer time. In general, the band-like Hamiltonian matrix and the localization features of Landau-level wave functions are available in greatly enhancing the efficiency of numerical calculations. And then, how to classify a lot of magneto-absorption peaks and to propose the conceive physical pictures becomes critically important. Under the gradient approximation [Eq. (3.4)], the dipole matrix element is mainly determined by the variations of the Hamiltonian matrix elements with the specific wave vector, in which their magnitudes are related to the intralayer and interlayer hopping integrals. One can first examine the specific matrix elements (the specific two sublattices) which create the largest atomic interaction. Second, the initial Landau-level state associated with a certain one sublattice is excited to the final Landau-level state related to another sublattice. If the inner product is sufficiently strong, this channel is available during the vertical excitations. Third, the similar examinations are conducted on the second-largest terms, and the test results are compared with the above-mentioned ones. Finally, the magneto-optical selection rules could be obtained from the delicate analyses. In addition, any optical absorption rules might be absent, as a result of the thorough destructions in the spatial distribution symmetries of Landau levels. \\

The model calculations of magneto-absorption spectra are worthy of a closer examination, since they account for the existence of diverse selection rules. After solving the magneto-energy spectra and Landau-level wave functions, the matrices associated with them are directly utilized in evaluating the magneto-optical properties, e.g., the $1\times 16R_B$ and $16R_B\times 16R_B$ matrices for bilayer AB-bt silicene system.
Apparently, the accurate magneto-optical data consume very much computer time. In general, the band-like Hamiltonian matrix and the localization features of Landau-level wave functions are available in greatly enhancing the efficiency of numerical calculations. And then, how to classify a lot of magneto-absorption peaks and propose the conceive physical pictures becomes critically important. Under the gradient approximation [Eq. (3.4)], the dipole matrix element is mainly determined by the variations of the Hamiltonian matrix elements with the specific wave vector, in which their magnitudes are related to the intralayer and interlayer hopping integrals. One can first examine the specific matrix elements (the specific two sublattices) which create the largest atomic interaction. Second, the initial Landau-level state associated with a certain one sublattice is excited to the final Landau-level state related to another sublattice. If the inner product is sufficiently strong, this channel is available during the vertical excitations. Third, the similar examinations are conducted on the second-largest terms, and the test results are compared with the above-mentioned ones. Finally, the magneto-optical selection rules could be obtained from the delicate analysis. In addition, any optical absorption rules might be absent, as a result of the thorough destructions in the spatial distribution symmetries of Landau levels. \\

The magneto-optical absorption spectrum of monolayer graphene is very suitable for illustrating a specific selection rule of $\Delta = n^c-n^v=\pm 1$. This gapless system, with the linearly isotropic Dirac-cone structure,  exhibits the symmetric magneto-electronic energy spectrum about the Fermi level, as shown in Fig. 3.3(a) at $B_z=40$ T. For the specific $(k_x=0, k_y=0)$ state, there are four-fold degenerate states being, respectively, localized at 1/6, 2/6, 4/6 and 5/6 [in the unit of a periodical length corresponding to an enlarged rectangle in Fig. 3.2]. Of course, each localization state makes the same contribution to any physical properties. For example, the 2/6-localization Landau levels in Fig. 3.3(b) possess the well-behaved spatial distributions, being similar to those an oscillator [Refs]. The quantum number is characterized from the dominating mode of the B sublattice, since its amplitude is finite at the Fermi level [$E_F=0$]. As a result, the low-lying Landau levels are initial from the Dirac point, i.e., the ordering of quantum is $n^{c,v}=0$, 1, 2, 3... The similar results are revealed for the 1/6-localization Landau levels, but they are dominated by the A sublattice. Most important, the number difference of zero points is just $\pm 1$ for the B and A sublattices, directly reflecting their equivalence under a honeycomb lattice. The above-mentioned Landau-level features create a lot of delta-function-like prominent peaks due to the inter-Landau-level transitions from the valence to conduction states [Fig. 3.4]. The Landau-level energy spacing is proportional to the square root of the magnetic-field strength. Such symmetric peaks have a uniform intensity in the low-frequency range because of the identical group velocity/the Fermi velocity. Moreover, the available absorption peaks only arise from the $n^v$ valence Landau level and the $n^c=n^v +1$ [$n^c=n^v -1$] conduction Landau level for the 2/6 [1/6] localization center. These selection rules are associated with the B- and A-sublattice wavefunctions, respectively, corresponding to the initial and final Landau-level states. That only the nearest-neighbor hopping integral between the A and B sublattices is the main reason. \\

\subsection{Transport theories}\label{ch3.3}

\subsubsection{Quantum Hall effects}\label{ch3.3.1}

Here the linearly static Kubo formula\cite{3IOPBook;CY,3PCCP19;29525}, which directly combines with the generalized tight-binding model, is very suitable in fully exploring investigate the rich and unique quantum Hall conductivities of few-layer graphene systems with the high- and low-symmetry stacking configurations. The developed model will be very useful in the accurate identifications of the magneto-electronic selection rules under the static case; that is, the available scattering channels of the magneto-transport properties can be thoroughly examined and determined by the delicately numerical analysis\cite{3PCCP19;29525}. The dependencies of quantum conductivity on the Fermi energy and magnetic-field strength are investigated in detail. The feature-rich Landau levels can create the extraordinary magneto-transport properties, such as, those in the sliding bilayer graphenes, the trilayer ABC stacking, and AAB-stacked graphene. The unusual Hall conductivities, discussed later in Chap. 8, are predicted to cover the integer and non-integer conductivities, the zero and non-vanishing conductivities at the neutral point, the well-like, staircase, abnormal and composite/complex quantum structures. These results will be deduced to be dominated by the frequent crossing and anti-crossing energy spectra, and the spatial oscillation modes. Apparently, the low-energy perturbation approximation cannot generate the above-mentioned important phenomena\cite{3PCCP19;29525}. \\

Within the linear response, the transverse Hall conductivity is calculated from the static Kubo formula \cite{3IOPBook;CY,3PCCP19;29525}.
\begin{equation}
\sigma_{xy} = \frac {ie^2 \hbar} {S} \sum_{\alpha} \sum_{\alpha \neq \beta} (f_{\alpha} - f_{\beta})
\frac {\langle \alpha  |\bf{\dot{u}}_{x}| \beta\rangle  \langle \beta |\bf{\dot{u}}_{y}|\alpha \rangle} {(E_{\alpha}-E_{\beta})^2}.
\tag{3.5}
\end{equation}
$|\alpha >$/${|\beta >}$] is the initial/final Landau-level state with energy $E_\alpha$/$E_\beta$, S the area of the $B_z$-enlarged unit cell, $f_\alpha$ the Fermi-Dirac distribution functions, and $\bf{\dot{u}}_{x}$ the velocity operator along the $x$-direction. The matrix elements of the velocity operators, which can determine the available inter-Landau-level transitions, are solved under the gradient approximation, as discussed earlier in Eq. (4)\cite{3PCCP19;29525} :
\begin{align}
\langle \alpha  |\bf{\dot{u}}_{x}| \beta\rangle = \frac {1}{\hbar} \langle \alpha  |\frac {\partial H} {\partial k_x} | \beta\rangle \nonumber \\
\langle \alpha  |\bf{\dot{u}}_{y}| \beta\rangle = \frac {1}{\hbar} \langle \alpha  |\frac {\partial H} {\partial k_y} | \beta\rangle.
\tag{3.6}
\end{align}
As for few-layer graphene systems, such matrix elements are dominated by the intralayer nearest-neighbor intercation \cite{3PCCP19;29525} so that the quantized mode of the initial state on the A$^l$ sublattice must be identical to that of the final state on the B$^l$ sublattice. Apparently, the well-behaved, perturbed and undefined Landau levels of layered graphenes\cite{3IOPBook;CY}, accompanied with various magnetic selection rules\cite{3IOPBook;CY,3PCCP17;15921}, are expected to exhibit the rich and unique quantum conductivities.\\

Equation (3.5) is used to calculate the Hall conductivities of monolayer graphene, clearly indicating the typical quantum phenomenon. The Fermi energy-dependent $\sigma_{xy}$, as shown in Fig. 3.5(a), exhibits a lot of uniform-height plateau structures, but with the non-homogenous widths being inversely proportional to the square-root of $E_F$. The quantum strictures directly reflects the fact that the Landau-level states become fully occupied or unoccupied at their initial and final positions. The same height of ${4e^2/h}$ should be closely related to all the available scattering events. By the detailed analysis on the enumerator in Eq. (3.5), the velocity matrix elements, which are determined by the initial/occupied and final/unoccupied Landau-level states, strongly depend on  the well-behaved symmetries of spatial distributions on two equivalent A and B sublattices. As a result, the static selection rule of $\Delta n=\pm 1$ is identical to the dynamic ones\cite{3PCCP19;29525}. Specifically, there exists a staircase of ${4E^2/h}$ height at the neutral point. This result is due to $n^v=0 \rightarrow n^c=1$, $n^v=1 \rightarrow n^2=1$, $n^v=2 \rightarrow n^c=3$ ... etc, in which the deeper-energy valence Landau levels would make the small contributions. It is consistent with the analytical calculations from the effective-mass approximation\cite{3PRB83;165443}. The similar results are revealed in the magnetic-field-dependent quantum Hall conductivities, e.g., $\sigma_{xy}$ in Fig. 3.5(b). The widths of plateau structures are proportional to the square-root of $B_z$, directly reflecting the key feature of the magneto-electronic energy spectra\cite{3IOPBook;CY}. \\

\subsubsection{Magneto-heat capacity}\label{ch3.3.2}

The electronic heat capacities in graphene-related systems are very sensitive to the changes in the magnetic-field strength and temperatures.
The $T$-dependent total mean energy of an electronic spectrum under a uniform perpendicular magnetic field is characterized as
\begin{equation}
U(T,B_z)=\frac{3\sqrt{3}b^2}{4}N_A\sum_{\sigma,h,n}\int_{1stB.Z.}\frac{d^2k}{(2\pi)^2}[E^h_n(\vec{k})-\mu]f((E^h_n(\vec{k})-\mu)). \tag{3.7}
\end{equation}
Furthermore, the electronic specific heat, being defined as the variation of the total energy with temperature, is
\begin{align}
C(T,B_z)&=\frac{\partial {U(T, B_z)}}{\partial T} \nonumber\\
&= \frac{3\sqrt{3}b^2}{4}\frac{N_A}{\beta T} \sum_{\sigma,h,n}\beta^2(E^h_n-\mu)^2\frac{exp[\beta(E^h_n-\mu)]}{\{1+exp[\beta(E^h_n-\mu)]\}}\int_{1stB.Z.}\frac{d^2k}{(2\pi)^2}.\tag{3.8}
\end{align}
$N_A$ is the Avogadro number. $h(=c \& v)$ represents the conduction and valence Landau levels, making the important contributions to the thermal property. $\beta =1/k_BT$ and $k_B$ is the Boltzmann constant.\\

The low-temperature magneto-electronic specific heat strongly depends on the Landau-level energy spectrum and the Zeeman spitting effect, even if the latter is very small [$\sim 1$ meV] under $B_z\sim 10$ T. Each Landau-level state has a total energy, as expressed by
\begin{equation}
E^h_n(\sigma,B_z) = E^{c,v}_n(B_z)+E_z(\sigma,B_z), \tag{3.9}
\end{equation}
and
\begin{equation}
E_z(\sigma,B_z) = (g\sigma \pi)/m^{\ast}(B_z/\phi_0). \tag{3.10}
\end{equation}.
The Zeeman splitting energy, $E_z(\sigma, B_z)$, plays a critical role in the low-energy free carrier distributions and thus the magneto specific heat. The $g$ factor ($\sim$2) is identical to that of pure graphite\cite{3PRB67;045405}, $\sigma$ ($=\pm 1/2$) stands for the electron spin, $\phi_0$ ($=hc/e$) corresponds to the magnetic flux quantum, and $m^\ast$ denotes the bare electron mass. For example, the specific heat in monolayer graphene is almost vanishing at $T\le 10$ K in the absence of the Zeeman splitting [details in Chap. 8.3]; furthermore, there is no special structure at room temperature.\\

\subsection{Modified Random-phase approximation for magneto-electronic Coulomb excitations}\label{ch3.4}

As to magneto-Coulomb excitations, monolayer germanene is chosen for a model study, since this system possesses an important spin-orbital coupling [SOC], a significant buckling structure, and the single-orbital-dominated low-lying energy bands. Furthermore, the external electric and magnetic fields could be applied to greatly diversify the single-particle excitations and the magnetoplasmon modes. Similar to graphene, germanene/silicene, as clearly shown in Fig. 3.6, consists of a honeycomb lattice with  A and B sublattices. The latter presents the non-planar structure, in which the two sublattice planes are separated by a distance of $2\ell$ ($\ell=0.33$ $\AA$), as clearly illustrated in Fig. 3.6. The low-energy Hamiltonian, which is built from the spin-dependent $4p_z$-orbital tight-binding functions of Ge atoms, is given by\cite{3SciRep7;40600}.
\begin{align}
H=&-\gamma_0\sum_{\langle I,J\rangle,\alpha}c^{\dag}_{I\alpha}c_{J\alpha}+i\frac{\lambda_{SOC}}{3\sqrt{3}}\sum_{\langle\langle I,J\rangle\rangle,\alpha,\beta}v_{IJ}c^{\dag}_{I\alpha}\sigma^{z}_{\alpha\beta}c_{J\beta}\nonumber \\
&-i\frac{2}{3}\lambda_{R2}\sum_{\langle\langle I,J\rangle\rangle,\alpha,\beta}u_{IJ}c^{\dag}_{I\alpha}(\vec{\sigma}\times \hat{d}_{IJ})^{z}_{\alpha\beta}c_{J\beta}  \nonumber  \\
&+ \ell\sum_{I,\alpha}\mu_{I}E_{z}c^{\dag}_{I\alpha}c_{I\alpha}  \  . \tag{3.11}
\end{align}
Equation (11) could be obtained from Eq. (1) by ignoring the layer-dependent Coulomb potential energies, spin0orbital interactions and hopping integrals. The sums are carried out for nearest-neighbors $\langle I,J\rangle$ or next-nearest-neighbor  lattice site pairs $\langle\langle I,J \rangle\rangle$. The first term (I) in Eq. (\ref{eq:1}) accounts  for  nearest-neighbor hopping with energy transfer $\gamma_0=0.86$ eV.
The second term (II) describes the effective spin-orbital coupling for parameter $\lambda_{SOC}=46.3$ meV. Additionally, we chose $v_{IJ}=\pm 1$ if the next-nearest-neighbor hopping is anticlockwise/clockwise  with respect to the positive $z$-axis. In the third term (III), the intrinsic Bychkov-Rashba spin-orbital coupling is included in the next-nearest neighbor hopping through $\lambda_{R2}=10.7$ meV, in which $u_{IJ}=\pm 1$ are for the A/B lattice sites, respectively. $\hat{d}_{IJ}=\vec{d}_{IJ}/|d_{IJ}|$ is a unit vector joining two sites $I$ and $J$ on the same sublattice  as shown in Fig. 1(a). The staggered sublattice potential energy produced by the external electric field is characterized by the fourth term (IV), where $\mu_I=\pm 1$ for the A/B sublattice sites and  $\ell=0.33$ $\AA$, referring to Fig. 1(b).\\

Monolayer germanene is assumed to be in a uniform perpendicular magnetic field. The magnetic flux, the product of the field strength and the hexagonal area, is $\Phi= [3\sqrt{3}b^{2}B_{z}/2]/\phi_{0}$, where $b$ (=2.32 $\AA$) is the lattice constant. The vector potential, $\vec{A} = (B_{z}x)\hat{y}$, leads to a new period along the armchair direction, since it can create an extra magnetic Peierls phase, i.e., $\exp\{i[\frac{2\pi}{\phi_{0}}\int \vec{A}\cdot d\vec{r}] \}$. The unit cell is thus enlarged and its dimension is determined by $R_{B}= 1/\Phi$. The reduced first Brillouin zone has an area of $1/(3\sqrt{3}b^{2}R_{B})$. The enlarged unit cell contains 4$R_{B}$ Ge atoms and the Hamiltonian matrix is a $8 R_B\times 8 R_{B}$ Hermitian matrix with the spin degree of freedom. Corresponding to $B_z=10$ T, where $R_{B}=3000$, the Hamiltonian has a dimension of $24000\times 24000$.\\

Only the fourth interaction in Eq. (\ref{eq:1}) is a diagonal matrix regardless of the Peierls phase. By the detailed calculations, the independent Hamiltonian matrix elements, which are due to the extra position-dependent Peierls phases, are expressed in the following Eqs. (12)-(20 [the sublattice-, site- and SOC-dependent  magnetic Hamiltonian matrix elements].
\begin{align}
[I]~~~~\langle B^\alpha_{J}|H|A^\beta_{I}\rangle & =\gamma_0\sum_{\langle I,J\rangle}{1\over
N}\exp{[i\vec{k}\cdot(\vec{R}_{A^\beta_{I}}-\vec{R}_{B^\alpha_{J}})]} \nonumber  \\
& \hspace{5mm} \times\exp{\{i[{2\pi\over \phi_0}\int ^{\vec{R}_{B^\alpha_{J}}}_{
\vec{R}_{A^\beta_{I}}}\vec{A}\cdot d\vec{r}]\}} \nonumber  \\
& =\gamma_0t_{1,I}\delta_{I,J+1}\delta_{\alpha,\beta}+\gamma_0s\delta_{I,J}\delta_{\alpha,\beta}  \  , \tag{3.12}
\end{align}
where $\vec{R}_{A^\beta_{I}}$ and $\vec{R}_{B^\alpha_{J}}$ denote  the lattice sites for the  $A$ and $B$ sublattices, with the spin polarizations $\beta$ and $\alpha$, respectively. Also, $I$ [$J$] corresponds to the initial [final] sublattice site. Concerning the  kinetic energy, the nearest-neighbor matrix element includes the phase-dependent $t_{1,I}=\exp\{i[-k_x{b\over 2}-k_y{\sqrt{3}b\over 2}+\pi{\Phi\over \phi_0}(I-1+{1\over 6})]\}+\exp\{i[-k_x{b\over 2}+k_y{\sqrt{3}b\over 2}-\pi{\Phi\over \phi_0}(I-1+{1\over 6})]\}$ and a specific term of $s=\exp[i(-k_{x}b)]$. The vector potential can create more complicated hopping phases in the SOC-related interactions, as presented in Eqs. (13)-(20). For example, $t_{2,1}$ ($t_{8,1}$) represents the phase term for the next-nearest-neighbor hopping from A$^{\beta}_{1}$ to A$^{\alpha}_{1}$ (A$^{\beta}_{1}$ to A$^{\alpha}_{2}$), corresponding to the effective SOC (intrinsic Rashba SOC). An illustration of the SOC-induced hopping phases is clearly indicated in Fig. 1(d). Next,
\begin{align}
[II]~~~~\langle A^\alpha_{J}|H|A^\beta_{I}\rangle & =\frac{\lambda_{SOC}}{3\sqrt{3}}\sum_{\langle\langle I,J\rangle\rangle}{1\over
N}\exp{[i\vec{k}\cdot({\vec{R}}_{A^\beta_{I}}-{\vec{R}}_{A^\alpha_{J}})]} \nonumber  \\
& \hspace{5mm} \times\exp{\{i[{2\pi\over \phi_0}\int ^{{\vec{R}}_{A^\alpha_{J}}}_{{\vec{R}}_{A^\beta_{I}}}{\vec{A}} \cdot d{\vec{r}}]\}}+E_{z}\ell \nonumber  \\
& =\frac{\lambda_{SOC}}{3\sqrt{3}}t_{2,I}\delta_{I,J}\delta_{\alpha,\beta}+E_{z}\ell  \ , \tag{3.13}
\end{align}
where $t_{2,I}=\exp i[k_{y}a+2\pi{\Phi\over\phi_0}(I-1)]-\exp i[-k_{y}a-2\pi{\Phi\over\phi_0}(I-1)]$.  Furthermore,
\begin{align}
\langle B^\alpha_{J}|H|B^\beta_{I}\rangle & =\frac{\lambda_{SOC}}{3\sqrt{3}}\sum_{\langle\langle I,J\rangle\rangle}{1\over
N}\exp{[i{\vec{k}}\cdot({\vec{R}}_{B^\beta_{I}}-{\vec{R}}_{B^\alpha_{J}})]} \nonumber  \\
& \hspace{5mm} \times\exp{\{i[{2\pi\over \phi_0}\int ^{{\vec{R}}_{B^\alpha_{J}}}_{{\vec{R}}_{B^\beta_{I}}}{\vec{A}}\cdot d{\vec{r}}]\}}-E_{z}\ell \nonumber  \\
& =\frac{\lambda_{SOC}}{3\sqrt{3}}t_{3,I}\delta_{I,J}\delta_{\alpha,\beta}-E_{z}\ell \ , \tag{3.14}
\end{align}
where $t_{3,I}=\exp i\{-k_{y}a-2\pi{\Phi\over\phi_0}[(I-1)+\frac{1}{3}]\}-\exp i\{k_{y}a+2\pi{\Phi\over\phi_0}[(I-1)+\frac{1}{3}]\}$.
\begin{align}
\langle A^\alpha_{J}|H|A^\beta_{I}\rangle & =\frac{\lambda_{SOC}}{3\sqrt{3}}\sum_{\langle\langle I,J\rangle\rangle}{1\over
N}\exp{[i{\vec{k}}\cdot({\vec{R}}_{A^\beta_{I}}-{\vec{R}}_{A^\alpha_{J}})]} \nonumber  \\
& \hspace{5mm} \times\exp{\{i[{2\pi\over \phi_0}\int ^{{\vec{R}}_{A^\alpha_{J}}}_{{\vec{R}}_{A^\beta_{I}}}{\vec{A}}\cdot d{\vec{r}}]\}} \nonumber  \\
& =\frac{\lambda_{SOC}}{3\sqrt{3}}t_{4,I}\delta_{I,J-1}\delta_{\alpha,\beta} \ ,  \tag{3.15}
\end{align}
where $t_{4,I}=\exp i\{k_{x}\frac{3}{2}b-k_{y}\frac{a}{2}-\pi{\Phi\over\phi_0}[(I-1)+\frac{1}{2}]\}-
\exp i\{k_{x}\frac{3}{2}b+k_{y}\frac{a}{2}+\pi{\Phi\over\phi_0}[(I-1)+\frac{1}{2}]\}$. In addition,
\begin{align}
\langle B^\alpha_{J}|H|B^\beta_{I}\rangle & =\frac{\lambda_{SOC}}{3\sqrt{3}}\sum_{\langle\langle I,J\rangle\rangle}{1\over
N}\exp{[i{\vec{k}}\cdot({\vec{R}}_{B^\beta_{I}}-{\vec{R}}_{B^\alpha_{J}})]} \nonumber  \\
& \hspace{5mm} \times\exp{\{i[{2\pi\over \phi_0}\int ^{{\vec{R}}_{B^\alpha_{J}}}_{{\vec{R}}_{B^\beta_{I}}}{\vec{A}}\cdot d{\vec{r}}]\}} \nonumber  \\
& =\frac{\lambda_{SOC}}{3\sqrt{3}}t_{5,I}\delta_{I,J-1}\delta_{\alpha,\beta} \ ,  \tag{3.16}
\end{align}
where $t_{5,I}=\exp i\{k_{x}\frac{3}{2}b-k_{y}\frac{a}{2}-\pi{\Phi\over\phi_0}[(I-1)+\frac{5}{6}]\}-\exp i\{k_{x}\frac{3}{2}b+k_{y}\frac{a}{2}+\pi{\Phi\over\phi_0}[(I-1)+\frac{5}{6}]\}$.
\begin{align}
[III]~~~~\langle A^\alpha_{J}|H|A^\beta_{I}\rangle_{\alpha\neq\beta} & =\frac{2}{3}\lambda_{R2}\sum_{\langle\langle I,J\rangle\rangle}{1\over
N}\exp{[i{\vec{k}}\cdot({\vec{R}}_{A^\beta_{I}}-{\vec{R}}_{A^\alpha_{J}})]} \nonumber  \\
& \hspace{5mm} \times\exp{\{i[{2\pi\over \phi_0}\int ^{{\vec{R}}_{A^\alpha_{J}}}_{{\vec{R}}_{A^\beta_{I}}}{\vec{A}}\cdot d{\vec{r}}]\}}  \nonumber  \\
& =\frac{2}{3}\lambda_{R2}t_{6,I}\delta_{I,J} \ , \tag{3.17}
\end{align}
where $t_{6,I}=\exp i[k_{y}a+2\pi{\Phi\over\phi_0}(I-1)-\frac{\pi}{2}]+\exp i[-k_{y}a-2\pi{\Phi\over\phi_0}(I-1)+\frac{\pi}{2}]$.
Similarly,
\begin{align}
\langle B^\alpha_{J}|H|B^\beta_{I}\rangle_{\alpha\neq\beta} & =\frac{2}{3}\lambda_{R2}\sum_{\langle\langle I,J\rangle\rangle}{1\over
N}\exp{[i{\vec{k}}\cdot({\vec{R}}_{B^\beta_{I}}-{\vec{R}}_{B^\alpha_{J}})]} \nonumber  \\
& \hspace{5mm} \times\exp{\{i[{2\pi\over \phi_0}\int ^{{\vec{R}}_{B^\alpha_{J}}}_{{\vec{R}}_{B^\beta_{I}}}{\vec{A}}\cdot d{\vec{r}}]\}} \nonumber  \\
& =\frac{2}{3}\lambda_{R2}t_{7,I}\delta_{I,J} \ , \tag{3.18}
\end{align}
where $t_{7,I}=\exp i\{k_{y}a+2\pi{\Phi\over\phi_0}[(I-1)+\frac{1}{3}]-\frac{\pi}{2}\}+\exp i
\{-k_{y}a-2\pi{\Phi\over\phi_0}[(I-1)+\frac{1}{3}]+\frac{\pi}{2}$\}.
\begin{align}
\langle A^\alpha_{J}|H|A^\beta_{I}\rangle_{\alpha\neq\beta}& =\frac{2}{3}\lambda_{R2}\sum_{\langle\langle I,J\rangle\rangle}{1\over
N}\exp{[i{\vec{k}}\cdot({\vec{R}}_{A^\beta_{I}}-{\vec{R}}_{A^\alpha_{J}})]} \nonumber  \\
& \hspace{5mm} \times\exp{\{i[{2\pi\over \phi_0}\int ^{{\vec{R}}_{A^\alpha_{J}}}_{{\vec{R}}_{A^\beta_{I}}}{\vec{A}}\cdot d{\vec{r}}]\}} \nonumber  \\
& =\frac{2}{3}\lambda_{R2}t_{8,I}\delta_{I,J-1} \ , \tag{3.19}
\end{align}
where $t_{8,I}=\exp i\{k_{x}\frac{3}{2}b+k_{y}\frac{a}{2}+\pi{\Phi\over\phi_0}[(I-1)+\frac{1}{2}]
-\frac{\pi}{6}\}+\exp i\{k_{x}\frac{3}{2}b-k_{y}\frac{a}{2}-\pi{\Phi\over\phi_0}[(I-1)+\frac{1}{2}]+\frac{\pi}{6}$\}.
Finally, we have
\begin{align}
\langle B^\alpha_{J}|H|B^\beta_{I}\rangle_{\alpha\neq\beta} & =\frac{2}{3}\lambda_{R2} \sum_{\langle\langle I,J\rangle\rangle}{1\over
N}\exp{[i{\vec{k}}\cdot({\vec{R}}_{B^\beta_{I}}-{\vec{R}}_{B^\alpha_{J}})]} \nonumber  \\
& \hspace{5mm} \times\exp{\{i[{2\pi\over \phi_0}\int ^{{\vec{R}}_{B^\alpha_{J}}}_{{\vec{R}}_{B^\beta_{I}}}{\vec{A}}\cdot d{\vec{r}}]\}} \nonumber  \\
& =\frac{2}{3}\lambda_{R2}t_{9,I}\delta_{I,J-1} \ , \tag{3.20}
\end{align}
where $t_{9,I}=\exp i\{k_{x}\frac{3}{2}b+k_{y}\frac{a}{2}+\pi{\Phi\over\phi_0}[(I-1)+\frac{5}{6}]-
\frac{\pi}{6}\}+\exp i\{k_{x}\frac{3}{2}b-k_{y}\frac{a}{2}-\pi{\Phi\over\phi_0}[(I-1)+\frac{5}{6}]+\frac{\pi}{6}$\}. To obtain the Landau-level energy spectra and wave functions efficiently,  it needs to  diagonalize a giant magnetic Hamiltonian matrix under the exact method.\\

The longitudinal Coulomb excitations of monolayer germanene is characterized by the magneto-dielectric function [$\epsilon (q, \omega, B_z)$] and the energy loss function [$-1/\epsilon (q, \omega, B_z)$], in which the former and the latter, respectively, represent the bare and screened response functions of valence and conduction Landau-level electrons. According to the definition, the ratio of the electric displacement [the external Coulomb potential] and the effective field [the effective Coulomb potential] is the longitudinal dielectric function of $\epsilon (q, \omega, B_z)$. When a 2D condensed-matter system experiences the external perturbation of the time-dependent Coulomb potential [e.g., the incident electron (light) beam], all the occupied electronic states simultaneously exhibit very complicated screening phenomena. Apparently, the dynamic electron-electron Coulomb interactions belong to the inelastic many-particle scattering. It is impossible to exactly solve the effective Coulomb potentials between two electrons under the various theoretical models, mainly owing to the infinite scattering processes. The previous studies show that a lot of approximate methods have been developed to investigate the induced Coulomb potential due to the screening charges, such as, the random-phase\cite{3PRB34;2,3PRB74;085406,3PLA352;446}, Hubbard\cite{3PRSL243;336} and Singwi-Sjolander approximations\cite{3PR176;589,3PR6;875}. And then, the total Coulomb potential is only the summation of the external and induced ones, or the dielectric function is easily obtained from their relations.\\

In general, the specific random-phase approximation [RPA], which only covers the bubble-like electron-hole pair excitations [a series of similar Feynman diagrams] in the absence of the vertex corrections\cite{3PRB34;2,3PRB74;085406,3PLA352;446}, is frequently utilized to thoroughly explore the Coulomb excitations even for the low-dimensional systems with the strong correlation effects. Furthermore, it is very suitable for studying the magnet-electronic excitation spectra in the presence of a uniform perpendicular magnetic field, e.g., the various  magnetoplasmon modes, the electric-field-enriched Coulomb excitations, and the diverse (momentum, frequency)-phase diagrams. Under this approximation, one can get
\begin{equation}
\epsilon(q,\omega, B_z)=\epsilon_{1}(q,\omega, B_z)+i\epsilon_{2}(q,\omega, B_z)=\epsilon_{0}-v_{q}\chi^{0}(q,\omega, B_z). \tag{3.21}
\end{equation}
$v_{q}=2\pi e^{2}/q$ represents the in-plane Fourier transformation of the bare Coulomb potential energy, and $\epsilon_{0}$=2.4 (taken from that of graphite \cite{3PRB34;2}) is the background dielectric constant due to the deep-energy electronic states. It should be noticed that the variation of $\epsilon_{0}$ will lead to  a vertical shift of the real part of the dielectric function and thus alter the peak intensity and position in the energy loss function. However, the main features of collective excitations keep the same. Such a form of dielectric function is frequently revealed  in many research studies on two-dimensional systems, theoretically\cite{3PLA352;446,3PRB74;085406} and experimentally \cite{3Carbon50;183,3PRB83;161403,3NanoLett14;3827}. Furthermore, the induced Coulomb potential is proportional to the effective one within the linear response approximation, in which the coefficient is $-v_q\chi^0 (q, \omega, B_z)$. The 2D bare response function\cite{3PRB34;2} is expressed in the second term of Eq. (21). The bra and ket account for the initial and final Landau state wavefunctions with the respective quantum numbers of $n$ and $m$, where the wave vectors are $\vec{k}$ and $\vec{k}+\vec{q}$. In the presence of $B_z \hat{z}$, the electronic states in a 2D system becomes fully quantized; therefore, the summation is done for all the available inter-Landau-level transitions at any temperature
\begin{align}
\chi^{0}(q,\omega, B_z) & =\frac{1}{3\sqrt{3}b^{2}R_{B}}
\sum_{n,m}|\langle n;\vec{k}+\vec{q}|e^{i\vec{q}\centerdot\vec{r}}|m;\vec{k}\rangle|^{2} \nonumber\\
& \hspace{5mm} \times \frac{f(E_{n})-f(E_{m})}{E_{n} -E_{m}-(\omega +i\Gamma)}\ . \tag{3.22}
\end{align}
The pre-factor $1/(3\sqrt{3}b^2 R_{B})$ is a normalization constant, meaning the same contributions arising from the highly degenerate Landau-level states in the reduced first Brillouin zone. The equilibrium Fermi-Dirac distribution function is $f(E)=1/[1+exp(E-\mu/k_{B}T)]$. $\Gamma$ is an energy broadening parameter induced by the various de-excitation mechanisms. $\mu$ is the temperature-dependent chemical potential, where the $T$-dependence might be negligible under the specific magnetic energy range in this work. The bare response function only relies on the magnitude of the momentum transfer under the isotropic Landau-level energy spectrum. The energy broadening of the Landau levels associated with the lattice structure is almost vanishing at low temperature. In addition, this work is focused on the low-energy magneto-electronic excitations. $q$ is much smaller than the reciprocal lattice vector; that is, the local-field effects can be ignored\cite{3ACSNano5;1026}. \\

The detailed calculations for the magneto-Coulomb matrix elements are are clearly illustrated as
\begin{align}
\langle n;\vec{k}+\vec{q}|e^{i\vec{q}\centerdot\vec{r}}|m;\vec{k}\rangle  \hspace{10cm}
\nonumber\\
= \sum_{s=\alpha,\beta}\sum_{I=1-8R_{B_{0}}}\langle \phi_{z}(\vec{r}-
\vec{R}_{I})|e^{-i\vec{q}\cdot(\vec{r}-\vec{R}_{I})}|\phi_{z}(\vec{r}-\vec{R}_{I})\rangle
[u_{nsI}(\vec{k}+\vec{q})u_{msI}^{*}(\vec{k})]  \ .  \tag{3.23}
\end{align}
Here, $\vec{R}_{I}$ defines the positions of atoms in a unit cell. $u_{msI}(\vec{k})$ ($u_{nsI}(\vec{k}+\vec{q})$) are the coefficients for the TB wavefunctions derived from Eqs. (11)-(20).
$\langle\phi_{z}(\vec{r}-\vec{R}_{I})|e^{-i\vec{q}\cdot(\vec{r}-\vec{R}_{I})}|\phi_{z}(\vec{r}-\vec{R}_{I})\rangle =C(q)=[1+[\frac{qa_{0}}{Z}]^{2}]^{-3}$ was calculated by using hydrogenic wavefunction, where $a_{0}$ is the Bohr radius and $Z$ is an effective core charge\cite{3PRB34;2}. We note that the overlapping integrals between neighboring atoms are neglected, an approximation made originally in the 2D model by Blinowski et al\cite{3JPP41;47}. For the small transferred momenta,, $C(q)$ is very close to 1. Since all the $\pi$-electronic states are included in the calculations, the strength and frequency of the resonances in $Im[-1/\epsilon(q,\omega)]$ can be correctly defined. Moreover, the calculations would be reliable in a wide range of the field strength and the chemical potential.\\

The screened response functions, the effective energy loss functions, will become very complicated when a 2D $N$-layer system is taken into account. All the perturbed Coulomb potentials on the different layers need to be included in the theoretical calculations, and so do the layer-dependent screened charge distributions [indicated Coulomb potentials]. The linear response of RPA is utilized to establish a specific relation between the induced and effective Coulomb potentials [details in Ref.\cite{3PLA352;446}]. The dielectric function is no longer a scaler function, while it is replaced by an $N \times N$ dielectric tensor under the Dyson equation. It should be noticed that this function is characterized by the layer indices, but not the energy subband ones being used in the previous studies for the multi-band systems\cite{3PLA352;446}. And then, the effective energy loss spectra, being defined for the layered systems, are accurately derived under the inelastic scattering approximation.  Moreover, the modified layer-dependent RPA is consistent with the layer- and sublattice-based generalized tight-binding model [discussed earlier in Sec. 3.1. That is to say, it is also suitable for the magneto-electronic single-particle and collective excitations. Such formula has been successfully utilized to fully explore the electronic excitations of few-layer graphene systems without/with the magnetic and electric fields, e.g., the diversified electron-hole excitation regions and plasmon/magnetoplasmon modes due to the number of layers and stacking configurations\cite{3PRB74;085406,3PRB98;041408}.\\

\newpage
\begin{center}
\section{Concluding remarks}\label{ch12}
\author{ Shih-Yang Lin,\textit{$^{e,\S}$} Thi-Nga Do,\textit{$^{b,c,\ast}$} Chiun-Yan Lin,\textit{$^{a}$} Jhao-Ying Wu,\textit{$^{d,\dag}$}\\ Po-Hsin Shih,\textit{$^{a,\ddag}$} Ching-Hong Ho,\textit{$^{d,\P}$} Ming-Fa Lin\textit{$^{a,\sharp}$}}
\end{center}
\vskip 1.0 truecm

Apparently, this current book presents the delicately theoretical frameworks about the diverse quantization phenomena, especially for those due to a uniform perpendicular magnetic field and lattice symmetries in layered condensed-matter systems. The generalized tight-binding model, the dynamic Kubo formula, the static one, and the sublattice- and layer-dependent random-phase approximation are developed/modified to thoroughly explore the electronic properties, optical absorption spectra, quantum transports, and Coulomb excitations under the magnetic quantization, respectively. Furthermore, the first method is closely combined with the second/third/fourth one. The existence of many-particle excitonic effects in absorption spectra of layered materials and the modified magneto-optical theory needs to be thoroughly clarified in the near-future works. All the critical mechanisms, the planar or buckling structures, the uniform or nonuniform geometries, the different layer numbers, the distinct stacking configurations, the atom-induced ionization potentials, the strong electron/hole dopings, the significant guest-atom substitutions, the important intralayer and interlayer hopping integrals, the various spin-orbital couplings, the intrinsic intralayer and interlayer Coulomb interactions, and the electric and magnetic fields, are included in the calculations simultaneously using the very efficient manners. Part of calculated results are consistent with the high-resolution measurements {Zhang;156801}, e.g., the low-lying Hall conductivities of the monolayer, AB- and ABC-graphene systems \cite{Zhang;156801}, and the magneto-optical selection rules of few-layer AB-stacked graphene systems. However, most of them require a series of experimental examinations, such as, the spin-orbital-diversified Landau levels and magneto-optical absorption spectra in bilayer silicene of AB and AA stackings, four kinds of Landau levels and selection rules in Si-substituted graphene, and the doping- and field-modulated plasmon modes in monolayer germanene. The above-mentioned models are very useful in understanding the essential physical properties of emergent materials, such as, few-layer Si \cite{SRp3;1075,PRL95;226801}, Ge \cite{PRB87;235426,PRB89;085429}, Sn \cite{NJP21;023010}, Pb \cite{Brunt;5942}, GaAs \cite{Akiyama;L843}, P \cite{PRB92;165409,EPL119;37005,NanoLett18;229}, Sb \cite{Singh;6386}, Bi \cite{Chen;7290}, MoS$_2$ \cite{PRL110;066803}, and topological insulators \cite{Luo;29516}.\\

The analytic derivations and numerical evaluations are available in fully exploring the diverse quantization phenomena. We have successfully established the generalized tight-binding model which is based on the layer- and sublattice-dependent subenvelope functions [Refs]. The developed theoretical framework is suitable for the various geometric structures, the uniform/non-uniform external fields, the multi-orbital hybridizations in chemical bonds, and the environment-induced spin-orbital couplings [Refs]. It can analytically deal with the dramatic changes in the extra phases and magnitude of the neighboring hopping integrals. The geometry- and field-created unit cell is directly reflected in the  Hamiltonian matrix as a giant Hermitian one, being efficiently solved by the exact diagonalization method. The necessary calculation parameters in the calculations are examined from the well fitting of the low-lying energy bands at the high-symmetry points with the first-principles results [Refs]. The main features of electronic properties can be utilized under the very effective ways to comprehend the absorption spectra, quantum Hall conductivities,  and electronic Coulomb excitations, such as, the same contributions from the highly degenerate Landau levels, the localized probability distributions of the magnetic wave functions in largely reducing the computer time, the specific oscillation modes and strong couplings of the initial and final states for the well decision of magneto-optical selection rules, and the Coulomb matrix elements due to the subenvelope functions for determining the bare and screened response behaviors.\\

The twisted bilayer graphenes show the diversified fundamental properties for the Moire superlattices with the non-uniform physical environments; furthermore, such systems are quite different from the sliding ones or the normal stackings. Most important, the generalized tight-binding model can well characterize the complex subgroups of the highly degenerate Landau levels, further illustrating that this model is very suitbale for exploring the diverse magnetization phenomena even under the various geometric structures and intrinsic interactions. The electronic energy spectra, van Hove singularities, optical transition spectra, and magnetic subenvelop functions strongly depend on the multi-combined effects arising from the stacking symmetries/twisted angles, gate voltages and magnetic fields. There are many pairs of ${2p_Z}$-induced valence and conduction energy subbands, mainly owing to the zone folding effect. However, two pairs appear in the bilayer AA nd AB stackings. Among of them, the lowest two pairs, which cover the Fermi level, present the monolayer-like Dirac-cone structures being initiated from the K/K$^\prime$ valley. All the ${\theta\neq\,0^\circ}$ and ${60^\circ}$ twisted bilayer graphenes are zero-gap semiconductors because of the vanishing density of states at ${E_F}$. Specifically, the energy ranges of Dirac-cone structures quickly decline as $\theta$ decreases from ${30^\prime}$ to 0$^\circ$ [$\theta$ increases from 30$^\circ$ to 60$^\prime$]. The degenerate Dirac cones become split and thus create the finite Fermi momenta [free electrons and holes simultaneously], when gate voltage starts from zero; that is, the semiconductor-semimetal transition occurs during the variation of $V_z$. With the increasing energy of ${|E^{c.v}|}$, the stable valleys could form near the M, $\Gamma$ and K points. In addition to the extreme band-edge states, the M-point valleys frequently come to exist as the saddle forms. A lot of critical points in the energy-wave-vector space are created by the zone-folding effect of the Moire superlattice and the layer-dependent Coulomb potentials, in which three kinds of van Hove singularities, V-shape/a plateau plus two temple-like cusp structures,  prominent symmetric peaks and shoulders [measurements from the Fermi level], respectively, correspond to the K/K$^\prime$-point,  M-point, and M-, $\Gamma$- and K-related valleys. Apparently, a lot of optical absorption structures are created the vertical excitations of the rich band-edge states. According to the ranges of optical  excitation frequencies, there exist (I) the gapless, featureless and  linear-$\omega$ spectral functions [the K-valley Dirac cone], [II] three symmetric absorption peaks [the saddle points near the M ones], and (III) the  shoulder and peak structures  [near the K, M and $\Gamma$ points]. The gate voltages leads to an optical gap and the dramatic changes in the number, intensity, and frequency of absorption spectra. Very interesting, the magneto-electronic properties of the twisted bilayer graphene systems, being based on many subenvelops functions of multi-sublattices,  present the hybridized  phenomena arising from monolayer graphene, and AA $\&$ AB bilayer stackings. They directly reflect the robust relations  between the neighboring A and B lattice sites on the same layer, and for the projections of them on the upper/lower layer. All the oscillation modes in the layer-dependent distinct sublatices are well characterized through  the delicate analysis, and they belong to the perturbed ones in the presence of complicated interlayer hopping integrals. These results suggest that the effective-mass approximation might be not suitable for the current topic.
Moreover, the significant differences between the twisted and sliding bilayer graphenes lie in the main features of electronic structures,  absorption spectra, magneto-electronic properties and quantum transports  [discussed later in in Chap. 9.1], mainly owing to the various stacking configurations in the Moire superlattices of the former. On the experimental side, the high-resolution measurements from angle-resolved photoemission spectroscopy \cite{12NatMat12;887},  scanning tunneling spectroscopy \cite{12PRL109;196802} and Hall-bar equipment \cite{12NanoLett12;3833} have confirmed the low-lying linear and parabolic valence subbands. the V-shape structure and the $\theta$-dependent prominent symmetric peaks [the K and M points], and the quasi-monolayer Hall conductivity. The optical examinations seem to be absent up to now. How to examine the gate-voltage-enriched essential properties require the further experiments, especially for those related to the diverse magnetic quantizations.

Obviously, the stacking modulation, gate voltage, and magnetic field greatly diversify the physical phenomena of bilayer graphene. Furthermore, their essential properties are very different from those in twisted \cite{Mele;161405} and sliding \cite{Son;155410} bilayer systems, especially for the important differences in the 1D and 2D characteristics. The geometric modulation along the $x$-direction induces a periodical boundary condition and thus the dramatic change from 2D into 1D behaviors. The semi-metallic systems presents a plenty of 1D valence and conduction subbands, accompanied by the splitting of state degeneracy, the creation of non-monotonous energy dispersions, more band-edge states, and the distorted or irregular standing waves. Specifically, the layer-dependent Coulomb potential destroys a specific relation between the [A$^1$, A$^2$] and [B$^1$, B$^2$] subenvelope functions, even changes the zero-point number, and create the  A lot of van Hove  singularities in density of states cover the double- and single-peak in the square-root divergent forms, a pair of very prominent peaks near $E_F$, and a $V_g$-induced plateau across the Fermi level. The threshold absorption frequency/optical gap is vanishing in any systems. A pristine bi-layer AB stacking exhibits featureless optical spectrum at lower frequencies. Concerning the geometry-modulated systems, the observable absorption peaks, corresponding to the vertical transitions of band-edge states, appear only under the destruction of the symmetric/anti-symmetric linear superposition associated with the [A$^1$, A$^2$] and [B$^1$, B$^2$) sublattices, leading to the absence of optical selection rules. A simple dependence of absorption structures on the domain-wall width is absent. However, the reduced intensity and the enhanced number are revealed during the increase of gate voltage. The frequency, number, and form intensity of optical absorption peaks strongly depend on the modulation period and electric-field strength. In the presence of magnetic field, its strong competition with the stacking modulation leads to the complicated and non-homogenous Peierls phases and thus the rich magneto-electronic properties. The highly degenerate Landau levels of a normal stacking becomes many quasi-1D Landau subbands, in which the latter have the partially flat and oscillatory energy dispersions and present the frequent crossing and anti-crossing phenomena in the $k_y$-dependent magneto-electronic energy spectra. The greatly reduced state degeneracy and extra band-edge states create more symmetric and asymmetric peaks in the wider energy ranges of density of states. Three kinds of magnetic wave functions, the well-behaved, perturbed and seriously distorted ones, come to exist within the specific regions related to the stacking configurations; that is, each Landau-subband state might be the superposition of the various Landau-level ones. Based on the detailed and delicate analyses, the close relations among the geometric structures, the band-edge states, and the spatial probability distributions account for the main features of Landau subbands. In addition to the stacking modulation, the quantum confinement of 1D graphene nanoribbons \cite{Lin;161409}, the significant interlayer hopping integrals in 3D graphites \cite{Ohta;206802}, and the composite external fields \cite{Castro;216802}, create the Landau subbands, being quite different from one another. All the theoretical predictions on the essential properties require the high-resolution experimental verifications. For the stacking-, gate-voltage- and magnetic-field-manipulated bilayer graphene systems, The above-predicted essential properties, valence bands, van Hove singularities of electronic/magneto-electronic energy spectra and band property near the Fermi level, spatial distributions of subenvelope functions, and absorption spectra of vertical excitations, could be, respectively, verified from the ARPES, $V_z$-dependent STS, energy-fixed and various optical spectroscopies. \\

A AA-bt bilayer silicene exhibits the rich and unique electronic and optical properties under the specific buckling and stacking configuration. The M, K and $\Gamma$ valleys, as measured from the Fermi level, are responsible for the diversified phenomena. The significant differences with monolayer silicene and few-layer graphene, the main features of the low-lying Landau levels, lie in the absence of stable K-valleys near the Fermi level. Specifically, the first pair of energy bands in this system presents a concave-downward/cancave-upward K-point valleys at the higher/deeper energies for conduction/valence states, being accompanied with the similar and independent $\Gamma$-point valley; that is, both parabolic valleys are well established and separated.  It should be noticed that the unstable M-point valley belongs to the middle structure of the parabolic K-point one. The first and other two band-edge states, which correspond to the saddle and extreme points, respectively, reveal the prominent symmetric peaks and shoulder structures as the 2D van Hove singularities. Furthermore, the constant-energy valence and conduction loops cross ${E_F=0}$ with a very narrow spacing of ${\sim\,8}$ meV and thus create a temple-like cusp structures. The band gap is opened and gradually enhanced along the K$\Gamma$ direction as the gate voltage grows from zero. Also, it is characterized by a pair of prominent peaks in the square-root divergent form. However, the spin splitting never comes to exist, owing to the absence of the combined effects arising from the spin-orbital interaction and the sublattice-dependent Coulomb potentials. The above-mentioned unusual valley structure in energy spectrum can create the rich magnetic quantizations and optical absorption spectra. The magneto-electronic states are initiated from the K and $\Gamma$ valleys simultaneously, thus leading to the direct superposition of two Landau-level subgroups with the frequent crossing behaviors. Their Landau levels, respectively, have four- and two-fold degenerate localization centers [1/6, 2/6, 4/6, 5/6] and [1/2, 2/2]. Any magneto-electronic states, which are magnetically quantized from the first pair of energy bands, are dominated by the B$^l$ sublattices; that is, the equivalence of A$^l$ and B$^l$ sublattices on the same layer is destroyed by the critical interlayer hopping integral [Figs. 6.1(a) and 6.1(b)]. They display the almost monotonous Landau-level energy spacings and $B_z$-dependances. With the energy ranges covering the saddle M point and the Fermi level, the very large quantum numbers [oscillation modes] come to exist, especially for the $\Gamma$-dependent ones covering the larger carrier density. On the other hand, the magneto-electronic states, being magnetically quantized from the second pair of conduction and valence bands, exhibit the normal behavior. As to the vertical optical excitations, only the valence and conduction constant-emery loops display the observable absorption structure, the square-root asymmetric peak, while the $\Gamma$-, K- and M-dependent band-edge states hardly make any contributions. Such unusual optical property mainly come from the vanishing dipole matrix elements [the wave-vector-independent interlayer hopping integrals and the symmetric/anti-symmetric superposition of the tight-binding functions for  valence/conduction states]. The spectral characteristics strongly depend on the variation of gate voltage, especially for the enhanced threshold peak intensity and optical gap. Apparently, there exist certain important differences between the bilayer AA-bt and monolayer silicene systems in electronic and optical properties, such as, the well-separated parabolic K- and $\Gamma$-point valleys or the merged K-point Dirac cone and parabolic $\Gamma$-point valley, the independent valence/conduction constant-energy loops across the Fermi level or the roughly isotropic Dirac-cone structure, and the composite two Landau-level subgroups or the dramatic transformation of localization center and state degeneracy within the same group during the variation from the K-point valley/the Fermi level and then M- and  $\Gamma$-point valleys, the form, number, frequency and intensity of optical absorption structures spectra, and their strong dependences on the external fields. The high-resolution experimental measurements from ARPES, STS and optical spectroscopies are available in examining the valley-enriched fundamental physical properties due to the quite large interlayer hopping integrals. \\

The diversified essential properties are obviously revealed in AB-bt bilayer silicene, compared with AB-stacked bilayer graphene. The buckled honeycomb lattice, the large multi-interlayer hopping integrals, and the layer-dependent significant spin-orbital couplings are responsible for the electric-field-enriched phenomena. An indirect semiconducting system presents the specific semiconductor-semimetal transition, the drastic change of energy dispersions near the K and T valleys, and the dramatic formation of the distorted conduction and valence Dirac cones during the variation of electric-field strength. These are directly reflected in the main features of van Hove singularities and optical absorption spectra, such as, the frequency, form, number and intensity of special structures. Very important, the threshold absorption frequency is higher than the band gap, and its declining behavior is very sensitive to the ranges of electric field. Apparently, the magnetic quantization, which originates from to a magnetic field or a composite electric and magnetic one, is greatly enriched by the critical mechanisms. The main characteristics cover the double degeneracy for each $(k_x, k_y)$ state in the reduced first Brillouin zone. the significant K and T valleys with the totally different behaviors, the spin- and sublattice-dominated magneto-electronic properties, four Landau-level subgroups quantized from the first pair of valence and conduction bands [nearest to the Fermi level], the coexistence of the well-behaved and abnormal sublattice-related subenvelope functions, the frequent anti-crossing/crossing/non-crossing of two Landau levels in the magnetic-/electric-field-dependent energy spectra, and the dramatic transformation in the well separated energy ranges. spatial oscillation modes and occupation number of valence/conduction states during the variation of gate voltage. These clearly indicate the destruction in the equivalence of A$^i$ and B$^i$ on the same layer [A$^i$/B$^i$ and A$^j$/B$^j$ on the distinct layers], being absent in the normally stacked graphene systems. The unusual magneto-electronic states are further revealed in the magneto-absorption spectra with a lot of non-uniform delta-function-like peaks. Furthermore, they present the rich and unique magneto-optical properties: the sixteen categories for the available magneto-excitation channels/a plenty of absorption structures within a very narrow frequency range, the single-, double- $\&$ twin-peak symmetric structures, the dramatic changes through the gate voltage [e.g., the great decrease about the number of excitation categories and the optical gap], and the type-I and type-II absorption peaks, respectively, with $\&$ without the magneto-optical selection rules. The above-mentioned results that the layered silicene systems could be applied to novel designs of Si-based nanoelectronics \cite{12Zhao;24} and nanodevices with enhanced mobilities \cite{Akinwande;5678}. Their experimental examinations require the high-resolution measurements from the ARPES \cite{Vogt;155501}, STS \cite{Feng;3507} and optical spectroscopies \cite{Zhuang:161409}. Both silicene and graphene are in sharp contrast with each other in the essential physical properties, directly reflecting the significant differences in geometric symmetries and intrinsic interactions. \\

The Si-substituted graphene systems, which belong to the emergent 2D binary materials, have been thoroughly investigated for their electronic and optical properties in the absence/presence of magnetic quantization. They create the various chemical $\&$ physical environments by changing the Si-distribution configuration and concentration, leading to the diverse phenomena. There exist the non-uniform  site energies and hopping integrals; therefore, the A$_i$ and B$_i$ sublattices in the substitution-enlarged unit cell might be highly non-equivalent or quasi-equivalent. The former present the well separated parabolic valence and conduction bands and belong to the finite-gap semiconductors, e.g., the  Si-A$_i$-dressed configurations. However, the latter have the anisotropic and distorted Dirac-cone structures and exhibit the zero-gap semiconducting behaviors [the vanishing density of states at the Fermi level], such as, the Si-[A$_i$, B$_i$]-decorated systems. Their van Hove singularities are revealed as the shoulder structures and the asymmetric V-shape forms. The zero-field wave functions clearly show that the low-lying conduction and valence states near the K valleys are. respectively, dominated by the Si-guest and C-host atoms. Furthermore, the certain-sublattice envelope functions are negligible there and thus the localized states come to exist. Apparently, the optical spectra of the finite- and zero-gap systems, respectively, show the shoulder and featureless absorption structures. Only a pristine system displays a simple linear relation between the spectral intensity and excitation frequency, while the non-linear behaviors appear in the other Si-substituted systems. Such results are principally determined by whether the dominating dipole matrix elements depend on wave vectors/energies under the joint density of states proportional to frequency. There are four kinds of Landau levels, according to the spatial probability distributions $\&$ oscillation modes on the distinct sublattices, and the concise relations between A$_i$ and B$_i$ sublattices. The typical magneto-electronic states cover (I) the significant B$_i$ sublattices of valence Landau levels $\&$ $A_i$ sublattices of conduction Landau levels with the same mode, and the similar relation between the B$_i$ and A$_i$ sublattices of the conduction Landau levels; (II) the important/observable [A$_i$, B$_i$] sublattices with a mode difference of $\pm 1$, the serious deviations of localization centers $\&$ the highly asymmetric distributions composed of the main and side modes; (III) the same modes for valence B$_i$ and conduction A$_i$ sublattices, the vanishing or $\pm 1$ zero-point differences between conduction A$_i$ and B$_i$ sublattices $\&$ the perturbed multi-modes in most of conduction and valence B$_i$; (IV) the oscillator-like oscillation modes with the equivalent A and B sublattices. Such Landau levels lead to the unusual magneto-optical selection rules: the dominating $\Delta n=0$ (the first kind of Landau levels), the coexistent $\Delta n=1$ $\&$ 0 with strong competitions (the second and third kinds), and the specific $\Delta n=1$ (the fourth kind). Obviously, the above-mentioned features, respectively, correspond to the specific concentration and distribution configuration: the Si-A$_i$-dressed graphene with an enough high concentration, the [A$_i$, B$_i$]-decorated graphene, the low-concentration A$_i$-doped system, and the pristine one. Moreover, the non-uniform bond lengths, site energies $\&$ hopping integrals, and the effects due to the external fields are simultaneously taken into account within the generalized theoretical framework without the perturbation forms. This method is expected to be very useful in fully understanding the diversified fundamental properties of the adatom-adsorbed graphene systems [details in Chap. 13], e.g., the magnetic quantization phenomena in graphene oxides [Refs], hydrogenated/halogenated/alkalized graphene systems [Refs]. \\

The direct combination of generalized tight-binding model and the static Kubo formula, being based on the same framework, is successfully developed to explore to investigate the unusual magneto-transport properties in high- and low-symmetry bilayer and trilayer graphenes, especially for the detailed identifications of the selection rules during the Landau-level scatterings. Such delicate method is very suitable for fully exploring the rich Hall conductivities of the emergent 2D materials, such as, silicene \cite{SRp3;1075,PRL95;226801}, germanene \cite{PRB87;235426,PRB89;085429}, tinene \cite{NJP21;023010}, MoS$_2$ \cite{PRL110;066803}, and phosphorene \cite{PRB92;165409,EPL119;37005,NanoLett18;229}[discussed later in Chap. 13]. The Fermi-energy- and magnetic-field-dependent quantum Hall conductivities are greatly diversified under the various stacking configurations, since they could create the intragroup and intergroup Landau-level transitions with/without the monolayer-like and other selection rules. Generally speaking, three kinds of Landau levels, the non-crossing/crossing/anti-crossing behaviors, and the split magneto-electronic energy spectra account for the diverse quantum transport phenomena, e.g., the integer and non-integer conductivities at zero temperature, the splitting-created reduction and complexity of quantum conductivity, the vanishing or non-zero conductivities at the neutrality point, the well-like, staircase and composite plateau structures, the uniform/distinct step heights, and the simple or irregular $B_z$-dependences of step widths. Concerning the quantum Hall conductivities of the sliding bilayer graphenes, the AA, AB and AA$^\prime$ stackings are dominated by two categories of intragroup transition channels under the normal selection rule of $\Delta n=\pm 1$, while the extra intergroup ones and selection rules [${\Delta\,n=0}$ $\&$ ${\pm\,2}$] are clearly revealed in the other low-symmetry systems. The non-integer quantum conductivities only present in the AA$^\prime$ stacking with the energy-dependent localization center, covering the step heights of $3.8e^2/h$ and $4.2e^2/h$. Most of sliding bilayer systems exhibit the monolayer-like step height of $4e^2/h$ even for the undefined Landau levels, such as AA, AB, $\delta =6b_0/8$ and $\delta =11b_0/8$ stackings. Specifically, the exclusive step heights of $e^2/h$, $3e^2/h$, $5e^2/h$ $\&$ $7e^2/h$ are observed in the $\delta =1/8$ stacking. Moreover a non-negligible conductivity appears near the neutrality point, while there exist Landau levels very close to zero energy, e.g., those of $\delta =b/8$, $6b/8$ and $8b/8$. Only the AA stacking shows four types of magnetic-field-dependent plateau structures [the wells, monolayer-like $\&$ non-monotonous staircases, and composite ones], mainly owing to the significant overlap of two Dirac-cone structures in the Landau-level energy spectra. On the other side, the AAB-stacked trilayer graphene exhibits the complex plateau structures with $2e^2/h$ height within a wide energy range and the irregular $B_z$-dependence, directly reflecting the splitting and anti-crossing energy spectra and the localization-split modes. Apparently, such features are absent in AAA-, ABA- and ABC-stacked trilayer systems. Only the ABA system presents the insulating behavior near zero Fermi energy within a very narrow window. The numbers of intragroup and integroup inter-Landau-level channels, corresponding to the AAA-, ABA-, ABC- and AAB-stacked graphene systems,  are [3, 0], [3, 2], [3, 6] and [3, 6]. Also, the selection rules of static scatterings, respectively, cover ${\Delta\,n=\pm\,1}$, ${\Delta\,n=\pm\,1}$ [0 $\&$ ${\pm\,2}$] for few anti-crossings], ${\Delta\,n=\pm\,1}$, 0 $\&$ ${\pm\,2}$, and ${\Delta\,n=\pm\,1}$, 0, ${\pm\,2}$, ${\pm\,3}$ $\&$ ${\pm\,4}$. In general, the above-mentioned theoretical predictions, being excepted for the AB stackings, could be verified from the high-resolution quantum transport measurements [details in Chap. 2.3]. Such examinations are very useful in identifying the diverse quantum transport phenomena, as well as in clarifying the close relations/the one-to-one correspondence between the zero-field band structure and the main features of the highly-degenerate Landau levels.  \\

The thermal properties of monolayer graphene clearly illustrate another magnetic quantization phenomenon. The specific heat strongly depends on the temperature, magnetic-field strength, Zeeman effect and doping density of conduction electrons/valence holes. An unusual $T^2$ dependence is revealed at zero ${B_z}$, directly reflecting the linear and isotropic Dirac-coe structure. Such behavior is in sharp contrast with the different $T$-dependences of AAA-, ABA- and ABC-stacked graphites \cite{12Lin;074603}, mainly owing to the distinct dimensions and stacking configurations/interlayer hopping integrals. Magnetic fields dramatically change the temperature dependence, where ${C(T, B_z)}$ presents a composite form of ${1/T^2}$ and exponential function. An obvious peak, which is induced by the unoccupied spin-up and occupied spin-down Landau levels across the chemical potential, comes to exist at the critical temperature. Also, the $B_z$-dependent specific heat shows a single peak at the critical field strength [$B_{zc}$]. By the delicate calculations, a simple linear relation is identified to exist between $T_c$ and $B_z$ [${B_{zc}}$ and $T$]. Under doping cases, the specific heat might display an extra shoulder at higher temperature. This structure mainly comes from the second lowest unoccupied Landau levels and the second highest occupied ones. The different filling cases determine whether it would survive or not. There are certain important differences between graphene and single-walled carbon nanotubes regarding specific heat as a result of the dimension-dominated van Hove singularities in density of states. The magneto-electronic specific heats of 1D carbon nanotubes demonstrate four kinds of temperature dependences which corresponds to three metallic and one semiconducting band structures. The main features of thermal properties are closely related to the low-lying band structures; therefore, the high-resolution calorimeter measurements could be utilized to test them, e.g., the .ratio of ${g/m^\ast}$ in determining the Zeeman splitting energy. \\

As for monolayer germanene, the many-body Coulomb excitations have been thoroughly explored by the direct association of the modified random-phase approximation and the generalized tight-binding model [formulas in Chap. 3.4], especially for the bare and screened response functions due to dynamic charge screenings. Apparently, the critical factors, the hopping integrals, the spin-orbital couplings, the directions $\&$ magnitudes of transferred momenta, gate voltages and magnetic fields, are responsible for the diverse single-particle and collective excitations. The low-frequency excitation properties directly reflect the characteristics of the low-lying bands, the strong wave-vector dependence, the anisotropic behavior, the SOC-induced separation of Dirac points, the doping levels,  and the $V_z$-created  destruction of spin-configuration degeneracy. There exists a forbidden excitation region between the intraband and interband electron-hole boundaries, being ascribed to the Fermi-momentum and band-edge states. The undamped acoustic plasmons could survive within this region with a prominent peak intensity. All the plasmon modes, being purely due to  free conduction electrons, belong to 2D acoustic modes at small transferred momenta, as observed in an electron gas [Refs]. With the increasing momentum, they might experience the interband damping and become another kind of undamped plasmons, change into the seriously suppressed modes in the heavy intraband electron-hole damping, remain the same undamped plasmons, or gradually vanish during the enhanced interband damping, i.e., there exist four kinds of plasmon modes after free carrier dopings. Specifically, the first kind only appears in germanene with the stronger spin-orbital couplings. The fourth kind of plasmon modes in monolayer germanene are dominated by gate voltage, while they are frequently revealed in few-layer extrinsic graphenes without external fields \cite{SR7;2017}. Moreover, even at the undoped case, the magneto-electronic Coulomb excitation spectra could exhibit several interband magnetoplasmon modes with the observable intensities under low transferred momenta. Most important, their collective oscillation frequencies possess the non-monotonous momentum-dependences, in which the first and second critical momenta represent the dramatic variation in the propagating direction of magneto-plasma wave and the disappearance due to the very strong Landau dopings. The main reason is the strong competition between the longitudinal Coulomb interactions and the transverse cyclotron forces. The free-carrier dopings can lead to a very prominent intraband magnetoplasmon [a 2D-EGS-like mode]; furthermore, they have strong effects on the reduced frequencies and intensities of interband magnetoplasmons. Also, the former presents the discontinuous dependence in the increase of the magnetic-field strength, mainly owing to the quick enhancement in Landau-level energies. General speaking, the significant differences and similarities among germanene, silicene and graphene lie in the existence of plasmon/magnetoplasmon modes and excitation gaps. The high-resolution experimental examinations for the reflection EELS and the inelastic X-ray scatterings are very useful in
providing the the significant intralayer hopping integrals and spin-orbital couplings. Similar studies could be extended to other few-layer IV-group monatomic systems, e.g., bilayer Si \cite{PRB97;125416}, and  single-layer Sn \cite{PRB94;045410} $\&$ Pb \cite{JPCC119;11896} materials.

The Landau levels (LLs) of $3$D layered systems of topological matter with $2$D massless Dirac fermions have also been considered in this book. Two typical systems were selected into the demonstration and investigation. One is rhombohedral graphite (RG), as a representative for spinless nodal-line topological semimetals (TSM) protected by the chiral symmetry (CS); the other is Bi$_{2}$Se$_{3}$ or the similar, for spinful strong topological insulators (TIs) protected by the time reversal symmetry (TRS). Remember that nontrivial topological phase of the normal integer quantum Hall effect (QHE) can occur at any $2$D plane of a system in trivial topological phase. By contrast, the QHE or even the Landau quantization in a system originally being nontrivial could be intriguing. Among others, the two selected systems both show up the half-integer QHE to some extent, upon the presence of $2$D massless Dirac fermions. Both of them have a topologically robust zero-mode LL that is a constant function of the external perpendicular magnetic field. It has been concretized here that within the minimal model, the topological phase of nodal-line TSM in RG is protected by the CS, where $2$D Dirac fermions are hosted along the spiralling nodal lines. The characteristics of the zero-mode LL and $3$D half-integer QHE have analogs in $2$D graphene, as an evidence of the same topological classification, to which both graphene and RG belong. The minimal model for RG might be argued due to the nonnearest-neighbor interlayer coupling. Thanks to the material setting, where the nearest-neighbor interlayer hopping dominates, this is not problematic within the single-particle or even the weak-interacting approximations. The solid finding of a complete dimension scale between the $3$D half-integer QHE in RG and the $2$D half-integer QHE in graphene yielded a proof of this aspect. The scaling results from numerical cauculation of magneto-optical plasmons in graphene and the minimal model for RG are also recognized. For the strong TI, Bi$_{2}$Se$_{3}$ or the similar, there are spreading experimental and numerical investigations involving TRS breaking means. As well realized, the surface Dirac cone is gapped by acquiring a mass term due to magnetic doping or a proximate magnetic material. However, in the case of applying an external perpendicular magnetic field, it is odd that the surface Dirac cone remains to a significant extent. To this problem, a unifying solution based on the CS is given here, in the light of a recent research. \\

\newpage
\begin{center}
\section{Future perspectives and open issues}\label{ch13}
\author{ Ching-Hong Ho,\textit{$^{d,\P}$} Po-Hsin Shih,\textit{$^{a,\ddag}$} Chiun-Yan Lin,\textit{$^{a}$} Thi-Nga Do,\textit{$^{b,c,\ast}$}\\ Jhao-Ying Wu,\textit{$^{d,\dag}$} Shih-Yang Lin,\textit{$^{e,\S}$} Ming-Fa Lin\textit{$^{a,\sharp}$}}
\end{center}
\vskip 1.0 truecm

Very interesting, the current book proposes future perspectives and issues, being very useful in exploring the emergent condensed-matter systems, especially for the layered materials. The main-stream open topics cover the diversified essential properties in the physical, chemical and material sciences. They are closely related to the dimension crossover   [3D$\rightarrow$0D], the stacking modulation, the composite uniform-nonuniform fields, the deformed/curved surfaces, the chemical absorptions/substitutions, the various defects, the amorphous structures. the significant combinations of multi-orbital hybridizations and spin-orbital interactions, the magneto-excitonic effects, and the many-particle e-e Coulomb interactions. Furthermore, the theoretical models would be modified to include the many-body effects in the complex scattering processes of optical absorptions, quantum transports and Coulomb decay rates. \\

The dimension crossover phenomena will become one of the main-stream topics since the first discovery of few-layer graphene systems by the mechanical exfoliation in 2004 \cite{Novoselov;666}. The 2D layered materials can dramatically change into the 3D bulk systems, the 1D nanomzaterials, or evn the 0D quantum dots. Up to now, the hexagonal honeycomb lattices can form 3D graphites [AA-, AB- and ABC-stacked ones; Fig. 13.1], 2D layered graphenes, 1D graphene nanoribbons [Fig. 13.2], 1D carbon nanotubes [Fig. 13.3], and 0D graphene disks, in which the ${sp^2}$ bondings dominate the optimal geometric structures [the planar/curved/folded/cylindrical geometries].  In general, the $\pi$ bondings, being due to the carbon-${2p_z}$ orbitals, are sufficiently in explaining the  low-energy physical properties except for the highly deformed surfaces with the strong hybridization of ${(2s, 2p_x, 2p_y, 2p_z)}$ orbitals, e.g., those in the large- and small-radius carbon nanotubes. Apparently, only three nearest neighbors are revealed in these systems, being quite different from the ${sp^3}$ bondings in the well-known diamond. Very interesting, the 3D layered  structures are never successfully synthesized in experimental laboratories using the other  group-IV elements. This clearly indicates that the dramatic structure transformations might come to exist during the growth of Si/Ge/Sn/Pb layers. Among the significant mechanisms, the buckled geometries, with the very strong competitions between the ${sp^2}$ and ${sp^3}$ bondings, are expected to play the critical factors. The ${2D\rightarrow\,3D}$ structural changes can only be done through  the first-principles methods [discussed later in Chap. 13]. There are more complicated geometric structures even under the few-layer cases, compared with those of graphene systems. For example, the bilayer silicene systems are predicted to present four kinds of typical geometries, covering the AA-bb, AA-bt, AB-BB and AB-bt ones. Obviously, the distinct geometric symmetries can greatly diversify the essential physical properties. How to get the structure-complicated site energies, intralayer $\&$ interlayer hopping integrals, and spin-orbital couplings is the key evaluation in the further developments of the phenomenological models. On the other side, the ${2D\rightarrow\,1D}$ and ${2D\rightarrow\,0D}$ transformations have been achieved in graphene-related systems, 1D/0D graphene \cite{ACSNano6;6930,APL110;051601}, silicene \cite{SSR67;1} and germanene nanoribbons/quantum dots \cite{MSS93;92,MRE4;114005}. The quantum confinement and open/terminated edge structures lead to very close cooperations with the other intrinsic interactions. The various fundamental properties will exhibit the unusual behaviors, being worthy of systematic investigations. \\

There are five typical methods in creating the non-uniform magnetic quantization environments, covering the stacking modulation \cite{13IOP;SC,13CRCPress;CY}, the reduced or enhanced dimensions \cite{13CRCPress;CY,13IOPBook;CY}, the composite external fields \cite{13PRB83;195405,13IOP;SC}, the deformed/curved surfaces \cite{13PCCP18;7573,13JPSJ81;064719}, and the hybridized structures \cite{Zhu;494}. In general, the quasi-1D magneto-electronic states have strong energy dispersions, at least, along one direction of wave vector, in which such relations are responsible for the unusual transport conductivities and absorption spectra. For example, the previous theoretical predictions show that it is difficult to observe the quantum Hall effects in Bernal graphite, as a result of the significant overlaps of various Landau subbands \cite{13Nature438;201}. Furthermore, the ABC-stacked graphite presents the well-separated Landau subbands \cite{13NatPhys7;953}, directly indicating the possibilities in observing the quantum Hall effects. In the stacking-modulated bilayer graphene, the Landau subbands possess the lower state degeneracy [details in Chap. 5], and the $k_y$-dependent electronic states will make different contributions to the magnetic-field- and carrier-density-dependent transport properties. Equation (3.3.1) needs to be accurately calculated for each $k_y$ state. This fact not only greatly enhances the numerical barriers, but also largely induces the analytic difficulties in proposing the concise and full scattering mechanisms. The existence of selection rule, which appears during the static scatterings between the occupied and unoccupied Landau-subband states, is the near-future studying focus. The quantized conductivities are expected to exhibit the fractal structures or be thoroughly absent under the specific geometry-modulated configurations. The similar researches could be conducted on the rich magneto-optical properties \cite{13IOPBook;CY}. \\

The composite external fields, which consists of a uniform magnetic field and a spatially modulated electric/magnetic field [Fig. 13.4], can create the unusual fundamental properties. The previous theoretical studies on monolayer graphene \cite{13CRCPress;CY,13IOPBook;CY,13IOP;SC} show that their strong effects on magneto-electronic [optical properties] cover the one-dimensional Landau subbands with the significant energy dispersions, many band-edge states strong anisotropy, lower state degeneracy, seriously  distorted/perturbed/well-behaved spatial probability distributions; a transformation of oscillation modes, and a lot of square-root-form asymmetric peaks in density of states [many asymmetric absorption peaks, blue or red shifts, the reduced spectral intensities, and the specific rule of ${\Delta n=1}$ and the extra ones of ${\Delta n=0, 2; 3}$. Up to now, only few studies are conducted on few-layer graphene systems, and the similar works on monolayer silicene and germanene are absent. The complex effects, being due to the buckled structures, the intralayer $\&$ interlayer hopping integrals, the spin-orbital interactions, and the composite fields, are deduced to create the unique physical phenomena, especially for those from the two latter factors. For example, the spin- and wave-vector-complicated magnetic wave functions might be composed of more oscillation modes and present the highly active behaviors in optical transitions, Coulomb excitations and transport scatterings. \\

The various curved  structures have become one the main stream-main-stream topics, e.g., the folded graphene nanoribbons \cite{13NatComm5;3189,Prada;106802}, curved ones \cite{13ACSNano4;1362,Martins;075710}, carbon nanoscrolls \cite{13NanoLett9;3766,Viculis;1361}, carbon toroids \cite{13PRB70;075411,Haddon;388}, and carbon onions \cite{13CPL305;225,Banhart;433} purely due to the dominant ${sp^2}$ chemical bondings. For the diverse magnetic quantization phenomena, carbon allotropes are very ideal condensed-matter systems in fully exploring the curvature effects [the significant multi-orbital hybridizations of C-${[2s, 2p_x, 2p_y, 2p_z]}$ orbitals], the critical interactions of the non-uniform interlayer hopping integrals \cite{Yan;2159}, the strong competition between the quantum confinement and the magnetic localization , and the diamagnetism, paramagnetism, ferromagnetism and anti-ferromagnetism [arising from the circulating charges and spin distributions; Refs. \cite{13NatComm5;3189,13ACSNano4;1362,13NanoLett9;3766,13PRB70;075411,13CPL305;225}]. Apparently, the systematic investigated will be made on whether the highly degenerate Landau levels, Landau subbands, or quantum discrete states could survive on the closed/open surfaces. They could be done by the generalized tight-binding model which is developed in this book. The important mechanisms should cover the scale of surface area, the degree of curvature, and the strength of magnetic field. Also, the geometry-enriched layered materials are very suitable for observing the diversified magneto-optical selection rule, abnormal quantum Hall effects, and the momentum-dependent inter-Landau-level magnetoplasmon modes. \\

The hybridized condensed-matter materials, which are composed of the lower-dimensional systems with the same element, have been successfully synthesized in experimental laboratories by the various methods, such as, a 1D multi-walled carbon nanotube [a superposition of some single-walled ones; \cite{Monteiro;377}], a 3D multi-shell carbon onion [due to the various fullerens, a 3D carbon nanotube bundle \cite{Schaper;73}, a 1D carbon nanotube-graphene nanoribbon hybrid [a CNT-GNR system; Ref.\cite{Yang;3996}];  a 3D compound of carbon nanotubes and graphitic layers. Up to now, the theoretical predictions on the magneto-electronic and optical properties are  only done for the specific CNT-GNR  hybridized materials \cite{Khoeini;1315}, clearly indicating that the geometric symmetries/asymmetries dominate the formation of the quasi-Landau subbands. That is to say, the composite effects, arising from the 1D quantum confinement, the transverse cyclotron motion, and the complex interlayer atomic interactions, are responsible for  the non-unifrom magnetic quantization phenomena. The main features of Landau subbands are very sensitive to the width of graphene nanoribbon, the radius and position of carbon nanotube, and the strength of magnetic field. Apparently, the diverse magnetic properties are expected to be  easily observed/predicted in the other hybridized carbon-related systems. For example, the almost identical single-walled carbon nanotubes could be intercalated into the bulk Bernal graphite [the few-layer graphene systems] because of the weak van Der Waals intercations between two neighboring graphene layers. Such systems are very suitable for the model studies of the 3D or 2D Landau subbands. How to derive the independent magnetic Hamiltonian matrix elements through the generalized tight-binding model, would become an interesting work, since they strongly depend on the nanotube radius, the intertube distance, and the relative position between carbon nanotube and graphene. \\

The developed methods for efficiently solving the various Hamiltonians are one of the basic challenges in theoretical physics. The engineering of an energy gap, being closely related to semiconductor applications of layered material, can diversify the main features of magneto-electronic properties. A very effective way is to create the various defects. Different defects are frequently produced during the experimental syntheses, e.g., the vacancies [the studying focus; Refs. \cite{13PRB80;033407}], modified lattices [e.g, hexagons replaced by pentagons and heptagons in Fig. 13.5; Ref. \cite{Terrones;351}], and adatom impurities \cite{Sahin;115432,He;3766}. There are only a few theoretical predictions on the defect-enriched essential properties for the layered group-IV systems \cite{Li;7881}. The defect-related silicene, germanene and graphene will be very suitable for a model study. For such systems, the defects can induce the destruction of lattice symmetry, the distinct site energies [the change of the ionization potential] and the non-uniform atomic interactions [the position-dependent hopping integrals]. The dependences of electronic properties on the type, distribution and concentration of defects are worthy of a systematic investigation. For example, whether the   semiconductor-semimetal transition, the localization of wave functions, and the valley- and spin-split states come to exist under the various defect configurations could be explored thoroughly. The complicated cooperative/competitive relations among the combined effects are proposed to comprehend the diverse electronic properties. Such effects are greatly enhanced by the external electric and magnetic fields, e.g, the drastic changes on the magnetic quantization in terms of the distortions and mixing of the Landau-level wave functions, the changes of the localization ranges, the reduced  state degeneracy, and the irregular field-dependent energy spectra covering the crossing and anti-crossing behaviors. \\

In addition to the guest-atom substitutions [discussed in Chap. 8], the chemical absorptions are very efficient in drastically changing the fundamental physical properties, as thoroughly investigated in the previous two books using the first-principles method for adatom-adsorbed graphene-related systems [e.g., the hydrogenated, oxidized and alkalized graphene systems in Figs. 13.6(a), 13.6(b), and 13.6(c), respectively; Ref. \cite{13CRCPress;SY}]. They might provide another effective ways to greatly diversify the magnetic quantization. According to the detailed analyses, the critical factors, which determines the diverse phenomena, lie in the optimal geometric structures and the host-guest atomic interactions. Apparently, the guest-adatom positions in the quasi-sublattices also play an important role in the significant contributions of the magnetic wave functions. The corresponding adatom lattice needs to be taken into consideration simultaneously. Whether guest adatoms could exhibit the localized spatial distributions under a uniform perpendicular magnetic field is one of the studying focuses. The distribution configuration and concentration are expected to play important roles in any physical quantities, mainly owing to the strong effects on the single-orbital or multi-orbital hybridizations in adatom-host, adatom-adatom, host-host chemical bonds, and the significant   competition between the guest-adatom and magnetic periods. Up to date,  the magneto-electronic energy spectra and wave functions can only be solved by the generalized tight-binding model, but not the effective-mass approximation or the first-principles calculations [discussed in Chap. 3]. The various non-negligible hopping integrals and the   atom-ionization-induced site energies could be obtained only by fitting the low-energy valence and conduction bands from the first-principles results. For many adatom chemisorptions on graphene, the key orbital-orbital interactions have been successfully identified from the atom-dominated energy bands, the spatial charge distributions before and after chemical modifications, and atom- and orbital-projected density of states. However, the delicate fittings and the very complicated magnetic Peierls phases will come to exist and become the high barriers in fully exploring the diversified quantization phenomena. In general, the magnetic Hamiltonian is a giant Hermitian matrix with many independent and imaginary elements. The critical roles of the  chemically adsorbed adatoms  on the main features of highly degenerate Landau levels will be very interesting, e.g., their effects on the magnetically quantized energy spectra, the magneto-state degeneracy, the destruction of the well-behaved oscillation modes, and the very close relations between the original and adsorption-created lattices. Apparently, whether  the Landau wave functions could built by the guest adatoms is one of the studying focuses; that is, two kinds of magnetic wave functions, which are, respectively, due to host atoms and guest adatoms, might co-exist in different layers. \\

The emergent layered materials are expected to exhibit the rich magnetic quantizations in optical properties under the single-particle scheme characteristics of magneto-optical excitations, such as, the stacking-modulated bilayer graphenes, the twisted ones, the AA- and AB-stacked bilayer silicene/germanene/stanene/bismuthene. and so on \cite{ACSNano4;1465,SRp4;7509,SRP9;624,OLt43;6089,PLA383;68}. To fully comprehend the key characteristics, the very close associations between the generalized tight-binding mode and the linear Kubo formula are necessary in overcoming the rather high barriers induced by the numerical calculations. For example, the Landau subbands, which mainly originate from the stacking modulations, possess the wave-vector-dependent  energy spectra and magnetic wave functions, but not the highly degenerate and well-behaved Landau levels. The numerical calculations need to evaluate for the continuous wave vectors; therefore, the giant and ${\bf k}$-dependent magnetic Hamiltonian matrix, being accompanied with the more complicated dipole moments. Apparently, the completely numerical calculations will be almost impossible. How to greatly enhance the efficiency of computation should be the near-future studying focus. As for the twisted bilayer graphene systems, a lot of Landau-level subgroups, being induced by the Moire superlattice, might show the unique magneto-optical properties, e.g., more inter-Landau-level vertical transition categories and extra selection rules \cite{13IOPBook;CY}. The unusual phenomena are under the current investigations, since the computational problems could be solved for the highly degenerate magnet-electronic states. \\

The optical and magneto-optical properties, with the many-body effects [the excitonic effects], are very interesting topics in the fundamental researches and potential applications. The excitations, which are created during the optical excitation processes, belong to the strongly bound or metastable states of electron-hole pairs. They are formed by the attractive Coulomb forces between the negative and positive charges. Such quasiparticles could be condensed together and clearly reveal the macroscopic phenomena in some semiconductors at very low temperatures \cite{neil;4356,Brinkman;1508}, being quite different from the Cooper pairs in all the superconductors \cite{Nakamura;786}. Any semiconductors might exhibit the excitonic effects under the suitable conditions, e.g., the red shift of threshold absorption frequency and the enhanced spectral intensity \cite{13NanoLett14;3353}. Apparently, the full theoretical framework, which covers the screened Coulomb interactions in the nonlinear scattering events [the standard Kubo formula], is required to display the basic many-particle characteristics in various optical spectra. Up to now, some previous theoretical works are suitable for the undoped/doped 3D semiconductors \cite{Van;8154;Dong;2361}, while those for 2D emergent materials seem to be absent. The difficulties in the theoretical calculations lie in the non-parabolic energy dispersions in the whole first Brilloun zone \cite{Lewiner;2347}, the optical vertical excitations due to several pairs of low-lying valence and conduction bands \cite{Huang;859}, the dynamic charge scatterings in the electron-hole pairs [the random-phase approximation; Ref. \cite{Sarma;1936}], and the greatly enhanced complexity arising from the external fields \cite{Gusynin;157402}. Apparently, that only the lowest energy of bound-state excitations is regarded as the threshold excitation frequency through the Schrodinger-like equation under the effective-mass approximation is not suitable in explaining the full excitonic effects. If the magnetic quantization could be included in the new theoretical model, the magneto-exciton effects are worthy of the systematic studies. Obviously, the exchange self-energies of Landau levels are not sufficient in calculating the many-body excitation frequencies \cite{13IOPBook;CY}. It should be noticed that only few experimental cases on 2D few-layer systems might show the weak evidences of magneto-excitonic effects \cite{13IOP;SC,13CRCPress;CY,13IOPBook;CY}. The specific excitation frequencies due to the prominent inter-Landau-level transitions are available in optical detectors \cite{Futia;1626}, sensors \cite{Kimura;178}, and attenuators \cite{Li;826}. In short, how to combine the generalized tight-binding model \cite{13IOP;SC,13CRCPress;CY,13IOPBook;CY}, the modified random-phase approximation \cite{13PRB34;2,13PRB74;085406,13PLA352;446}, and  the nonlinear dynamic Kubo formula \cite{13REVSCIINS83;093108,13APPOPT48;5713} is the key point of view in constructing the exact and reliable theoretical framework. \\

Obviously, the layered condensed systems will show the diverse quantum Hall conductivities, and the modifed static Kubo formula might be necessary in thoroughly understaning the delicate transport phenomena due to the various scatterings/excitations [e.g., the electron-electron, electron-phonon and electron-impurity interactions, and the thermal effects]. For example, both twisted bilayer graphenes and AB- and AA-stacked silicene/germanene [details in Chaps. 4, 6 $\&$ 7] present the rich and unique magneto-electronic energy spectra and wave functions, clearly illustrating the existence of unusual magneto-transport properties. The splittings of K/K$^\prime$ valleys and/or spin configurations and the complex crossing $\&$ anti-crossing behaviors, and the irregular magnetic wave functions with the multi-oscillation modes could play the critical roles in greatly diversifying the quantum transport phenomena, such as, the abnormal integer conductivities, the non-integer ones, and the special or random plateau structures in the Fermi-energy- and magnetic-field-dependences. Furthermore, the distinct selection rules for the static              inter-Landau-level transitions are worthy of the full explorations through the detailed analysis. On the other side, only few theoretical investigations are conducted on the complex effects arising from the and inelastic scatterings \cite{Novikov;245435,Mak;1341}. Specifically, it is very difficult to directly cover the many-particle interactions in the transport formula; that is, a complete theoretical model should be further developed to identify the novel results from the high-resolution measurements [details in Chap, 2.3]. A possible algorithm is the close associations of the generalized tight-binging mode, the modified random-phase approximation and the Kubo formula. \\

The electron-electron Coulomb interactions are one of the main-stream topics in condensed-matter physics, since they play critical roles in the essential physical properties. Apparently, the many-body phenomena, with the rich and unique behaviors, are expected to be revealed in the emergent layered materials. To fully explore the diversified single-particle and collective excitations, a lot of open issues need to be overcome. The theoretical framework, which covers the close association of the generalized   tight-binding model, the modified random-phase approximation, and the screened exchange self-energy, might be suitable in thoroughly investigating the quasiparticle states. In general, the intralayer $\&$ interlayer hopping integrals in the band-structure calculations are obtained from the numerical fittings with the first-principles results near the high-symmetry ${\bf k}$-points. However, it would become very difficult to the get reliable parameters if the buckled structures, stacking configurations, multi-orbital hybridizations, interlayer atomic interactions, and spin-orbital couplings in few-layer systems are taken into account simultaneously. For example, a bilayer AB-bt silicene is predicted to present the layer-dependent spin-orbital interactions and the largest interlayer hopping integral for the first pair of valence and conduction bands with the unusual energy dispersions [details in Ref. \cite{Yaokawa;10657}]. The previous study clearly shows that a set of good parameters for two pairs of energy bands is absent up to now. The similar cases appear in other bilayer systems, e.g., germanene \cite{13SciRep7;40600}, stanene \cite{13PRB94;045410}, antimonene \cite{13PRB98;115117} and bismuthene \cite{13NJP20;062001} without the full parameters in the published papers. When the random-phase approximation is used to include the band-structure effects, one must solve the Coulomb matrix elements [Eq, (3.23)] arising from the orbital-dominated tight-binding function. How to get rid of the significant calculation errors in such elements will be the studying focus. Also, the delicate Coulomb excitation spectra can provide the full information in the further topics, such as, the dominance on the quasiparticle energies and lifetimes \cite{13PRB97;2018}, the time-dependent propagations of plasma waves in the ${\bf r}$-space, the screened Coulomb potentials and charge density distributions in planar structures, the simultaneous combinations of electron-electron and electron-phonon interactions. Furthermore, the analytic formulas for distinct many-body properties have to be modified/derived for the highly efficient calculations. \\

In addition to the above-mentioned phenomenological models, the first-principles methods using the Vienna ab initio simulation packages are very reliable for certain essential physical properties, e.g., the optimal geometric structures \cite{Kostelnik;1574,Hafner;6}, electronic properties \cite{Hafner;6,Neaton;1298}, spin-related magnetic configurations \cite{Eelbo;136804} and phonon spectra \cite{Barabash;155704}. Very important, they can provide the suitable parameters [site energies, hoppig integrals and spin-orbital couplings] for the generalized tight-binding model. Apparently, such methods are useless in  the magnetic quantization phenomena as a result of the very large unit cell due to the vector potential. The excellent point of view, being based on the basic chemistry and physics, is proposed through the developed theoretical framework: the critical multi-hybridizations are accurately examined from the host-atom- $\&$ guest-atom-dominated energy bands, the spatial charge density $\&$ the difference after chemisorption/bonding, and atom- , orbital- $\&$ spin-projected density of states; furthermore, the diverse magnetic configurations are delicately identified from the strong competition between the boundary host atoms and adatoms, the spin-split/spin-degenerate electronic structures, the  spin-induced net magnetic moment, the spatial spin distributions around two kinds of atoms, and the spin-split density of states. This has been successfully developed to fully explore and comprehend the geometric, electronic and magnetic properties for the 2D graphene-reflated systems [the 1D graphene-nanoribbon-related ones; details in Ref. \cite{13CRCPress;SY}], such as, the essential properties in the AA-, AB-, ABC- $\&$ AAB-stacked pristine graphenes, sliding graphenes, rippled graphenes, graphene oxides, hydrogenated graphenes,  halogen-, alkali-, Al- $\&$ Bi-doped graphene compoundss [armchair and zigzag graphene nanoribbons without/with hydrogen terminations, curved and zipped graphene nanoribbons, folded graphene nanoribbons, carbon nanoscrolls, bilayer graphene nanoribbons, edge-decorated graphene nanoribbons, alkali-, halogen-, Al-, Ti and Bi-absorbed graphene nanoribbons. The diverse physical phenomena, including the semiconductors, semimetals and metals the absence or recovery of the Dirac-cone structures \cite{Lin;207}, the linear, parabolic and oscillatory energy dispersions \cite{Tran;10623}, the critical points in the energy-wave-vector space [the minima, maxima and saddle points; Ref. \cite{13CRCPress;SY}], the semiconductor-metal transition \cite{Lin;26443,Lin;209}, the creation of band gap \cite{Huang;84}, and non-magnetic, ferromagnetic or anti-ferromagnetic spin configurations \cite{Lin;1722}, are clearly explained by the concise physical and chemical pictures. The close associations of the first-principles calculations and the generalized tight-binding model will be very useful for the emergent 2D materials in exploring the various magnetic quantization phenomena. For example, the diverse magneto-electronic properties, magneto-optical spectra, quantum Hall transports, and magneto-Coulomb excitations can be thoroughly investigated only after the reliable fitting for the intrinsic interactions.  A set of suitable parameters is strongly required for few-layer 2D materials, being purely composed by  group-IV and group-V atoms. These systems are expected to reveal the rich and unique physical phenomena. \\

The distinct magnetisms, which are induced by the host atoms/molecules, guest ones, and the external magnetic fields, are closely related to the emergent spintronics and can provide the full informations on potential applications. They cover the paramagnetism, diamagnetism, non-magnetism ferromagnetism, and anti-ferromagnetism. The former two phenomena mainly originate from the orbital currents of valence and conduction electrons, being created by a uniform magnetic field. According to the previous theoretical calculations, the diverse magnetic properties are predicted to survive in 1D carbon nanotubes \cite{Lu;1123}, 0D carbon tori \cite{Tsai;075411}, 0D C$_{60}$-related fullerenes \cite{Jonsson;572}, 2D graphenes and 3D Bernal graphite \cite{Moran;6746}. For example, the metallic/narrow-gap [one-third] and finite-gap [two-thirds] carbon nanotubes, respectively, exhibit the paramagnetic and diamagnetic phenomena \cite{Searles;017403}, when they are present in a uniform magnetic field parallel to the tubular axis. Also, the specific persistent currents exist in other low-dimension carbon-related systems with the sp$^2$-dominated surface structures \cite{Kibis;741}. Generally speaking, the magnetic response is very weak, so that the experimental examinations are required to be finished at very low temperature for a almost homogenous system \cite{Murmu;5069}. To get the magnetization field/the magnetic susceptibility, a very accurate derivative of the total energy versus the magnetic-field strength is required in the numerical calculations. However, the variation of the former might be rather small within a finite difference of the latter, especially for 2D and 3D condensed-matter systems. A lot of magneto-electronic states are taken into account simultaneously. Furthermore, the theoretical models are not so reliable in the higher-energy ones, such as, the effective-mass approximation even for monolayer graphene/silicene/germanene \cite{13RevModPhys81;109,13Rep76;056503,13PRB80;165409,13PRB77;115313,Tabert;085434,Matthes;395305}. As a result, the delicate investigations on the diverse magnetism of layered group-IV and group-V materials would be a significant challenge. The different magnetic behaviors, but the similar difficulties are expected to appear in the ferromagnetic and anti-ferromagnetic materials. The strong cooperations or competitions among the host-atom-, guest-atom-induced spin moments and magnetic field will be fully explored through the combination of the generalized tight-binding model and the Hubbard model. For example, how to simulate the magnetic-field effects on the  anti-ferromagnetic spin configurations at two open boundaries of zigzag graphene nanoribbons is one of studying focuses. \\


\begin{thebibliography}{00}

\bibitem{1IOP;SC}
Chen S C, Wu J Y, Lin C Y, and Lin M F 2017
Theory of Magnetoelectric Properties of 2D Systems
\emph{IOP Concise Physics.} San Raefel, CA, USA: Morgan $\&$ Claypool Publishers.

\bibitem{1CRCPress;CY}
Lin C Y, Chen R B, Ho Y H, and Lin M F 2018
Electronic and optical properties of graphite-related systems,
\emph{CRC Press} Boca Raton, Florida.

\bibitem{1IOPBook;CY}
C. Y. Lin, T. N. Do, Y. K. Huang, and M. F. Lin 2017
Electronic and optical properties of graphene in magnetic and electric fields
\emph{IOP Concise Physics.} San Raefel, CA, USA: Morgan $\&$ Claypool Publishers.

\bibitem{1PCCP18;7573}
Chung H C, Chang C P, Lin C Y, and Lin M F 2016
Electronic and Optical Properties of Graphene Nanoribbons in External Fields,
\emph{Phys. Chem. Chem. Phys.} \textbf{18} 7573.

\bibitem{1JPCC116;8271}
J. H. Wong, B. R. Wu, M. F. Lin,
Strain Effect on the Electronic Properties of Single Layer and Bilayer Graphene
J. Phys. Chem. C. 116, 8271-8277 (2012).

\bibitem{1PRB83;195405}
Ou Y C, Sheu J K, Chiu Y H, Chen R B, and Lin M F 2011
Influence of modulated fields on the Landau level properties of graphene
\emph{Phys. Rev. B} \textbf{83} 195405.

\bibitem{1CPC184;1821}
Ou Y C, Chiu Y H, Lu J M, Su W P, and Lin M F 2013
Electric modulation effect on magneto-optical spectrum of monolayer graphene
\emph{Comput. Phys. Commun.} \textbf{184} 1821.

\bibitem{1PRB91;155428}
Cherkez V, Trambly de Laissardiere G, Mallet P, Veuillen J Y 2015
Van Hove singularities in doped twisted graphene bilayers studied by
scanning tunneling spectroscopy \emph{Phys. Rev. B} \textbf{91}
155428.

\bibitem{1PRL121;037702}
Huang S, Kim K, Efimkin D K, Lovorn T, Taniguchi T, Watanabe K, MacDonald A H, Tutuc E, and LeRoy B J 2018 Topologically Protected Helical States in Minimally Twisted Bilayer Graphene
\emph{Phys. Rev. Lett.} \textbf{121} 037702.

\bibitem{1AdvMater22;3723}
Fan Z, Yan J, Zhi L, Zhang Q, Wei T, Feng J, Zhang M, Qian W, and Wei F 2010
A three-dimensional carbon nanotube/graphene sandwich and its application as electrode in supercapacitors
\emph{Adv. Mater.} \textbf{22} 3723.

\bibitem{1PRX3;021018}
Vaezi A, Liang Y, Ngai D H, Yang L, and Kim E A 2013
Topological Edge States at a Tilt Boundary in Gated Multilayer Graphene
\emph{Phys. Rev. X} \textbf{3} 021018.

\bibitem{1SciRep9;859}
Huang B L, Chuu C P, and Lin M F 2019
Asymmetry-enriched electronic and optical properties of bilayer graphene
\emph{Sci. Rep.} \textbf{9} 859.

\bibitem{1SciRep4;7509}
Huang Y K, Chen S C, Ho Y H, Lin C Y, and Lin M F 2014
Feature-Rich Magnetic Quantization in Sliding Bilayer Graphenes
\emph{Sci. Rep.} \textbf{4} 7509.

\bibitem{1PRB97;125416}
Do T N, Shih P H, Gumbs G, Huang D, Chiu C W, and Lin M F 2018
Diverse magnetic quantization in bilayer silicene
\emph{Phys. Rev. B} \textbf{97} 125416.

\bibitem{1SciRep7;40600}
Shih P H, Chiu Y H, Wu J Y, Shyu F L, and Lin M F 2017
Coulomb excitations of monolayer germanene
\emph{Sci. Rep.} \textbf{7} 40600.

%%%%%%%%%%%%%%%%%%%%%%%%%%%%%%%%%%%%%%%%%%%%%%%%%%%%

%%%%%%%%%%%%%%%%%%%%%%%%%%%%%%%%%%%%%%%%%%%%

\bibitem{1RMP82;3045}
Hasan M Z, and Kane C L 2010
Colloquium: Topological insulators
\emph{Rev. Mod. Phys.} \textbf{82} 3045.

\bibitem{1PCCP19;29525}
Do T N, Chang C P, Shih P H, Lin  M F 2017
Stacking-enriched magneto-transport properties of few-layer graphenes
\emph{Phys. Chem. Chem. Phys.} \textbf{19} 29525.

\bibitem{1PRB74;085406}
Ho J H, Lu C L, Hwang C C, Chang C P, and Lin M F 2006
Coulomb excitations in AA- and AB-stacked bilayer graphites
\emph{Phys. Rev. B} \textbf{74} 085406.

\bibitem{1PRB98;041408}
Lin C Y, Lee M H, and Lin M F 2018
Coulomb excitations in trilayer ABC-stacked graphene
\emph{Phys. Rev. B Rapid communication} \textbf{98} 041408.

\bibitem{1PCCP17;26008}
Lin C Y, Wu J Y, Ou Y J, Chiu Y H, and Lin M F 2015
Magneto-electronic properties of multilayer graphenes
\emph{Phys. Chem. Chem. Phys.} \textbf{17} 26008.

\bibitem{1NanoLett12;3833}
Wang Z F, Liu F, and Chou M Y  2012
Fractal Landau-Level Spectra in Twisted Bilayer Graphene
\emph{Nano Lett.} \textbf{12} 3833.

\bibitem{1PRB90;205434}
Lin C Y, Wu J Y, Chiu Y H, and Lin M F 2014
Stacking-dependent magneto-electronic properties in multilayer graphenes
\emph{Phys. Rev. B} \textbf{90} 205434.

\bibitem{1PRB70;075411}
Tsai C C, Shyu F L, Chiu C W, Chang C P, Chen R B, and M. F. Lin,
Magnetization of armchair carbon tori
\emph{Phys. Rev. B} \textbf{70} 075411.

%%%%%%%%%%%%%%%%%%%%%%%%%%%%%%%%%%%%%%%%

\bibitem{PRB95;115411} Wu J Y, Chen S C, G Gumbs and Lin M F 2017 Field-created diverse quantizations in monolayer and bilayer black phosphorus \emph{Phys. Rev. B} \textbf{95} 115411.

\bibitem{SRp8;13303} Wu J Y, Chen S C, Do T N, Su W P, Gumbs G and Lin M F 2018 The diverse magneto-optical selection rules in bilayer black phosphorus \emph{Sci. Rep.} \textbf{8} 13303.

\bibitem{PRB94;205427} Wu J Y, Chen S C, Gumbs G, and Lin M F 2016 Feature-rich electronic excitations of silicene in external fields \emph{Phys. Rev. B} \textbf{94} 205427.

\bibitem{PRB97;125416} Do T N, Shih P H, Gumbs G, Huang D, Chiu C W and Lin M F 2018 Diverse magnetic quantization in bilayer silicene \emph{Phys. Rev. B} \textbf{97} 125416.

\bibitem{PRB88;085434} Tabert C J and Nicol E J 2013 Magneto-optical conductivity of silicene and other buckled honeycomb lattices \emph{Phys. Rev. B} \textbf{88} 085434.

\bibitem{PRL110;197402} Tabert C J and Nicol E J 2013 Valley-Spin Polarization in the Magneto-Optical Response of Silicene and Other Similar 2D Crystals \emph{Phys. Rev. Lett.} \textbf{110} 197402.

\bibitem{PRB91;035423} Tabert C J, Carbotte J P and Nicol E J 2015 Magnetic properties of Dirac fermions in a buckled honeycomb lattice \emph{ Phys. Rev. B} \textbf{91} 035423.

\bibitem{IOP2017} Chen S C, Wu J Y, Lin C Y and Lin M F 2017 Theory of Magnetoelectric Properties of 2D Systems \emph{IOP Concise Physics. San Raefel} CA, USA: Morgan \& Claypool Publishers.

\bibitem{PRB94;045410} Chen S C, Wu C L, Wu J Y and Lin M F 2016 Magnetic quantization of sp3 bonding in monolayer gray tin \emph{Phys. Rev. B} \textbf{94} 045410.

\bibitem{NJP20;062001} Chen S C, Wu J Y and Lin M F 2018 Feature-rich magneto-electronic properties of bismuthene \emph{New Jour. Phys. Fast track communication} \textbf{20} 062001.

\bibitem{PRB98;115117} Yu J, Katsnelson M I and Yuan S 2018 Tunable electronic and magneto-optical properties of monolayer arsenene: From GW0 approximation to large-scale tight-binding propagation simulations \emph{Phys. Rev. B} \textbf{98} 115117.

\bibitem{SRp9;2332} Chung H C, Chiu C W and Lin M F 2019 Spin-polarized magneto-electronic properties in buckled monolayer GaAs \emph{Sci. Rep.} \textbf{9} 2332.

\bibitem{RSCAdv5;20858} Ho Y H, Su W P and Lin M F 2015 Hofstadter spectra for d-orbital electrons: a case study on MoS2 \emph{RSC Adv.} \textbf{5} 20858.

\bibitem{PRB89;55316} Ho Y H, Wang Y H and Chen H Y 2014 Magnetoelectronic and optical properties of a MoS$_{2}$ monolayer \emph{Phys. Rev. B} \textbf{89} 55316.

%%%%%%%%%%%%%%%%

\bibitem{1NatComm5;3189}
Vo T H, Shekhirev M, Kunkel D A, Morton M D, Berglund E, Kong L, Wilson P M, Dowben P A, Enders A, Sinitskii A 2014
Large-scale solution synthesis of narrow graphene nanoribbons
\emph{Nat. Comm.} \textbf{5} 3189.

\bibitem{1JPSJ81;064719}
Lin C Y, Chen S C, Wu J Y, and Lin M F 2012
Curvature Effects on Magnetoelectronic Properties of Nanographene Ribbons
\emph{J. Phys. Soc. Jpn.} \textbf{81} 064719.

\bibitem{1NanoLett9;3766}
Patra N, Wang B, and Kral P 2009
Nanodroplet Activated and Guided Folding of Graphene Nanostructures
\emph{Nano Lett.} \textbf{9} 3766.

\bibitem{1PRB75;125408}
Yazyev O V, and Helm L 2007
Defect-induced magnetism in graphene
\emph{Phys. Rev. B } \textbf{75} 125408.

\bibitem{1PRL106;105505}
Kotakoski J, Krasheninnikov A V, Kaiser U, and Meyer J C 2011
From Point Defects in Graphene to Two-Dimensional Amorphous Carbon
\emph{Phys. Rev. Lett. } \textbf{106} 105505.

\bibitem{1CRCPress;SY}
S. Y. Lin, N. T. T. Tran, S. L. Chang, W. P. Su, and M. F. Lin 2018
Structure- and adatom-enriched essential properties of graphene nanoribbons
\emph{CRC Press} Boca Raton, Florida.

\bibitem{1PRB54;2896}
Lin M F, and Shung K W-K 1996
The electronic specific heat of single-walled carbon nanotubes
\emph{Phys. Rev. B } \textbf{54} 2896.

\bibitem{1PhysicaE11;356}
Chiu C W, Lin M F, and Shyu F L 2001
Electronic specific heat of nanographite ribbons
\emph{Physica E} \textbf{11} 356.

\bibitem{1ACSNano5;1026}
Wu J Y, Chen S C, Roslyak O, Gumbs G, Lin M F 2011
Plasma Excitations in Graphene: Their Spectral Intensity and Temperature Dependence in Magnetic Field
\emph{ACS Nano} \textbf{5} 1026.

\bibitem{1PRL94;226403}
Matsui T, Kambara H, Niimi Y, Tagami K, Tsukada M, and Fukuyama H 2005
STS Observations of Landau Levels at Graphite Surfaces
\emph{Phys. Rev. Lett.} \textbf{94} 226403.

\bibitem{1NatPhys3;623}
Li G, and Andrei E Y 2007
Observation of Landau levels of Dirac fermions in graphite
\emph{Nat. Phys.} \textbf{3} 623.

\bibitem{1PRL94;226403}
Matsui T, Kambara H, Niimi Y, Tagami K, Tsukada M, and Fukuyama H 2005
STS observations of Landau levels at graphite surfaces
\emph{Phys. Rev. Lett.} \textbf{94} 226403.

\bibitem{1NatPhys10;815}
Fu Y S, Kawamura M, Igarashi K, Takagi H, Hanaguri T, and Sasagawa T 2014
Imaging the two-component nature of Dirac¡VLandau levels in the topological surface state of Bi2Se3
\emph{Nat. Phys.} \textbf{10} 815.

%transmision
\bibitem{1PRL100;136403}
Orlita M, Faugeras C, Martinez G, Maude D K, Sadowski M L, and Potemski M 2008
Dirac fermions at the H Point of graphite: magnetotransmission studies
\emph{Phys. Rev. Lett.} \textbf{100} 136403.

\bibitem{1SCIENCE322;1529}
Keppler H, Dubrovinsky L S, Narygina O, and Kantor I 2008
Optical absorption and radiative thermal conductivity of silicate perovskite
to gigapascals \emph{Science} \textbf{322} 1529.

%magneto-reflection
\bibitem{1PRB15;4077}
Toyt W W, and Dresselhaus M S 1977
Minority carriers in graphite and the H-point magnetoreflection spectra
\emph{Phys. Rev. B} \textbf{15} 4077.

\bibitem{1NanoLett7;2711}
Casiraghi C, Hartschuh A, Lidorikis E, Qian H, Harutyunyan H, Gokus T, Novoselov K S, and Ferrari A C 2007 Rayleigh Imaging of Graphene and Graphene Layers
\emph{Nano Lett.} \textbf{7} 2711.

\bibitem{1Science315;5817}
Novoselov K S, Jiang Z, Zhang Y, Morozov S V, Stormer H L, Zeitler U, Maan J C, Boebinger G S, Kim P, Geim A K 2007 Room-Temperature Quantum Hall Effect in Graphene
\emph{Science} \textbf{315} 1379.

\bibitem{1Nature438;201}
Zhang Y, Tan Y W, Stormer H L, and Kim P 2005 Experimental observation of the quantum Hall effect and Berry's phase in graphene
\emph{Nature} \textbf{438} 201.

\bibitem{1NatPhys7;953}
Zhang L, Zhang Y, Camacho J, Khodas M, and Zaliznyak I 2011
The experimental observation of quantum Hall effect of l=3 chiral quasiparticles in trilayer graphene
\emph{Nature} \textbf{7} 953.

\bibitem{1PRL54;1820}
Gornik E, Lassnig R, Strasser G, Stormer H L, Gossard A C, and Wiegmann W 1985
Specific Heat of Two-Dimensional Electrons in GaAs-GaAlAs Multilayers
\emph{Phys. Rev. Lett.} \textbf{54} 1820.

\bibitem{1PRB33;2893}
Mendez E E, Esaki L, and Wang W I 1986
Resonant magnetotunneling in GaAlAs-GaAs-GaAlAs heterostructures
\emph{Phys. Rev. B} \textbf{33} 2893(R).

%IXS
\bibitem{1Winfried}
W. Sch\"{u}lke 2007 Electron dynamics by inelastic x-Ray scattering
\emph{Oxford University Press} Oxford.

\bibitem{1PRB77;115313}
Koshino M and Ando T 2008 Magneto-optical properties of multilayer
graphene \emph{Phy. Rev. B } \textbf{77} 115313.

\bibitem{1PRB80;165409}
Koshino M and McCann E 2009 Trigonal warping and Berry's phase N pi
in ABC-stacked multilayer graphene \emph{Phy. Rev. B} \textbf{80}
165409.

\bibitem{1Rep76;056503}
McCann E and Koshino M 2013 The electronic properties of bilayer
graphene \emph{Rep. Prog. Phys.} \textbf{76} 056503.

\bibitem{1RevModPhys81;109}
Castro Neto A H, Guinea F, Peres N M R, Novoselov K S and Geim A K
2009 The electronic properties of graphene \emph{Rev. Mod. Phys.}
\textbf{81} 109-162.

\bibitem{1PRB77;235430}
Chan K T, Neaton J B, and Cohen M L 2008
First-principles study of metal adatom adsorption on graphene
\emph{Rev. Mod. Phys.} \textbf{77} 235430.

\bibitem{1PRB75;041401}
Yao Y, Ye F, Qi X L, Zhang S C, and Fang Z 2007
Spin-orbit gap of graphene: First-principles calculations
\emph{Phys. Rev. B} \textbf{75} 041401(R).

\bibitem{1PRB86;165108}
Feng W, Yao Y, Zhu W, Zhou J, Yao W, and Xiao D 2012
Intrinsic spin Hall effect in monolayers of group-VI dichalcogenides: A first-principles study
\emph{Phys. Rev. B} \textbf{86} 165108.

\bibitem{1PRL109;055502}
Ezawa M 2012
Valley-Polarized Metals and Quantum Anomalous Hall Effect in Silicene
\emph{Phys. Rev. Lett.} \textbf{109} 055502.

\bibitem{1PRB93;075408}
Ostahie B, and Aldea A  2016
Phosphorene confined systems in magnetic field, quantum transport, and superradiance in the quasiflat band
\emph{Phys. Rev. B} \textbf{93} 075408.

\bibitem{1SciRep5;12295}
Zhou X Y, Zhang R, Sun J P, Zou Y L, Zhang D, Lou W K, Cheng F, Zhou G H, Zhai F, Chang K 2015 Landau levels and magneto-transport property of monolayer phosphorene
\emph{Sci. Rep. } \textbf{5} 12295.

\bibitem{1PRB78;085401}
Berman O L, Gumbs G, and Lozovik Y E 2008 Magnetoplasmons in layered graphene structures
\emph{Phys. Rev. B } \textbf{78} 085401.

\bibitem{1PRB77;125417}
Bychkov Yu A, and Martinez G 2008
Magnetoplasmon excitations in graphene for filling factors $\nu\leq6$
\emph{Phys. Rev. B } \textbf{77} 125417.

\bibitem{1Science302;425}
Mao W L, Mao H k, Eng P J, Trainor T P, Newville M, Kao C, Heinz D L, Shu J, Meng Y, Hemley R J 2003
Bonding Changes in Compressed Superhard Graphite
\emph{Science} \textbf{302} 425.

\bibitem{1PRL93;245502}
Guo W, Zhu C Z, Yu T X, Woo C H, Zhang B, and Dai Y T 2004
Formation of sp3 Bonding in Nanoindented Carbon Nanotubes and Graphite
\emph{Phys. Rev. Lett.} \textbf{93} 245502.

\bibitem{1JPCC115;2705}
Lee D W, and Seo J W 2011 sp$^{2}$/sp$^{3}$ Carbon Ratio in Graphite Oxide with Different Preparation Times
\emph{J. Phys. Chem. C} \textbf{115} 2705.

\bibitem{1ChemMater16;1786}
Guerin K, Pinheiro J P, Dubois M, Fawal Z, Masin F, Yazami R, and Hamwi A 2004
Synthesis and Characterization of Highly Fluorinated Graphite Containing sp$^{2}$ and sp$^{3}$ Carbon
\emph{Chem. Mater.} \textbf{16} 1786.

\bibitem{1APL91;131906}
Hu A 2007
Direct synthesis of sp-bonded carbon chains on graphite surface by femtosecond laser irradiation
\emph{Appl. Phys. Lett.} \textbf{91} 131906.

\bibitem{1PRSLSA106;749}
Bernal J D 1924 The Structure of Graphite
\emph{Proc. R. Soc. London, Ser. A } \textbf{106} 749.

\bibitem{1Nature438;197}
Novoselov K S, Geim A K, Morozov S V, Jiang D, Katsnelson M I, Grigorieva I V, Dubonos S V, and Firsov A A 2005 Two-dimensional gas of massless Dirac fermions in graphene
\emph{Nature} \textbf{438} 197.

\bibitem{1NatComm5;3189}
Vo T H, Shekhirev M, Kunkel D A, Morton M D, Berglund E, Kong L, Wilson P M, Dowben P A, Enders A, Sinitskii A 2014
Large-scale solution synthesis of narrow graphene nanoribbons
\emph{Nat. Comm.} \textbf{5} 3189.

\bibitem{1NanoLett12;844}
Peng J, Gao W, Gupta B K, Liu Z, Romero-Aburto R, Ge L, Song L, Alemany L B, Zhan X, Gao G, Vithayathil S A, Kaipparettu B A, Marti A A, Hayashi T, Zhu J J, and Ajayan P M 2012
Graphene Quantum Dots Derived from Carbon Fibers
\emph{Nano Lett.} \textbf{12} 844.

\bibitem{1Nature354;56}
Iijima S 1991 Helical microtubules of graphitic carbon
\emph{Nature} \textbf{354} 56.

\bibitem{1Nature363;685}
Taylor R, and Walton D R M 1993
The chemistry of fullerenes
\emph{Nature} \textbf{363} 685.

\bibitem{1CPL305;225}
Tomita S, Fujii M, Hayashi S, and Yamamoto K 1999
Electron energy-loss spectroscopy of carbon onions
\emph{Chem. Phys. Lett.} \textbf{305} 225.

\bibitem{1Nature391;59}
Wilder J W G, Venema L C, Rinzler A G, Smalley R E, and Dekker C 1998
Electronic structure of atomically resolved carbon nanotubes
\emph{Nature} \textbf{391} 59.

\bibitem{1PRL79;2093}
Crespi V H, Cohen M L, and Rubio A 1997
In Situ Band Gap Engineering of Carbon Nanotubes
\emph{Phys. Rev. Lett.} \textbf{79} 2093.

\bibitem{1PRL82;3520}
Rubio A, Sanchez-Portal D, Artacho E, Ordejon P, and Soler J M 1999
Electronic States in a Finite Carbon Nanotube: A One-Dimensional Quantum Box
\emph{Phys. Rev. Lett.} \textbf{82} 3520.

\bibitem{1PRL109;126801}
Yan W, Liu M, Dou R F, Meng L, Feng L, Chu Z D, et al. 2012
Angle-dependent van Hove singularities in a slightly twisted
graphene bilayer \emph{Phys. Rev. Lett.} \textbf{109} 126801.

\bibitem{1NanoLett12;3162}
Havener R W, Zhuang H, Brown L, Hennig R G, and Park J 2012
Angle-Resolved Raman Imaging of Interlayer Rotations and Interactions in Twisted Bilayer Graphene
graphene bilayer \emph{Nano Lett.} \textbf{12} 3162.

\bibitem{1NatComm5;5309}
Wu J B, Zhang X, Ijas M, Han W P, Qiao X F, Li X L, Jiang D S, Ferrari A C, and Tan P H 2014
Resonant Raman spectroscopy of twisted multilayer graphene
graphene bilayer \emph{Nat. Comm.} \textbf{5} 5309.

\bibitem{1PRL97;187401}
Ferrari A C, Meyer J C, Scardaci V, Casiraghi C, Lazzeri M, Mauri F, Piscanec S, Jiang D, Novoselov K S, Roth S, and Geim A K 2006
Raman Spectrum of Graphene and Graphene Layers
\emph{Phys. Rev. Lett.} \textbf{97} 187401.

\bibitem{1PRL109;196802}
Brihuega I, Mallet P, Gonzalez-Herrero H, Trambly de Laissardiere G, Ugeda M M, Magaud L, Gomez-Rodriguez J M, Yndurain F, and Veuillen J-Y 2012
Unraveling the Intrinsic and Robust Nature of van Hove Singularities in Twisted Bilayer Graphene by Scanning Tunneling Microscopy and Theoretical Analysis
\emph{Phys. Rev. Lett.} \textbf{109} 196802.

\bibitem{1NanoLett14;3353}
Havener R W, Liang Y, Brown L, Yang L, and Park J 2014
Van Hove Singularities and Excitonic Effects in the Optical Conductivity of Twisted Bilayer Graphene
\emph{Nano Lett.} \textbf{14} 3353.

\bibitem{1PRB85;195458}
Moon P, and Koshino M 2012
Energy spectrum and quantum Hall effect in twisted bilayer graphene
\emph{Phys. Rev. B} \textbf{85} 195458.

\bibitem{1PRB81;161405}
Mele E J 2010
Commensuration and interlayer coherence in twisted bilayer graphene
\emph{Phys. Rev. B} \textbf{81} 161405.

\bibitem{1NatMat12;887}
Kim K S, Walter A L, Moreschini L, Seyller T, Horn K, Rotenberg E, and Bostwick A 2013
Coexisting massive and massless Dirac fermions in symmetry-broken bilayer graphene
\emph{Nat. Mater.} \textbf{12} 887.

\bibitem{1PRB87;205404}
Moon P, and Koshino M 2013
Optical absorption in twisted bilayer graphene
\emph{Phys. Rev. B} \textbf{87} 205404.

\bibitem{1PRL108;076601}
Sanchez-Yamagishi J D, Taychatanapat T, Watanabe K, Taniguchi T, Yacoby A, and Jarillo-Herrero P 2012
Quantum Hall Effect, Screening, and Layer-Polarized Insulating States in Twisted Bilayer Graphene
\emph{Phys. Rev. Lett.} \textbf{108} 076601.

\bibitem{1Nature556;43}
Cao Y, Fatemi V, Fang S, Watanabe K, Taniguchi T 2018
Efthimios Kaxiras $\&$ Pablo Jarillo-Herrero Unconventional superconductivity in magic-angle graphene superlattices \emph{Nature} \textbf{556} 43.


%%%%%%%%%%%%%%%%

\bibitem{1NatNanotech13;204}
Jiang L, Wang S, Shi Z, Jin C, Utama M I B, Zhao S, Shen Y R, Gao H J, Zhang G, and Wang F 2018
Manipulation of domain-wall solitons in bi- and trilayer graphene
\emph{Nat. Nanotech.} \textbf{13} 204.

\bibitem{1Nature520;650}
Ju L, Shi Z, Nair N, Lv Y, Jin C, Velasco Jr J, Ojeda-Aristizabal C, Bechtel H A, Martin M C, Zettl A, Analytis J, and Wang F 2015
Topological valley transport at bilayer graphene domain walls
\emph{Nature} \textbf{520} 650.

\bibitem{1Science312;1191}
Berger C, Song Z, Li X, Wu X, Brown N, Naud C, Mayou D, Li T, Hass J, Marchenkov A N, Conrad E H, First P N, de Heer W A 2006
Electronic Confinement and Coherence in Patterned Epitaxial Graphene
\emph{Science} \textbf{312} 1191.

\bibitem{1Nature408;440}
Pavesi L, Negro L D, Mazzoleni C, Franzo G, and Priolo F 2000
Optical gain in silicon nanocrystals
\emph{Nature} \textbf{408} 440.

\bibitem{1Science257;1906}
McConnell H M, Owicki J C, Parce J W, Miller D L, Baxter G T, Wada H G, Pitchford S 1992
The cytosensor microphysiometer: biological applications of silicon technology
\emph{Science} \textbf{257} 1906.

\bibitem{1Nature460;974}
Madar R 2004
Silicon carbide in contention
\emph{Nature} \textbf{430} 974.

\bibitem{1Science291;851}
Cui Y, Lieber C M 2001
Functional Nanoscale Electronic Devices Assembled Using Silicon Nanowire Building Blocks
\emph{Science} \textbf{291} 851.

\bibitem{1NanoLett12;3507}
Feng B, Ding Z, Meng S, Yao Y, He X, Cheng P, Chen L, and Wu K 2012
Evidence of Silicene in Honeycomb Structures of Silicon on Ag(111)
\emph{Nano Lett.} \textbf{12} 3507.

\bibitem{1PRL108;155501}
Vogt P, Padova P D, Quaresima C, Avila J, Frantzeskakis E, Asensio M C, Resta A, Ealet B, and Lay G L 2012 Silicene: Compelling Experimental Evidence for Graphenelike Two-Dimensional Silicon
\emph{Phys. Rev. Lett.} \textbf{108} 155501.

\bibitem{1NanoLett13;685}
Meng L, Wang Y, Zhang L, Du S, Wu R, Li L, Zhang Y, Li G, Zhou H, Hofer W A, and Gao H J 2013
Buckled Silicene Formation on Ir(111)
\emph{Nano Lett.} \textbf{13} 685.

\bibitem{1PRL108;245501}
Fleurence A, Friedlein R, Ozaki T, Kawai H, Wang Y, and Yamada-Takamura Y 2012
Experimental Evidence for Epitaxial Silicene on Diboride Thin Films
\emph{Phys. Rev. Lett.} \textbf{108} 245501.

\bibitem{1NatNanotechnol10;227}
Tao L, Cinquanta E, Chiappe D, Grazianetti C, Fanciulli M, Dubey M, Molle A, and Akinwande D 2015
Silicene field-effect transistors operating at room temperature
\emph{Nat. Nanotechnol. } \textbf{10} 227.

\bibitem{1NanoLett12;227}
Ni Z, Liu Q, Tang K, Zheng J, Zhou J, Qin R, Gao Z, Yu D, and Lu J 2012
Tunable Bandgap in Silicene and Germanene
\emph{Nano Lett.} \textbf{12} 113.

\bibitem{1JPCC119;3818}
Padilha J E, and Pontes R B 2015
Free-Standing Bilayer Silicene: The Effect of Stacking Order on the Structural, Electronic, and Transport Properties \emph{J. Phys. Chem. C} \textbf{119} 3818.

\bibitem{1JPCC116;4163}
Zhang C, and Yan S  2012
First-Principles Study of Ferromagnetism in Two-Dimensional Silicene with Hydrogenation
\emph{J. Phys. Chem. C} \textbf{116} 4163.

\bibitem{1JPCC116;22916}
Osborn T H, and Farajian A A 2012
Stability of Lithiated Silicene from First Principles
\emph{J. Phys. Chem. C} \textbf{116} 22916.

\bibitem{1PRB88;245408}
Cai Y, Chuu C P, Wei C M, and Chou M Y 2013
Stability and electronic properties of two-dimensional silicene and germanene on graphene
\emph{Phys. Rev. B} \textbf{88} 245408.

\bibitem{1PRB97;125416}
Do T N, Shih P H, Gumbs G, Huang D, Chiu C W, and Lin M F 2018
Diverse magnetic quantization in bilayer silicene
\emph{Phys. Rev. B} \textbf{97} 125416.

\bibitem{1PRB94;205427}
Wu J Y, Chen S C, Gumbs G, Lin M F 2016
Feature-rich electronic excitations of silicene in external fields
\emph{Phys. Rev. B} \textbf{94} 205427.


%%%%%%%%%%%%%%%%%%%%%%%%%%

\bibitem{1PCCP17;992}
Kulish V V, Malyi O I, Persson C, and Wu P 2015
Adsorption of metal adatoms on single-layer phosphorene
\emph{Phys. Chem. Chem. Phys.} \textbf{17} 992.

\bibitem{1Catal2;54}
Machado B F, and Serp P 2012
Graphene-based materials for catalysis
\emph{Catal. Sci. Technol.} \textbf{2} 54.

\bibitem{1arXiv1806;0529}
S. Y. Lin, M. F. Lin 2018
Metal-adsorbed graphene nanoribbons
\emph{arXiv} \textbf{1806} 05290.

\bibitem{1ChemMater22;2790}
Leventis N, Sadekar A, Chandrasekaran N, and Sotiriou-Leventis C 2010
Synthesis of Monolithic Silicon Carbide Aerogels from Polyacrylonitrile-Coated 3D Silica Networks
\emph{Chem. Mater.} \textbf{22} 2790.

\bibitem{1JPCC110;10645}
McCarthy M C, Apponi A J, and Thaddeus P 1999
Rhomboidal SiC$^{3}$
\emph{J. Chem. Phys.} \textbf{110} 10645.

\bibitem{1CST67;2390}
Fan S, Zhang L, Xu Y, Cheng L, Lou J, Zhang J, and Lin Y 2007
Microstructure and properties of 3D needle-punched carbon/silicon carbide brake materials
\emph{Composites Science and Technology} \textbf{67} 2390.

\bibitem{1Adv16;561}
Shi Y F, Meng Y, Chen D H, Cheng S J, Chen P, Yang H F, Wan Y, and Zhao D Y 2006
Highly Ordered Mesoporous Silicon Carbide Ceramics with Large Surface Areas and High Stability
\emph{Adv. Funct. Mater.} \textbf{16} 561.

\bibitem{1APL89;013105}
Zhou W, Yan L, Wang Y, and Zhang Y 2006
SiC nanowires: A photocatalytic nanomaterial
\emph{Appl. Phys. Lett.} \textbf{89} 013105

\bibitem{1APL85;2932}
Hua J Q, Bando Y, Zhan J H, and Golberg D 2004
Fabrication of ZnS¢ASiC nanocables, SiC-shelled ZnS nanoribbons (and sheets), and SiC nanotubes (and tubes)
\emph{Appl. Phys. Lett.} \textbf{85} 2932.

\bibitem{1NanoLett7;570}
Romo-Herrera J M, Terrones M, Terrones H, Dag S, and Meunier V 2007
Covalent 2D and 3D Networks from 1D Nanostructures: Designing New Materials
\emph{Nano Lett.} \textbf{7} 570.

\bibitem{1PRB81;174301}
Togo A, Chaput L, Tanaka I, and Hug G 2010
First-principles phonon calculations of thermal expansion in
Ti$_{3}$SiC$_{2}$,Ti$_{3}$AlC$_{2}$, and Ti$_{3}$GeC$_{2}$
\emph{Phys. Rev. B} \textbf{81} 174301.

\bibitem{1PRB53;4458}
Fukumoto A
First-principles calculations of p-type impurities in cubic SiC
\emph{Phys. Rev. B} \textbf{53} 4458.

\bibitem{1ComStruc67;115}
Chisholm N, Mahfuz H, Rangari V K, Ashfaq Ad, and Jeelani S 2005
Abrication and mechanical characterization of carbon/SiC-epoxy nanocomposites
\emph{Composite Structures} \textbf{67} 115.

\bibitem{1NanoLett6;1581}
Mpourmpakis G, and Froudakis G E, Lithoxoos G P, and Samios J 2006
SiC Nanotubes: A Novel Material for Hydrogen Storage
\emph{Nano Lett.} \textbf{6} 1581.

\bibitem{1APL82;3107}
Do\'{a}n S, Teke A, Huang D, and Morkoc H 2003
4H¡VSiC photoconductive switching devices for use in high-power applications
\emph{Appl. Phys. Lett.} \textbf{82} 3107.

\bibitem{1NatMater6;479}
Melinon P, Masenelli B, Tournus F, and Perez A 2007
Playing with carbon and silicon at the nanoscale
\emph{Nat. Mater.} \textbf{6} 479.

\bibitem{1PRB82;035431}
Pochet P, Genovese L, Caliste D, Rousseau I, Goedecker S, and Deutsch T 2010
First-principles prediction of stable SiC cage structures and their synthesis pathways
\emph{Phys. Rev. B} \textbf{82} 035431.

%%%%%%%%%%%%%%%%%%%%%%%%%%%%%%%%%%%%%%

\bibitem{1RMP67;357}
Huckestein B 1995
Scaling theory of the integer quantum Hall effect
\emph{Rev. Mod. Phys.} \textbf{67} 357.

\bibitem{1PRL111;136804}
Xu Y, Yan B, Zhang H J, Wang J, Xu G, Tang P, Duan W, and Zhang S C 2013
Large-Gap Quantum Spin Hall Insulators in Tin Films
\emph{Phys. Rev. Lett.} \textbf{111} 136804.

\bibitem{1PRL106;126803}
Brune C, Liu C X, Novik E G, Hankiewicz E M, Buhmann H, Chen Y L, Qi X L, Shen Z X, Zhang S C, and Molenkamp L W 2011
Quantum Hall Effect from the Topological Surface States of Strained Bulk HgTe
\emph{Phys. Rev. Lett.} \textbf{106} 126803.

\bibitem{1PRB70;041303}
Inoue J, Bauer G E W, and Molenkamp L W 2004
Suppression of the persistent spin Hall current by defect scattering
\emph{Phys. Rev. B} \textbf{70} 041303(R).

\bibitem{1JPCC117;6049}
Lian K Y, Ji Y F, Li X F, Jin M X, Ding D J, and Luo Y 2013
Big Bandgap in Highly Reduced Graphene Oxides
\emph{J. Phys. Chem. C} \textbf{117} 6049.

\bibitem{1Carbon94;619}
Do T N, Lin C Y, Lin Y P, Shih P H, and Lin M F 2015
Configuration-enriched magnetoelectronic spectra of AAB-stacked trilayer graphene
\emph{Carbon} \textbf{94} 619.

\bibitem{1PRL73;2744}
Moler K A, Baar D J, Urbach J S, Liang R, Hardy W N, and Kapitulnik A 1994
Magnetic Field Dependence of the Density of States of YBa$_{2}$Cu$_{3}$O$_{6.95}$ as Determined from the Specific Heat \emph{Phys. Rev. Lett.} \textbf{73} 2744.

\bibitem{1NatNanotechnol6;418}
Lee J H, Jang J, Choi J, Moon S H, Noh S, Kim J, Kim J G, Kim I S, Park K I, and Cheon J 2011
Exchange-coupled magnetic nanoparticles for efficient heat induction
Specific Heat \emph{Nat. Nanotechnol. } \textbf{6} 418.

\bibitem{1PRL54;1820}
Gornik E, Lassnig R, Strasser G, Stormer H L, Gossard A C, and Wiegmann W 1985
Specific Heat of Two-Dimensional Electrons in GaAs-GaAlAs Multilayers
Specific Heat \emph{Phys. Rev. Lett. } \textbf{54} 1820.

\bibitem{1PRL59;1341}
Eisenstein J P, Gossard A C, and Narayanamurti V 1987
Quantum oscillations in the thermal conductance of GaAs/AlGaAs heterostructures
Specific Heat \emph{Phys. Rev. Lett. } \textbf{59} 1341.

\bibitem{1Polymer45;4227}
Thermal and flammability properties of polypropylene/carbon nanotube nanocomposites
Kashiwagia T, Grulkeb E, Hildingb J, Grotha K, Harrisa R, Butlera K, et. al., 2001
\emph{Polymer } \textbf{45} 4227.

\bibitem{1ACSAppl1;2256}
Higginbotham A L, Lomeda J R, Morgan A B, and Tour J M 2009
Graphite Oxide Flame-Retardant Polymer Nanocomposites
\emph{ACS Appl. Mater. Interfaces } \textbf{1} 2256.

\bibitem{1NanoLett15;2510}
Derivaz M, Dentel D, Stephan R, Hanf M-C, Mehdaoui A, Sonnet P, and Pirri C 2015
Continuous Germanene Layer on Al(111)
\emph{Nano Lett. } \textbf{15} 2510.

\bibitem{1NJP16;095002}
Davila M E, Xian L, Cahangirov S, Rubio A, and Lay G L 2014
Germanene: a novel two-dimensional germanium allotrope akin to graphene and silicene
\emph{New Jour. Phys.} \textbf{16} 095002.

\bibitem{1ACSNano11;3553}
Zhuang J, Gao N, Li Z, Xu X, Wang J, Zhao J, Dou S X, and Du Y 2017
Cooperative Electron¡VPhonon Coupling and Buckled Structure in Germanene on Au(111)
\emph{ACS Nano} \textbf{11} 3553.

\bibitem{1PRB83;161403}
Shin S Y,  Hwang C G, Sung S J, Kim N D, Kim H S, and Chung J W 2011
Observation of intrinsic intraband $\pi$-plasmon excitation of a single-layer graphene
\emph{Phys. Rev. B} \textbf{83} 161403.

\bibitem{1Carbon50;183}
Generalov A V, and Dedkov Yu S 2012
EELS study of the epitaxial graphene/Ni(1 1 1) and graphene/Au/Ni(1 1 1) systems
\emph{Carbon } \textbf{50} 183.

\bibitem{1PRB72;125117}

Gurtubay I G, Pitarke J M, Ku W, Eguiluz A G, Larson B C, Tischler J, Zschack P, and Finkelstein K D 2005 Electron-hole and plasmon excitations in 3d transition metals: Ab initio calculations and inelastic x-ray scattering measurements
\emph{Phys. Rev. B } \textbf{72} 125117.

\bibitem{1APL99;082110}
Shin S Y, Kim N D, Kim J G, Kim K S, Noh D Y, Kim K S, and Chung J W 2016
Control of the $\pi$ plasmon in a single layer graphene by charge doping
\emph{Appl. Phys. Lett.} \textbf{99} 082110.

\bibitem{1NatNanoTech6;630}
Ju L, Geng B, Horng J, Girit C, Martin M, Hao Z, Bechtel H A, Liang X, and Zettl A
Shen Y R, and Wang F 2011 Graphene plasmonics for tunable terahertz metamaterials
\emph{Nat. Nanotechnol.} \textbf{6} 630.

%%%%%%%%%%%%


%CHHo
\bibitem{volovik03}
Volovik G E 2003 Universe in a helium droplet \textit{Oxford University Press, New York}

\bibitem{chiu16}
Chiu C K, Teo J C Y, Schnyder A P and  Ryu S 2016 Classification of topological quantum matter with symmetries \textit{Rev. Mod. Phys.} \textbf{88} 035005

\bibitem{wen17}
Wen X G 2017 Colloquium: Zoo of quantum-topological phases of matter \textit{Rev. Mod. Phys.} \textbf{89} 041004

\bibitem{landau58}
Landau L D and Lifschitz E M 1958 Course of theoretical physics vol. 5 \textit{Pergamon, London}

\bibitem{klitzing80}
Klitzing K von, Dorda G and Pepper M 1980 New method for high-accuracy determination of the fine-structure constant based on quantized Hall resistance \textit{Phys. Rev. Lett.} \textbf{45} 494

\bibitem{thouless82}
Thouless D J, Kohmoto M, Nightingale M P and Nijs M den 1982 Quantized Hall conductance in a two-dimensional periodic potential \textit{Phys. Rev. Lett.} \textbf{49} 405

\bibitem{avron83}
Avron J E, Seiler R and Simon B 1983 Holonomy, the quantum adiabatic theorem, and Berry's phase \textit{Phys. Rev. Lett.} \textbf{51} 51

\bibitem{simon83}
Simon B 1983 Holonomy, the quantum adiabatic theorem, and Berry's phase \textit{Phys. Rev. Lett.} \textbf{51} 2167

\bibitem{novoselov05}
Novoselov K S, Geim A K, Morozov S V, Jiang D, Katsnelson M I and Grigorieva I V \textit{et al} 2005 Room-temperature quantum Hall effect in graphene \textit{Nature} \textbf{438} 197

\bibitem{zhang05}
Zhang Y, Tan Y W, Stormer H L and Geim A K 2005 Room-temperature quantum Hall effect in graphene \textit{Nature} \textbf{438} 201

\bibitem{zheng02}
Zheng Y and Ando T 2002 Hall conductivity of a two-dimensional graphite system \textit{Phys. Rev.} B \textbf{65} 245420

\bibitem{gusynin05}
Gusynin V P and Sharapov S G 2005 Unconventional integer quantum Hall effect in graphene \textit{Phys. Rev. Lett.} \textbf{95} 146801

\bibitem{ostrovsky08}
Ostrovsky P M, Gornyi I V and Mirlin A D 2008 Theory of anomalous quantum Hall effects in graphene \textit{Phys. Rev.} B \textbf{77} 195430

\bibitem{kawarabayashi09}
Kawarabayashi T, Hatsugai Y and Aoki H 2009 Quantum Hall plateau transition in graphene with spatially correlated random hopping \textit{Phys. Rev. Lett.} \textbf{103} 156804

\bibitem{konig12}
K\"{o}nig E J, Ostrovsky P M, Protopopov and Mirlin A D 2012 Metal-insulator transition in two-dimensional random fermion systems of chiral symmetry classes \textit{Phys. Rev.} B \textbf{85} 195130


\bibitem{kopelevich03}
Kopelevich Y, Torres J H S, da Silva R R, Mrowka F, Kempa H and Esquinazi P 2003 Reentrant metallic behavior of graphite in the quantum limit \textit{Phys. Rev. Lett.} \textbf{90} 156402

\bibitem{kempa06}
Kempa H, Esquinazi P and Kopelevich Y 2006 Integer quantum Hall effect in graphite \textit{Solid State Commun.} \textbf{138} 118

\bibitem{xu14}
Xu Y, Miotkowski I, Liu C, Tian J, Nam H and Alidoust N \textit{et al} 2014 Observation of topological surface state quantum Hall effect in an intrinsic three-dimensional topological insulator \textit{Nat. Phys.} \textbf{10} 956

\bibitem{yoshimi15}
Yoshimi R, Tsukazaki A, Kozuka Y, Falson J, Takahashi K S and Checkelsky J G \textit{et al} 2015 Quantum Hall effect on top and bottom surface states of topological insulator (Bi$_{1-x}$Sb$_{x}$)$_{2}$Te$_{3}$ films \textit{Nat. Commun.} \textbf{6} 6627

%%%%%%%%

\bibitem{1PRB80;033407}
Ma J, Alfe D, Michaelides A, and Wang E 2009
Stone-Wales defects in graphene and other planar sp$^{2}$-bonded materials
\emph{Phys. Rev. B} \textbf{80} 033407

\bibitem{1ACSNano4;1362}
Kim K, Sussman A, Zettl A. 2010
Graphene nanoribbons obtained by electrically unwrapping carbon nanotubes
\emph{ACS Nano} \textbf{4} 1362.

\bibitem{1JMCC3;3087}
Zhang W B, Song Z B, and Dou L M 2015
The tunable electronic structure and mechanical properties of halogenated silicene: a first-principles study
\emph{J. Mater. Chem. C} \textbf{3} 3087.

\end{thebibliography}

\newpage
\begin{thebibliography}{00}

\bibitem{2IBM30;355}
Binnig G, and Rohrer H 1986
Scanning tunneling microscopy
\emph{IBM Journal of Research and Development} \textbf{30} 355.

\bibitem{2APL110;051601}
Miccoli I, Aprojanz J, Baringhaus J, Lichtenstein T, Galves L A , Lopes J M J, et al., 2017 Quasi-free-standing bilayer graphene nanoribbons probed by electronic transport
\emph{Appl. Phys. Lett.} \textbf{110} 051601.

\bibitem{2JACS137;6097}
Liu J Z, Li B W, Tan Y Z, Giannakopoulos A, Sanchez-Sanchez C, Beljonne D, et al., 2015
Toward cove-edged low band gap graphene nanoribbons
\emph{J. Am. Chem. Soc.} \textbf{137} 6097.

\bibitem{2ACSNano4;1362}
Kim K, Sussman A, Zettl A. 2010
Graphene nanoribbons obtained by electrically unwrapping carbon nanotubes
\emph{ACS Nano} \textbf{4} 1362.

\bibitem{2NatComm5;3189}
Vo T H, Shekhirev M, Kunkel D A, Morton M D, Berglund E, Kong L, Wilson P M, Dowben P A, Enders A, Sinitskii A 2014
Large-scale solution synthesis of narrow graphene nanoribbons
\emph{Nat. Comm.} \textbf{5} 3189.

\bibitem{2NanoLett9;3766}
Patra N, Wang B, and Kral P 2009
Nanodroplet Activated and Guided Folding of Graphene Nanostructures
\emph{Nano Lett.} \textbf{9} 3766.

\bibitem{2NANOTECHNOLOGY27;055602}
Song J J, Zhang H J, Cai Y L, Zhang Y X, Bao S N, and He P M 2016
Bottom-up fabrication of graphene nanostructures on Ru(10\={1}0)
\emph{Nanotechnology} \textbf{27} 055602.

%ABA &ABC
\bibitem{2APL107;263101}
Que Y, Xiao W, Chen H, Wang D, Du S, and Gao H-J 2015
Stacking-dependent electronic property of trilayer graphene epitaxially grown on Ru (0001)
\emph{Appl. Phys. Lett.} \textbf{107} 263101.

%ABC
\bibitem{2PRB91;035410}
Xu R, Yin L J, Qiao J B, Bai K K, Nie J C, and He L 2015
Direct probing of the stacking order and electronic spectrum of
rhombohedral trilayer graphene with scanning tunneling microscopy
\emph{Phys. Rev. B} \textbf{91} 035410.

\bibitem{2ACSNano9;5432}
Pierucci D, Sediri H, Hajlaoui M, Girard J C, Brumme T, Calandra M , et al, 2015
Evidence for flat bands near the Fermi level in
epitaxial rhombohedral multilayer graphene
\emph{ACS Nano} \textbf{9} 5432.

\bibitem {2PRB48;17427}
Rong Z Y and Kuiper P 1993 Electronic effects in scanning tunneling
microscopy: Moire pattern on a graphite surface \textit{Phys. Rev.
B} \emph{48} 17427.

\bibitem {2PRB75;235449}
Campanera J M, Savini G, Suarez-Martinez I and Heggie M I 2007
Density functional calculations on the intricacies of Moire patterns
on graphite \emph{Phys. Rev. B} \textbf{75} 235449.

\bibitem{2PRB91;155428}
Cherkez V, Trambly de Laissardiere G, Mallet P, Veuillen J Y 2015
Van Hove singularities in doped twisted graphene bilayers studied by
scanning tunneling spectroscopy \emph{Phys. Rev. B} \textbf{91}
155428.

\bibitem{2PRL109;126801}
Yan W, Liu M, Dou R F, Meng L, Feng L, Chu Z D, et al. 2012
Angle-dependent van Hove singularities in a slightly twisted
graphene bilayer \emph{Phys. Rev. Lett.} \textbf{109} 126801.

\bibitem{2ACSNANO9;8997}
Simonov K A, Vinogradov N A, Vinogradov A S, Generalov A V, Zagrebina E M, Svirskiy G I,  et al., 2015
From graphene nanoribbons on Cu(111) to nanographene on Cu(110): Critical role of substrate
structure in the bottom-up fabrication strategy.
\emph{ACS Nano} \textbf{9} 8997.

\bibitem{2PRL100;056807}
de Parga A L V, Calleja F, Borca B, Passeggi M C G, Hinarejos J J, Guinea F, and Miranda R 2008
Periodically Rippled Graphene: Growth and Spatially Resolved Electronic Structure
\emph{Phys. Rev. Lett.} \textbf{100} 056807.

\bibitem{2PRB77;235412}
Varchon F, Mallet P, Veuillen J-Y, and Magaud L 2008
Ripples in epitaxial graphene on the Si-terminated SiC(0001) surface
\emph{Phys. Rev. B} \textbf{77} 235412.

\bibitem{2PRL110;136804}
Eelbo T, W$\acute{a}$niowska M, Thakur P, Gyamfi M, Sachs B, Wehling T O, Forti S, Starke U, Tieg C, Lichtenstein A I, and Wiesendanger R 2016
Adatoms and Clusters of 3d  Transition Metals on Graphene: Electronic and Magnetic Configurations
\emph{Phys. Rev. Lett.} \textbf{110} 136804.

\bibitem{2Carbon50;4633}
Xu P, Yang Y R, Barber S D, Schoelz J K, Qi D, Ackerman M L , et al., 2012
New scanning tunneling microscopy technique enables systematic study of the unique electronic transition from graphite to graphene
\emph{Carbon} \textbf{50} 4633.

\bibitem{2ChemPhys348;233}
Zeinalipour-Yazdi C D, and Pullman D P 2008
A new interpretation of the scanning tunneling microscope image of graphite
\emph{Chem. Phys.} \textbf{348} 233.

\bibitem{2PRL102;176804}
Li G, Luican A, and Andrei E Y 2009
Scanning tunneling spectroscopy of graphene on graphite
\emph{Phys. Rev. Lett.} \textbf{102} 176804.

\bibitem{2PRB91;115405}
Yin L J, Li S Y, Qiao J B, Nie J C, and He L 2015
Landau quantization in graphene monolayer, Bernal bilayer, and Bernal
trilayer on graphite surface
\emph{Phys. Rev. B} \textbf{91} 115405.

\bibitem{2Science324;924}
Miller D L, Kubista K D, Rutter G M, Ruan M, de Heer W A, First P N, et al., 2009
Observing the quantization of zero mass carriers in graphene
\emph{Science} \textbf{324} 924.

\bibitem{2PRL94;226403}
Matsui T, Kambara H, Niimi Y, Tagami K, Tsukada M, and Fukuyama H 2005
STS observations of Landau levels at graphite surfaces
\emph{Phys. Rev. Lett.} \textbf{94} 226403.

\bibitem{2NatPhys3;623}
Li G, and Andrei E Y 2007
Observation of Landau levels of Dirac fermions in graphite
\emph{Nat. Phys.} \textbf{3} 623.

\bibitem{2Nature391;59}
Wilder J W G, Venema L C, Rinzler A G, Smalley R E, and Dekker C 1998
Electronic structure of atomically resolved carbon nanotubes
\emph{Nature} \textbf{391} 59.

\bibitem{2PRL121;037702}
Huang S, Kim K, Efimkin D K, Lovorn T, Taniguchi T, Watanabe K, MacDonald A H, Tutuc E, and LeRoy B J 2018 Topologically Protected Helical States in Minimally Twisted Bilayer Graphene
\emph{Phys. Rev. Lett.} \textbf{121} 037702.

\bibitem{2CRCPress;CY}
Lin C Y, Chen R B, Ho Y H, and Lin M F 2018
Electronic and optical properties of graphite-related systems,
\emph{CRC Press} Boca Raton, Florida.

\bibitem{2SciRep4;7509}
Huang Y K, Chen S C, Ho Y H, Lin C Y, and Lin M F 2014
Feature-Rich Magnetic Quantization in Sliding Bilayer Graphenes
\emph{Sci. Rep.} \textbf{4} 7509.

\bibitem{5PRB90;205434}
Lin C Y, Wu J Y, Chiu Y H, and Lin M F 2014
Stacking-dependent magneto-electronic properties in multilayer graphenes
\emph{Phys. Rev. B} \textbf{90} 205434.

\bibitem{2PRB81;155413}
Kravets V G, Grigorenko A N, Nair R R, Blake P, Anissimova S, Novoselov K S, et al., 2010
Spectroscopic ellipsometry of graphene and an exciton-shifted van Hove peak in absorption
\emph{Phys. Rev. B} \textbf{81} 155413.

%5 eV saddle point
\bibitem{2PRL106;046401}
Mak K F, Shan J, and Heinz T F 2011
Seeing many-body effects in single- and few-layer graphene: Observation of two-dimensional
saddle-point excitons
\emph{Phys. Rev. Lett.} \textbf{106} 046401.

\bibitem{2APPOPT48;5713}
Martinez L F L, Garcia R C , Navarro R E B, and Martinez A L 2009
Microreflectance difference spectrometer based on a
charge coupled device camera: Surface distribution of
polishing-related linear defect density in GaAs (001)
\emph{Appl. Opt.} \textbf{48} 5713.

%Lamp
\bibitem{2ANALCHEM48;528}
Epstein M S, Rains T C 1976
Ealuation of a xenon-mercury arc lamp for background corrextion in atomic-absorption spectrometry
\emph{Anal. Chem.} \textbf{48} 528.

%0500-0505
\bibitem{2RNPFEB;33}
Stamm G L, Denningh R L, and Rockman A G 1969
Some operating characteristics of a xenon and a xenon-mercury short-arc lamp
immersed in water
\emph{Report of NRL progress} \textbf{33}.

\bibitem{2JMATSCI36;137}
Ferguson L G, and Dogan F 2001 Spectrally selective, matched emitters
for thermophotovoltaic energy conversion processed by tape casting
\emph{J. Mater. Sci.} \textbf{36} 137.

\bibitem{2CPL22;1806}
Zhen H L, Li N, Xiong D Y, Zhou X C, Lu W, and Liu H C 2005
Fabrication and investigation of an upconversion quantum-well
infrared photodetector integrated with a light-emitting diode
\emph{Chin. Phys. Lett.} \textbf{22} 1806.

\bibitem{2JRNBS59;405}
Stewart J E, and Richmond J C 1957
Infrared emission spectrum of silicon carbide heating elements
\emph{J. Res. Natl. Bur. Stand.} \textbf{59} 405.

\bibitem{2APPOPT54;4447}
Lu S H, Liu W C, and Liu J P 2015 High-axial-resolution,
full-field optical coherence microscopy using tungsten halogen lamp
and liquid-crystal-based achromatic phase shifter
\emph{Appl. Opt.} \textbf{54} 4447.

\bibitem{2APPOPT54;2289}
Wei J F, Hu X Y, Sun L Q, and Zhang K, and Chang Y 2015
Technology for radiation efficiency measurement of high-power
halogen tungsten lamp used in calibration of high-energy laser
energy meter
\emph{Appl. Opt.} \textbf{54} 2289.

%Furire spectrometer

\bibitem{2SCIENCE322;1529}
Keppler H, Dubrovinsky L S, Narygina O, and Kantor I 2008
Optical absorption and radiative thermal conductivity of silicate perovskite
to gigapascals \emph{Science} \textbf{322} 1529.

\bibitem{2FARDIS150;71}
Albert S, Albert K K, Lerch P , and Quack M 2011 Synchrotron-based
highest resolution Fourier transform infrared spectroscopy of
naphthalene (C$_{10}$H$_{8}$) and indole (C$_{8}$H$_{7}$N) and its
application to astrophysical problems
\emph{Farad. Discuss.} \textbf{150} 71.

\bibitem{2SPECTROSCOPY15;16}
Gasparian G A, and Lucht H 2000 Indium gallium arsenide NIR
photodiode array spectroscopy
\emph{Spectroscopy} \textbf{15} 16.

\bibitem{2JChromSci14;195}
Dessy R E, Nunn W G, Titus C A, and Reynolds W R 1976 Linear
photodiode array spectrometers as detector systems in automated
liquid chromatographs
\emph{J. Chromatogr. Sci.} \textbf{14} 195.

\bibitem{2REVSCIINS83;093108}
Podobedov V B, Miller C C, and Nadal M E 2012 Performance of the
NIST goniocolorimeter with a broad-band source and multichannel
charged coupled device based spectrometer
\emph{Rev. Sci. Instrum.} \textbf{83} 093108.

\bibitem{2PRB79;115441}
Kuzmenko A B, van Heumen E, van der Marel D, Lerch P, Blake P,
Novoselov K S, et al. 2009
Infrared spectroscopy of electronic bands in bilayer graphene
\emph{Phys. Rev. B} \textbf{79} 115441.

\bibitem{2PRB78;235408}
Zhang L M, Li Z Q, Basov D N, and Fogler M M 2008 Determination
of the electronic structure of bilayer graphene from infrared
spectroscopy \emph{Phys. Rev. B} \textbf{78} 235408.

\bibitem{2NatPhys7;944}
Lui C H, Li Z, Mak K F, Cappelluti E, and Heinz T F 2011
Observation of an electrically tunable band gap in trilayer graphene
\emph{Nat. Phys.} \textbf{7} 944.

\bibitem{2PRL104;176404}
Mak K F, Shan J, and Heinz T F 2010 Electronic Structure of Few-Layer Graphene: Experimental Demonstration of Strong Dependence on Stacking Sequence
\emph{Phys. Rev. Lett.} \textbf{104} 176404.

\bibitem{2PR138;A197}
Taft E A, and Philipp H R 1965 Optical properties of graphite
\emph{Phys. Rev. } \textbf{138} A197.

%5 eV saddle point
\bibitem{2PRL106;046401}
Mak K F, Shan J, and Heinz T F 2011 Seeing many-body effects in
single- and few-layer graphene: Observation of two-dimensional
saddle-point excitons
\emph{Phys. Rev. Lett.} \textbf{106} 046401.

\bibitem{2SUPSCITECH30;065001}
Awaji S, Watanabe K, Oguro H, Miyazaki H, Hanai S, Tosaka T,
et al., 2017 First performance test of a 25 T cryogen-free
superconducting magnet
\emph{Supercond. Sci. Technol.} \textbf{30} 065001.

\bibitem{2IEEETAS27;4603805}
Takahashi M, Iwai S, Miyazaki H, Tosaka T, Tasaki K, Hanai S,
et al., 2017 Design and test results of a cryogenic cooling system for a
25-T cryogen-free superconducting magnet
\emph{IEEE Trans. Appl. Supercond.} \textbf{27} 4603805.

\bibitem{2JLTP159;297}
Sakakura R, Matsuda Y H, Tokunaga M, Kojima E, and Takeyama S 2010
Application of an electro-magnetic induction technique for the
magnetization up to 100 T in a vertical single-turn coil system
\emph{J. Low Temp. Phys.} \textbf{159} 297.

\bibitem{2Science304;1129}
Zaric S, Ostojic G N, Kono J, Shaver J, Moore V C, Strano M S, et al., 2004 Optical signatures of the Aharonov-Bohm phase in single-walled carbon nanotubes
\emph{Science} \textbf{304} 1129.

%transmision
\bibitem{2PRL100;136403}
Orlita M, Faugeras C, Martinez G, Maude D K, Sadowski M L, and Potemski M 2008
Dirac fermions at the H Point of graphite: magnetotransmission studies
\emph{Phys. Rev. Lett.} \textbf{100} 136403.

%magneto-reflection
\bibitem{2PRB15;4077}
Toyt W W, and Dresselhaus M S 1977
Minority carriers in graphite and the H-point magnetoreflection spectra
\emph{Phys. Rev. B} \textbf{15} 4077.

%magneto-Raman
\bibitem{2NanoLett14;4548}
Berciaud S, Potemski M, and Faugeras C 2014 Probing electronic
excitations in mono- to pentalayer graphene by micro magneto-Raman
spectroscopy
\emph{Nano Lett.} \textbf{14} 4548.

\bibitem{2PR138;A197}
Taft E A, and Philipp H R 1965 Optical properties of graphite
\emph{Phys. Rev.} \textbf{138} A197.

\bibitem{2PRB80;161410}
Chuang K-C, Baker A M R, and Nicholas R J 2009 Magnetoabsorption
study of Landau levels in graphite
\emph{Phys. Rev. B} \textbf{80} 161410(R).

\bibitem{2PRL102;166401}
Orlita M, Faugeras C, Schneider J M, Martinez G, Maude D K, and Potemski M 2009 Graphite from the viewpoint of Landau level spectroscopy: an effective graphene bilayer and monolayer
\emph{Phys. Rev. Lett.} \textbf{102} 166401.

\bibitem{2PRB86;155409}
Goncharuk N A, N$\acute{a}$dvornik L, Faugeras C, Orlita M, and Smr$\breve{c}$ka
L 2012 Infrared magnetospectroscopy of graphite in tilted fields
\emph{Phys. Rev. B} \textbf{86} 155409.

\bibitem{2JAP117;112803}
Orlita M, Faugeras C, Barra A-L, Martinez G, Potemski M, Basko D M, et al., 2015 Infrared magneto-spectroscopy of two-dimensional and three-dimensional massless fermions: A comparison
\emph{J. App. Phys.} \textbf{117} 112803.

%2.3

\bibitem{Hall;287}
Hall E H 1879 On a new action of the magnet on electric currents \emph{American
Journal of Mathematics} \textbf{2} 287-292.

\bibitem{Nitta;1335}
Nitta J, Akazaki T, Takayanagi H, Enoki T 1997 Gate Control of Spin-Orbit Interaction in an Inverted In0.53Ga0.47As/In0.52Al0.48 As Heterostructure \emph{Physical Review Letters} \textbf{78} 1335.

\bibitem{Novoselov;177}
Novoselov K S, McCann E, Morozov S V, Fal'ko V I, Katsnelson M I, Zeitler U et al 2006 Unconventional quantum Hall effect and Berry's phase of 2£k in bilayer graphene. \emph{Nature physics} \textbf{2} 177.

\bibitem{Sanchez;076601}
Sanchez-Yamagishi J D, Taychatanapat T, Watanabe K, Taniguchi T, Yacoby A, Jarillo-Herrero P 2012 Quantum Hall effect, screening, and layer-polarized insulating states in twisted bilayer graphene \emph{Physical review letters} \textbf{108} 076601.

\bibitem{Henriksen;011004}
Henriksen E A, Nandi D, Eisenstein J P 2012 Quantum Hall effect and semimetallic behavior of dual-gated ABA-stacked trilayer graphene \emph{Physical Review X} \textbf{2} 011004.

\bibitem{Yuan;125455}
Yuan S, Roldan R, Katsnelson M I 2011 Landau level spectrum of ABA-and ABC-stacked trilayer graphene \emph{Physical Review B} \textbf{84} 125455.

\bibitem{Kumar;126806}
Kumar A, Escoffier W, Poumirol J M, Faugeras C, Arovas D P, Fogler M M, Raquet B 2011 Integer quantum hall effect in trilayer graphene \emph{Physical review letters} \textbf{107} 126806.

%specific heat

\bibitem{Hohne;dsc2013}
Hohne G, Hemminger W F, Flammersheim H J 2013 Differential scanning calorimetry. \emph{Springer Science \& Business Media}, New York.

\bibitem{Gill;931}
Gill P S, Sauerbrunn S R, Reading M 1993 Modulated differential scanning calorimetry. \emph{Journal of Thermal Analysis} \textbf{40} 931.

\bibitem{Schuller;142}
Schuller M, Shao Q, Lalk T 2015 Experimental investigation of the specific heat of a nitrate¡Valumina nanofluid for solar thermal energy storage systems. \emph{International Journal of Thermal Sciences} \textbf{91} 142.

\bibitem{Shinzato;413}
Shinzato K, Baba T 2001 A laser flash apparatus for thermal diffusivity and specific heat capacity measurements. \emph{Journal of thermal analysis and calorimetry} \textbf{64} 413.

\bibitem{Zhou;105}
Zhou K, Wang H P, Chang J, Wei B 2015 Experimental study of surface tension, specific heat and thermal diffusivity of liquid and solid titanium \emph{Chemical Physics Letters} \textbf{639} 105.

\bibitem{Kover;151}
Kover M, Behulova M, Drienovsky M, Motycka P 2015 Determination of the specific heat using laser flash apparatus \emph{Journal of Thermal Analysis and Calorimetry} \textbf{122} 151.


\bibitem{Van;1318}
Van der Hoeven Jr B J C, Keesom P H 1963 Specific heat of various graphites between 0.4 and 2.0 K \emph{Physical Review} \textbf{130} 1318.

\bibitem{Krumhansl;1663}
Krumhansl J, Brooks H 1953 The lattice vibration specific heat of graphite \emph{The Journal of Chemical Physics} \textbf{21} 1663.

\bibitem{DeSorbo;1660}
DeSorbo W, Tyler W W 1953 The specific heat of graphite from 13 to 300 K \emph{The Journal of Chemical Physics} \textbf{21} 1660.

\bibitem{Bowman;367}
Bowman J C, Krumhansl J A 1958 The low-temperature specific heat of graphite \emph{Journal of Physics and Chemistry of Solids}, \textbf{6} 367.

\bibitem{Yu;7565}
Yu A, Ramesh P, Itkis M E, Bekyarova E, Haddon R C 2007 Graphite nanoplatelet? epoxy composite thermal interface materials \emph{The Journal of Physical Chemistry C} \textbf{111} 7565.

\bibitem{Lin;295}
Lin C, Chung D D L 2009 Graphite nanoplatelet pastes vs. carbon black pastes as thermal interface materials \emph{Carbon}, \textbf{47} 295.

\bibitem{Gornik;1820}
Gornik E, Lassnig R, Strasser G, Stormer H L, Gossard A C, Wiegmann W 1985 Specific heat of two-dimensional electrons in GaAs-GaAlAs multilayers \emph{Physical review letters} \textbf{54} 1820.

\bibitem{Bayot;4584}
Bayot V, Grivei E, Melinte S, Santos M B, Shayegan M 1996 Giant low temperature heat capacity of GaAs quantum wells near Landau level filling $\nu= 1$ \emph{Physical review letters} \textbf{76} 4584.

\bibitem{Lin;7592}
Lin M F, Shung K W K 1995 Magnetoconductance of carbon nanotubes \emph{Physical Review B} \textbf{51} 7592.

%

\bibitem{2ZPhys243;229}
Zeppenfeld K 1971 Nonvertical interband transitions in graphite
by intrinsic electron scattering \emph{Z. Phys.} \textbf{243} 229.

\bibitem{2NanoRes10;234}
Dovbeshko G I, Romanyuk V R, Pidgirnyi D V, Cherepanov V V, Andreev E O, Levin V M,
Kuzhir P P, Kaplas T, and Svirko Y P 2015 Optical properties of pyrolytic carbon films
versus graphite and graphene
\emph{Nanoscale Res. Lett.} \textbf{10} 234.

\bibitem{2PRL89;076402}
Marinopoulos A G, Reining L, Olevano V, Rubio A, Pichler T, Liu X,
Knupfer M, and Fink J 2002
Anisotropy and interplane interactions in the dielectric response of graphite
\emph{Phys. Rev. Lett.} \textbf{89} 076402.

\bibitem{2PRB31;4773}
Fischer J E, Bloch J M, Shieh C C, Preil M E, and Jelley K 1985
Reflectivity spectra and dielectric function of stage-1 donor intercalation compounds of graphite
\emph{Phys. Rev.  B} \textbf{31} 4773.

\bibitem{2PRB88;075433}
Wachsmuth P, Hambach R, Kinyanjui M K, Guzzo M, Benner G, and Kaiser U 2013
High-energy collective electronic excitations in free-standing single-layer graphene
\emph{Phys. Rev.  B} \textbf{88} 075433.

\bibitem{2Carbon114;70}
Politanoa A. Radovi$\acute{c}$ I, Borkab D, Mi$\check{s}$kovi$\acute{c}$ Z L, Yu H K, Far$\acute{i}$as D, and Chiarello G 2017
Dispersion and damping of the interband $\pi$ plasmon in graphene grown on Cu(111) foils
\emph{Carbon} \textbf{114} 70.

\bibitem{2PRB91;045418}
Liou S C, Shie C-S, Chen C H, Breitwieser R, Pai W W, Guo G Y, and Chu M-W 2015
$\pi$-plasmon dispersion in free-standing graphene by momentum-resolved electron
energy-loss spectroscopy
\emph{Phys. Rev.  B} \textbf{91} 045418.

\bibitem{2ACSNano12;1837}
Hage F S, Hardcastle T P, Gjerding M N, Kepaptsoglou D M, Seabourne C R, Winther K T, Zan R, Amani J A, Hofsaess H C, Bangert U, Thygesen K S, and Ramasse Q M 2018
Local plasmon engineering in doped graphene
\emph{ACS Nano} \textbf{12} 1837.

\bibitem{2Carbon37;733}
Knupfera M, Pichlera T, Goldena M S, Finka J, Rinzlerb A, and Smalley R E,
Electron energy-loss spectroscopy studies of single wall carbon nanotubes
\emph{Carbon} \textbf{37} 733.

% ribbon
\bibitem{2Nature468;1088}
Suenaga K, and Koshino M 2010
Atom-by-atom spectroscopy at graphene edge
\emph{Nature} \textbf{468} 1088.

\bibitem{2PRB49;2888}
Lucas A A, Henrad L, and Lambin Ph 1994 Computation of the ultraviolet absorption and electron inelastic scattering cross section of multishell fullerenes
\emph{Phys. Rev. B} \textbf{49} 2888.

%onions
\bibitem{2CPL305;225}
Tomita S, Fujii M, Hayashi S, and Yamamoto K 1999
Electron energy-loss spectroscopy of carbon onions
\emph{Chem. Phys. Lett.} \textbf{305} 225.

%reflection
% bulk and surface plasmon

\bibitem{2SurSci602;2069}
Went M R, Vos M, and Werner W S M 2008
Extracting the Ag surface and volume loss functions from reflection electron energy loss spectra
\emph{Sur. Sci.} \textbf{602} 2069.

\bibitem{2APL89;213106}
Werner W S M 2006
Dielectric function of Cu, Ag, and Au obtained from reflection electron energy loss spectra, optical measurements, and density functional theory
\emph{Appl. Phys. Lett.} \textbf{89} 213106.

\bibitem{2SurSci601;L109}
Werner W S M, Went M R, and Vos M 2007
Surface plasmon excitation at a Au surface by 150¡V40000 eV electrons
\emph{Surf. Sci.} \textbf{601} L109.

%transmisiom

\bibitem{2Ultramicroscopy107;575}
Egerton R F 2007 Limits to the spatial, energy and momentum resolution of electron energy-loss spectroscopy \emph{Ultramicroscopy} \textbf{107} 575.

\bibitem{2HIbach}
Ibach H, and Mills D L 1982
Electron energy-loss spectroscopy and surface vibrations spectroscopy \emph{Academic}, New York.

\bibitem{2Ultramicroscopy96;367}
Brink H A, Barfels M M G, Burgner R P, and Edwards B N 2003
A sub-50 meV spectrometer and energy filter for use in combination with 200 kV monochromated
(S)TEMs
\emph{Ultramicroscopy} \textbf{367} 367.

\bibitem{2Micron34;235}
Su D S, Zandbergen H W, Tiemeijer P C, Kothleitner G,
Havecker M, Hebert C, Knop-Gericke A, Freitag B H, Hofer F, and Schlogl R 2003 High resolution EELS usingmonochromator and high performance spectrometer: Comparison
of V2O5 ELNES with NEXAFS and band structure calculations
\emph{Micron} \textbf{34} 235.

\bibitem{2JMicrosc194;203}
Terauchi M, Tanaka M, Tsuno K, and Ishida M 1999
Development of a high energy resolution electron energy-loss spectroscopy microscope
\emph{J. Microsc.} \textbf{194} 203.

\bibitem{2Ultramicroscopy106;1091}
Lazar S, Botton G A, and Zandbergen H W 2006
Enhancement of resolution in core-loss and low-loss spectroscopy in a monochromated microscope
\emph{Ultramicroscopy} \textbf{106} 1091.

%IXS
\bibitem{2Winfried}
W. Sch\"{u}lke 2007 Electron dynamics by inelastic x-Ray scattering
\emph{Oxford University Press} Oxford.

\bibitem{2PRB76;035439}
Mohr M, Maultzsch J, Dobard\v{z}i\'{c} E, Reich S, Milo\v{s}evi\'{c} I, Damnjanovi\'{c}q M, Bosak A, Krisch M, and Thomsen C 2007
Phonon dispersion of graphite by inelastic x-ray scattering, Phonon dispersion of graphite by inelastic x-ray scattering
\emph{Phys. Rev. B} \textbf{76} 035439.

%plasmon
\bibitem{2PRB89;014206}
Kimura K, Matsuda K, Hiraoka N, Fukumaru T, Kajihara Y, Inui M, and Yao M 2014
Inelastic x-ray scattering study of plasmon dispersions in solid and liquid Rb,
\emph{Phys. Rev. B} \textbf{89} 014206.

\bibitem{2PCM19;046207}
Tirao G, Stutz G, Silkin V M, Chulkov E V, and Cusatis C 2007
Plasmon excitation in beryllium: inelastic x-ray scattering experiments and first-principles
calculations
\emph{J. Phys.: Condens. Matter} \textbf{19} 046207.

\bibitem{2JPSJ84;084701}
Kimura K, Matsuda K, Hiraoka N, Kajihara Y, Miyatake T, Ishiguro Y, Hagiya T, Inui M, and Yao M 2015
Inelastic x-ray scattering study on plasmon dispersion in liquid Cs
\emph{J. Phys. Soc. Jpn.} \textbf{84} 084701.

\bibitem{2PRB71;060504}
Galambosi S, Soininen J A, Mattila A, Huotari S, Manninen S, Vanko Gy, Zhigadlo N D, Karpinski J, and Hamalainen K 2005
Inelastic x-ray scattering study of collective electron excitations in MgB$_{2}$
\emph{Phys. Rev. B} \textbf{71} 060504.

%graphite

\bibitem{2PRL101;266406}
Hambach R, Giorgetti C, Hiraoka N, Cai Y Q, Sottile F, Marinopoulos A G, Bechstedt F,
and Reining L 2008
Anomalous angular dependence of the dynamic structure factor near Bragg reflections: graphite
\emph{Phys. Rev. Lett.} \textbf{101} 266406.

\bibitem{2Kittel}
Charles Kittel 2004 Introduction to solid state physics 8th Edition.

\bibitem{2Ultramicroscopy96;367}
Brink H A, Barfels M M G, Burgner R P, and Edwards B N 2003
A sub-50 meV spectrometer and energy filter for use in combination with 200 kV monochromated
(S)TEMs
\emph{Ultramicroscopy} \textbf{96} 367.

\bibitem{2PRB55;13961}
Lin M F, Huang C S, and Chuu D S 1997 Plasmons in graphite and stage-1 graphite intercalation compounds
\emph{Phys. Rev. B} \textbf{55} 13961.

\bibitem{2PRB86;195424}
Scholz A, Stauber T, and Schliemann J 2012
Dielectric function, screening, and plasmons of graphene in the presence of spin-orbit interactions
\emph{Phys. Rev. B} \textbf{86} 195424.

\bibitem{2PRB38;2112}
Schulke W, Bonse U, Nagasawa H, Kaprolat A, and Berthold A 1988
Interband transitions and core excitation in highly oriented
pyrolytic graphite studied by inelastic synchrotron x-ray scattering:
Band-structure information
\emph{Phys. Rev. B} \textbf{38} 2112.

\bibitem{2PRB86;245430}
Zhang L, Schwertfager N, Cheiwchanchamnangij T, Lin X, Glans-Suzuki P-A, Piper L F J, Limpijumnong S, Luo Y, Zhu J F, Lambrecht W R L, and Guo J-H 2012
Electronic band structure of graphene from resonant soft x-ray spectroscopy: The role of core-hole effects
\emph{Phys. Rev. B} \textbf{86} 245430.

\bibitem{2RSI82;113108}
Gao X, Burns C, Casa D, Upton M, Gog T, Kim J, and Li C 2011
Development of a graphite polarization analyzer for resonant inelastic x-ray scattering
\emph{Rev. Sci. Instrum.} \textbf{82} 113108.

\bibitem{RSI77;053102}
Huotari S, Albergamo F, Vanko Gy, Verbeni R, and Monaco G 2006
Resonant inelastic hard x-ray scattering with diced analyzer crystals and positionsensitive
detectors
\emph{Rev. Sci. Instrum.} \textbf{77} 053102.

\bibitem{RMP73;203}
Kotani A, and Shin S 2001
Resonant inelastic x-ray scattering spectra for electrons in solids
\emph{Rev. Mod. Phys.} \textbf{73} 203.

\bibitem{RMP79;175}
Devereaux T P 2007
Inelastic light scattering from correlated electrons
\emph{Rev. Mod. Phys.} \textbf{79} 175.

\bibitem{RMP83;705}
Ament L J P, van Veenendaal M, Devereaux T P, Hill J P, and van den Brink J 2001
Resonant inelastic x-ray scattering studies of elementary excitations
\emph{Rev. Mod. Phys.} \textbf{83} 705.

%IXs

\bibitem{Strocov;631}
Strocov V N, Schmitt T, Flechsig U, Schmidt T, Imhof A, Chen Q Wang X 2010 High-resolution soft X-ray beamline ADRESS at the Swiss Light Source for resonant inelastic X-ray scattering and angle-resolved photoelectron spectroscopies. \emph{Journal of synchrotron radiation} \textbf{17} 631.

\bibitem{Qiao;033106}
Qiao R, Li Q, Zhuo Z, Sallis S, Fuchs O, Blum M, Brown A 2017 High-efficiency in situ resonant inelastic x-ray scattering (iRIXS) endstation at the Advanced Light Source \emph{Review of Scientific Instruments} \textbf{88} 033106.

\bibitem{Egerton;2011}
Egerton R F 2011 Electron energy-loss spectroscopy in the electron microscope \emph{Springer Science \& Business Media}, Plenum, New York and London.

\bibitem{Egerton;575}
Egerton R F 2007 Limits to the spatial, energy and momentum resolution of electron energy-loss spectroscopy \emph{Ultramicroscopy} \textbf{107} 575.

\end{thebibliography}

\newpage
\begin{thebibliography}{00}

\bibitem{Zhao;24}
Zhao J, Liu H, Yu Z, Quhe R, Zhou S, Wang Y, Yao Y 2016 Rise of silicene: A competitive 2D material \emph{Progress in Materials Science} \textbf{83} 24.

\bibitem{3PRB95;115411}
Wu J Y, Chen S C, G Gumbs and Lin M F 2017
Field-created diverse quantizations in monolayer and bilayer black phosphorus
\emph{Phys. Rev. B} \textbf{95} 115411.

\bibitem{3PRB97;125416}
Do T N, Shih P H, Gumbs G, Huang D, Chiu C W, Lin M F 2018
Diverse magnetic quantization in bilayer silicene
\emph{Phys. Rev. B} \textbf{97} 125416.

\bibitem{3APL104;131904}
Fu H X, Zhang J, Ding Z J, Li H, and Menga S 2014
Stacking-dependent electronic structure of bilayer silicene
\emph{Appl. Phys. Lett.} \textbf{104} 131904.

\bibitem{3JPCC119;3818}
Padilha J E, and Pontes R B 2015
Free-Standing Bilayer Silicene: The Effect of Stacking Order on the Structural, Electronic, and Transport Properties \emph{J. Phys. Chem. C} \textbf{119} 3818.

\bibitem{3PRB83;165443}
Koshino M, and, McCann E 2011
Landau level spectra and the quantum Hall effect of multilayer graphene
\emph{Phys. Rev. B} \textbf{83} 165443.

\bibitem{3PRL86;1062}
Koshino M, Aoki H, Kuroki K, Kagoshima S, and Osada T 2001
Hofstadter Butterfly and Integer Quantum Hall Effect in Three Dimensions
\emph{Phys. Rev. Lett.} \textbf{86} 1062.

\bibitem{3PRB72;165304}
Taut M, Eschrig H, and Richter M 2005
Skyrmion in a real magnetic film
\emph{Phys. Rev. B} \textbf{72} 165304.

\bibitem{3PRB90;205434}
Lin C Y, Wu J Y, Chiu Y H, and Lin M F 2014
Stacking-dependent magneto-electronic properties in multilayer graphenes
\emph{Phys. Rev. B} \textbf{90} 205434.

\bibitem{3PCCP17;15921}
Lin Y P, Lin C Y, Ho Y H, Do T N, and Lin M F 2015
Magneto-optical properties of ABC-stacked trilayer graphene
\emph{Phys. Chem. Chem. Phys.} \textbf{17} 15921.

\bibitem{3IOPBook;CY}
C. Y. Lin, T. N. Do, Y. K. Huang, and M. F. Lin 2017
Electronic and optical properties of graphene in magnetic and electric fields
\emph{IOP Concise Physics.} San Raefel, CA, USA: Morgan $\&$ Claypool Publishers.

\bibitem{3PCCP19;29525}
Do T N, Chang C P, Shih P H, Lin  M F 2017
Stacking-enriched magneto-transport properties of few-layer graphenes
\emph{Phys. Chem. Chem. Phys.} \textbf{19} 29525.

\bibitem{3PRB67;045405}
Shyu F L, Chang C P, Chen R B, Chiu C W, and Lin M F 2003
Magnetoelectronic and optical properties of carbon nanotubes
\emph{Phys. Rev. B} \textbf{67} 045405.

\bibitem{3SciRep7;40600}
Shih P H, Chiu Y H, Wu J Y, Shyu F L, Lin M F 2017
Coulomb excitations of monolayer germanene
\emph{Sci. Rep.} \textbf{7} 40600.

\bibitem{3PRB34;2}
Shung K W K 1986
Lifetime effects in low-stage intercalated graphite systems
\emph{Phys. Rev. B} \textbf{34} 2.

\bibitem{3PLA352;446}
Ho J H, Chang C P, and Lin M F 2006
Electronic excitations of the multilayered graphite
\emph{Phys. Lett. A} \textbf{352} 446.

\bibitem{3PRB74;085406}
Ho J H, Lu C L, Hwang C C, Chang C P, and Lin M F 2006
Coulomb excitations in AA- and AB-stacked bilayer graphites
\emph{Phys. Rev. B} \textbf{74} 085406.

\bibitem{3PRSL243;336}
Hubbard J 1963
The description of collective motions in terms of many-body perturbation theory.
II. The correlation energy of a free-electron gas
\emph{Proc. R. Soc. Lond.} \textbf{243} 336.

\bibitem{3PR176;589}
Singwi K S, Tosi M P, Land R H, and Sjolander A 1968
Electron correlations at metallic densities
\emph{Phys. Rev. } \textbf{176} 589.

\bibitem{3PR6;875}
Vashishta P, and Singwi K S 1972
Electron correlations at metallic densities. V
\emph{Phys. Rev. B} \textbf{6} 875.

\bibitem{3Carbon50;183}
Generalov A V, and Dedkov Yu S 2012
EELS study of the epitaxial graphene/Ni(1 1 1) and graphene/Au/Ni(1 1 1) systems
\emph{Carbon } \textbf{50} 183.

\bibitem{3PRB83;161403}
Shin S Y,  Hwang C G, Sung S J, Kim N D, Kim H S, and Chung J W 2011
Observation of intrinsic intraband $\pi$-plasmon excitation of a single-layer graphene
\emph{Phys. Rev. B} \textbf{83} 161403.

\bibitem{3NanoLett14;3827}
Nelson F J, Idrobo J.-C., Fite J D, Miskovic Z L, Pennycook S J, Pantelides S T, Lee J U, and Diebold A C 2014 Electronic excitations in graphene in the 1-50 eV Range: The $\pi$ and $\pi+\sigma$ peaks are not plasmons
\emph{Nano Lett.} \textbf{14} 3827.

\bibitem{3ACSNano5;1026}
Wu J Y, Chen S C, Roslyak O, Gumbs G,  Lin M F 2011
Plasma Excitations in Graphene: Their Spectral Intensity and Temperature Dependence in Magnetic Field
\emph{ACS Nano} \textbf{5} 1026.

\bibitem{3JPP41;47}
Blinowski J, Hau N H, Rigaux C, Vieren J P, Toullee R le, Furdin C, Herold A, and  Melin J 1980
\emph{J. Phys. (Paris)} \textbf{41} 47.

\bibitem{3PRB98;041408}
Lin C Y, Lee M H, and Lin M F 2018
Coulomb excitations in trilayer ABC-stacked graphene
\emph{Phys. Rev. B Rapid communication} \textbf{98} 041408.




\end{thebibliography}

\newpage

\begin{thebibliography}{00}

\bibitem{Zhang;156801}
Zhang F, Jung J, Fiete G A, Niu Q, MacDonald A H 2011 Spontaneous quantum Hall states in chirally stacked few-layer graphene systems \emph{Physical review letters} \textbf{106} 156801.


\bibitem{Brunt;5942}
Brunt T A, Rayment T, O'shea S J, Welland M E 1996 Measuring the surface stresses in an electrochemically deposited metal monolayer: Pb on Au (111) \emph{Langmuir} \textbf{12} 5942.

\bibitem{Akiyama;L843}
Akiyama M, Kawarada Y, Kaminishi K 1984 Growth of single domain GaAs layer on (100)-oriented Si substrate by MOCVD. \emph{Japanese Journal of Applied Physics} \textbf{23} L843.

\bibitem{Singh;6386}
Singh D, Gupta S K, Sonvane Y, Lukacevic I 2016 Antimonene: a monolayer material for ultraviolet optical nanodevices \emph{Journal of Materials Chemistry C} \textbf{4} 6386.

\bibitem{Chen;7290}
Chen C H, Kepler K D, Gewirth A A, Ocko B M, Wang J 1993 Electrodeposited bismuth monolayers on gold (111) electrodes: comparison of surface x-ray scattering, scanning tunneling microscopy, and atomic force microscopy lattice structures. \emph{The Journal of Physical Chemistry} \textbf{97} 7290.

\bibitem{Luo;29516}
Luo Z, Huang Y, Weng J, Cheng H, Lin Z, Xu B, Xu H 2013 1.06 £gm Q-switched ytterbium-doped fiber laser using few-layer topological insulator Bi2Se3 as a saturable absorber \emph{Optics express} \textbf{21} 29516.


\bibitem{12NatMat12;887}
Kim K S, Walter A L, Moreschini L, Seyller T, Horn K, Rotenberg E, and Bostwick A 2013
Coexisting massive and massless Dirac fermions in symmetry-broken bilayer graphene
\emph{Nat. Mater.} \textbf{12} 887.

\bibitem{12PRL109;196802}
Brihuega I, Mallet P, Gonzalez-Herrero H, Trambly de Laissardiere G, Ugeda M M, Magaud L, Gomez-Rodriguez J M, Yndurain F, and Veuillen J-Y 2012
Unraveling the Intrinsic and Robust Nature of van Hove Singularities in Twisted Bilayer Graphene by Scanning Tunneling Microscopy and Theoretical Analysis
\emph{Phys. Rev. Lett.} \textbf{109} 196802.

\bibitem{12NanoLett12;3833}
Wang Z F, Liu F, and Chou M Y  2012
Fractal Landau-Level Spectra in Twisted Bilayer Graphene
\emph{Nano Lett.} \textbf{12} 3833.

\bibitem{Mele;161405}
Mele E J 2010 Commensuration and interlayer coherence in twisted bilayer graphene \emph{Physical Review B} \textbf{81} 161405.


\bibitem{Son;155410}
Son Y W, Choi S M, Hong Y P, Woo S, Jhi S H 2011 Electronic topological transition in sliding bilayer graphene \emph{Physical Review B}, \textbf{84} 155410.

\bibitem{Lin;161409}
Lin Y M, Perebeinos V, Chen Z, Avouris P 2008 Electrical observation of subband formation in graphene nanoribbons \emph{Physical Review B} \textbf{78} 161409.

\bibitem{Ohta;206802}
Ohta T, Bostwick A, McChesney J L, Seyller T, Horn K, Rotenberg E 2007 Interlayer interaction and electronic screening in multilayer graphene investigated with angle-resolved photoemission spectroscopy \emph{Physical Review Letters} \textbf{98} 206802.

\bibitem{Castro;216802}
Castro E V, Novoselov K S, Morozov S V, Peres N M R, Dos Santos J L, Nilsson J, Neto A C 2007 Biased bilayer graphene: semiconductor with a gap tunable by the electric field effect \emph{Physical review letters} \textbf{99} 216802.

\bibitem{12Zhao;24}
Zhao J, Liu H, Yu Z, Quhe R, Zhou S, Wang Y, Yao Y 2016 Rise of silicene: A competitive 2D material \emph{Progress in Materials Science} \textbf{83} 24.

\bibitem{Akinwande;5678}
Akinwande D, Petrone N, Hone J 2014 Two-dimensional flexible nanoelectronics \emph{Nature communications} \textbf{5} 5678.

\bibitem{Vogt;155501}
Vogt P, De Padova P, Quaresima C, Avila J, Frantzeskakis E, Asensio M C, Le Lay G 2012 Silicene: compelling experimental evidence for graphenelike two-dimensional silicon \emph{Physical review letters} \textbf{108} 155501.

\bibitem{Feng;3507}
Feng B, Ding Z, Meng S, Yao Y, He X, Cheng P, Wu K 2012 Evidence of silicene in honeycomb structures of silicon on Ag (111) \emph{Nano letters} 12 3507.

\bibitem{Zhuang:161409}
Zhuang J, Xu X, Du Y, Wu K, Chen L, Hao W, Dou S X 2015 Investigation of electron-phonon coupling in epitaxial silicene by in situ Raman spectroscopy \emph{Physical Review B} \textbf{91} 161409.

%%

\bibitem{SRp3;1075} Tahir M and Schwingenschlogl U 2013 Valley polarized quantum Hall effect and topological insulator phase transitions in Silicene \emph{Sci. Rep.} \textbf{3} 1075.

\bibitem{PRL95;226801} Kane C L and Mele E J 2005 Quantum Spin Hall Effect in Graphene \emph{Phys. Rev. Lett.} \textbf{95} 226801.

\bibitem{PRB87;235426} Tabert C J and Nicol E J 2013 AC/DC spin and valley Hall effects in silicene and germanene \emph{Phys. Rev. B} \textbf{87} 235426.

\bibitem{PRB89;085429} Kim Y, Choi K and Ihm J 2014 Topological domain walls and quantum valley Hall effects in silicone \emph{Phys. Rev. B} \textbf{89} 085429.

\bibitem{NJP21;023010} Mu Y S, Xue Y, Zhou T and Yang Z Q 2019 Quantum anomalous Hall effects and various topological mechanisms in functionalized Sn monolayers \emph{New J. Phys.} \textbf{21} 023010.

\bibitem{PRL110;066803} Li X, Zhang F and Niu Q 2013 Unconventional Quantum Hall Effect and Tunable Spin Hall Effect in Dirac Materials: Application to an Isolated MoS2 Trilayer \emph{Phys. Rev. Lett.} \textbf{110} 066803.

\bibitem{PRB92;165409} Ghazaryan A and Chakraborty T 2015 Aspects of anisotropic fractional quantum Hall effect in phosphorene \emph{Phys. Rev. B} \textbf{92} 165409.

\bibitem{EPL119;37005} Ma R, Liu S W, Yang W Q and Deng M X 2017 Quantum Hall effect in monolayer and bilayer black phosphorus \emph{Europhys. Lett.} \textbf{119} 37005.

\bibitem{NanoLett18;229} Yang J, Tran S, Wu J, Che S, Stepanov P and Taniguchi T et al 2018 Integer and Fractional Quantum Hall effect in Ultrahigh Quality Few-layer Black Phosphorus Transistors \emph{Nano Lett.} \textbf{18} 229.




\bibitem{12Lin;074603}
Lin S Y, Ho Y H, Shyu F L, Lin M F 2013 Electronic Thermal Property of Graphite \emph{J. Phys. Soc. Jpn} \textbf{82} 074603.




\bibitem{SR7;2017} Shih P S, Chiu Y H, Wu J Y, Shyu F L and Lin M F 2017 Coulomb excitations of monolayer germanene \emph{Sci. Rep.} \textbf{7} 40600.

\bibitem{PRB97;125416} Do T N, Shih P H, Gumbs G, Huang D, Chiu C W and Lin M F 2018 Diverse magnetic quantization in bilayer silicene \emph{Phys. Rev. B} \textbf{97} 125416.

\bibitem{PRB94;045410}

\bibitem{JPCC119;11896} KaloniT P, Modarresi M, Tahir M, Roknabadi M R, Schreckenbach G and Freund M S 2015 Electrically Engineered Band Gap in Two-Dimensional Ge, Sn, and Pb: A First-Principles and Tight-Binding Approach \emph{J. Phys. Chem. C} \textbf{119} 11896.

\end{thebibliography}

\newpage
\begin{thebibliography}{00}


\bibitem{Novoselov;666}
Novoselov K S, Geim A K, Morozov S V, Jiang D, Zhang Y, Dubonos S V, Firsov A A 2004 Electric field effect in atomically thin carbon films \emph{science} \textbf{306} 666.

\bibitem{ACSNano6;6930} Ruffieux P, Cai J, Plumb N C, Patthey L, Prezzi D and Ferretti A et al 2012 Electronic structure of atomically precise graphene nanoribbons \emph{ACS Nano} \textbf{6} 6930.

\bibitem{APL110;051601} Miccoli I, Aprojanz J, Baringhaus J, Lichtenstein T, Galves L A and Lopes J M J et al 2017 Quasi-free-standing bilayer graphene nanoribbons probed by electronic transport \emph{Appl. Phys. Lett.} \textbf{110} 051601, .

\bibitem{SSR67;1} Kara A, Enriquez H, Seitsonen A P, Voon L C L Y, Vizzini S and Aufrayg B et al 2012 A review on silicene-New candidate for electronics \emph{Surf. Sci. Rep.} \textbf{67} 1.


\bibitem{MSS93;92} Salimiana F and Dideba D 2019 Comparative study of nanoribbon field effect transistors based on silicene and graphene \emph{Mat. Sci. Semicon. Proc.} \textbf{93} 92.

\bibitem{MRE4;114005} Abhinav E M, Sundararaj A, Gopalakrishnan C, Raja S V K and Chokhra S 2017 Impact of strain on electronic and transport properties of 6nm hydrogenated germanane nano-ribbon channel double gate field effect transistor \emph{Mater. Res. Express} \textbf{4} 114005.




\bibitem{13IOP;SC}
Chen S C, Wu J Y, Lin C Y, and Lin M F 2017
Theory of Magnetoelectric Properties of 2D Systems
\emph{IOP Concise Physics.} San Raefel, CA, USA: Morgan $\&$ Claypool Publishers.

\bibitem{13CRCPress;CY}
Lin C Y, Chen R B, Ho Y H, and Lin M F 2018
Electronic and optical properties of graphite-related systems,
\emph{CRC Press} Boca Raton, Florida.

\bibitem{13IOPBook;CY}
C. Y. Lin, T. N. Do, Y. K. Huang, and M. F. Lin 2017
Electronic and optical properties of graphene in magnetic and electric fields
\emph{IOP Concise Physics.} San Raefel, CA, USA: Morgan $\&$ Claypool Publishers.

\bibitem{13PRB83;195405}
Ou Y C, Sheu J K, Chiu Y H, Chen R B, and Lin M F 2011
Influence of modulated fields on the Landau level properties of graphene
\emph{Phys. Rev. B} \textbf{83} 195405.

\bibitem{13PCCP18;7573}
Chung H C, Chang C P, Lin C Y, and Lin M F 2016
Electronic and Optical Properties of Graphene Nanoribbons in External Fields,
\emph{Phys. Chem. Chem. Phys.} \textbf{18} 7573.

\bibitem{13JPSJ81;064719}
Lin C Y, Chen S C, Wu J Y, and Lin M F 2012
Curvature Effects on Magnetoelectronic Properties of Nanographene Ribbons
\emph{J. Phys. Soc. Jpn.} \textbf{81} 064719.

\bibitem{Zhu;494}
Zhu L, Wang J, Zhang T, Ma L, Lim C W, Ding F, Zeng X C 2010 Mechanically robust tri-wing graphene nanoribbons with tunable electronic and magnetic properties \emph{Nano letters} \textbf{10} 494.

\bibitem{13Nature438;201}
Zhang Y, Tan Y W, Stormer H L, and Kim P 2005 Experimental observation of the quantum Hall effect and Berry's phase in graphene
\emph{Nature} \textbf{438} 201.

\bibitem{13NatPhys7;953}
Zhang L, Zhang Y, Camacho J, Khodas M, and Zaliznyak I 2011
The experimental observation of quantum Hall effect of l=3 chiral quasiparticles in trilayer graphene
\emph{Nature} \textbf{7} 953.















\bibitem{Yan;2159}
Yan W, He W Y, Chu Z D, Liu M, Meng L, Dou R F, He L 2013 Strain and curvature induced evolution of electronic band structures in twisted graphene bilayer \emph{Nature communications} \textbf{4} 2159.

\bibitem{13ACSNano4;1362}
Kim K, Sussman A, Zettl A. 2010
Graphene nanoribbons obtained by electrically unwrapping carbon nanotubes
\emph{ACS Nano} \textbf{4} 1362.

\bibitem{13NatComm5;3189}
Vo T H, Shekhirev M, Kunkel D A, Morton M D, Berglund E, Kong L, Wilson P M, Dowben P A, Enders A, Sinitskii A 2014
Large-scale solution synthesis of narrow graphene nanoribbons
\emph{Nat. Comm.} \textbf{5} 3189.

\bibitem{13NanoLett9;3766}
Patra N, Wang B, and Kral P 2009
Nanodroplet Activated and Guided Folding of Graphene Nanostructures
\emph{Nano Lett.} \textbf{9} 3766.

\bibitem{13PRB70;075411}
Tsai C C, Shyu F L, Chiu C W, Chang C P, Chen R B, and M. F. Lin,
Magnetization of armchair carbon tori
\emph{Phys. Rev. B} \textbf{70} 075411.

\bibitem{1CPL305;225}
Tomita S, Fujii M, Hayashi S, and Yamamoto K 1999
Electron energy-loss spectroscopy of carbon onions
\emph{Chem. Phys. Lett.} \textbf{305} 225.


\bibitem{Prada;106802}
Prada E, San-Jose P, Brey L 2010 Zero Landau level in folded graphene nanoribbons \emph{Physical review letters} \textbf{105} 106802.

\bibitem{Martins;075710}
Martins B V C, Galvao D S 2010 Curved graphene nanoribbons: structure and dynamics of carbon nanobelts \emph{Nanotechnology}, 21 075710.

\bibitem{Viculis;1361}
Viculis L M, Mack J J, Kaner R B 2003 A chemical route to carbon nanoscrolls \emph{Science} \textbf{299} 1361.

\bibitem{Haddon;388}
Haddon R C 1997 Electronic properties of carbon toroids \emph{Nature} \textbf{388} 31.

\bibitem{Banhart;433}
Banhart F, Ajayan P M 1996 Carbon onions as nanoscopic pressure cells for diamond formation \emph{Nature} \textbf{382} 433.


\bibitem{Monteiro;377}
Monteiro-Riviere N A, Nemanich R J, Inman A O, Wang Y Y, Riviere J E 2005 Multi-walled carbon nanotube interactions with human epidermal keratinocytes \emph{Toxicology letters} \textbf{155} 377.

\bibitem{Schaper;73}
Schaper A K, Hou H, Greiner A, Schneider R, Phillipp F 2004 Copper nanoparticles encapsulated in multi-shell carbon cages \emph{Applied Physics A} \textbf{78} 73.


\bibitem{Yang;3996}
Yang Z, Liu M, Zhang C, Tjiu W W, Liu T, Peng H 2013 Carbon nanotubes bridged with graphene nanoribbons and their use in high-efficiency dye?sensitized solar cells. \emph{Angewandte Chemie International Edition} \textbf{52} 3996.

\bibitem{Khoeini;1315}
Khoeini F, Shokri A A 2011 Modeling of Transport in a Glider-Like Composite of GNR/CNT/GNR Junctions \emph{Journal of Computational and Theoretical Nanoscience} \textbf{8} 1315.


\bibitem{13PRB80;033407}
Ma J, Alfe D, Michaelides A, and Wang E 2009
Stone-Wales defects in graphene and other planar sp$^{2}$-bonded materials
\emph{Phys. Rev. B} \textbf{80} 033407

\bibitem{Terrones;351}
Terrones M, Botello-Mendez A R, Campos-Delgado J, Lopez-Urias F, Vega-Cantu Y I, Rodriguez-Macias F J, Terrones H 2010 Graphene and graphite nanoribbons: Morphology, properties, synthesis, defects and applications \emph{Nano today} \textbf{5} 351.

\bibitem{Sahin;115432}
Sahin H, Topsakal M, Ciraci S 2011 Structures of fluorinated graphene and their signatures \emph{Physical Review B} \textbf{83} 115432.

\bibitem{He;3766}
He Z, He K, Robertson A W, Kirkland A I, Kim D, Ihm J, Warner J H 2014 Atomic structure and dynamics of metal dopant pairs in graphene \emph{Nano letters} \textbf{14} 3766.

\bibitem{Li;7881}
Li S, Wu Y, Tu Y, Wang Y, Jiang T, Liu W, Zhao Y 2015 Defects in silicene: vacancy clusters, extended line defects, and di-adatoms \emph{Scientific reports} \textbf{5} 7881.

\bibitem{13CRCPress;SY}
S. Y. Lin, N. T. T. Tran, S. L. Chang, W. P. Su, and M. F. Lin 2018
Structure- and adatom-enriched essential properties of graphene nanoribbons
\emph{CRC Press} Boca Raton, Florida.





\bibitem{ACSNano4;1465} Ho Y H, Chiu Y H, Lin D H, Chang C P and Lin M F 2010 Magneto-optical selection rules in bilayer Bernal grapheme \emph{ ACS Nano} \textbf{4} 1465.

\bibitem{SRp4;7509} Huang Y K, Chen S C, Ho Y H, Lin C Y and Lin M F 2014 Feature-rich magnetic quantization in sliding bilayer graphenes \emph{Sci. Rep.} \textbf{4} 7509.

\bibitem{SRP9;624} Do T N, Gumbs G, Shih P H, Huang D, Chiu C W and Chen C Y et al 2019 Peculiar optical properties of bilayer silicene under the influence of external electric and magnetic fields \emph{Sci. Rep.} \textbf{9} 624.

\bibitem{OLt43;6089} Chen R B, Jang D J, Lin M C and Lin M F 2018 Optical properties of monolayer bismuthene in electric fields \emph{Opt. Lett.} \textbf{43} 6089.

\bibitem{PLA383;68} Shyu F L 2019 Field-induced spin polarized electronic and optical properties of armchair stanene nanoribbons \emph{Phys. Lett. A} \textbf{383} 68.


\bibitem{neil;4356}
O'neil M, Marohn J, McLendon G 1990 Dynamics of electron-hole pair recombination in semiconductor clusters \emph{Journal of Physical Chemistry} \textbf{94} 4356.

\bibitem{Brinkman;1508}
Brinkman W F, Rice T M 1973 Electron-hole liquids in semiconductors \emph{Physical Review B} \textbf{7} 1508.

\bibitem{Nakamura;786}
Nakamura Y, Pashkin Y A, Tsai J S 1999 Coherent control of macroscopic quantum states in a single-Cooper-pair box \emph{Nature} \textbf{398} 786.

\bibitem{13NanoLett14;3353}
Havener R W, Liang Y, Brown L, Yang L, and Park J 2014
Van Hove Singularities and Excitonic Effects in the Optical Conductivity of Twisted Bilayer Graphene
\emph{Nano Lett.} \textbf{14} 3353.

\bibitem{Van;8154}
Van de Walle C G, Martin R M 1987 Theoretical study of band offsets at semiconductor interfaces \emph{Physical Review B} \textbf{35} 8154.

\bibitem{Dong;2361}
Dong J, Sankey O F, Myles C W 2001 Theoretical study of the lattice thermal conductivity in Ge framework semiconductors \emph{Physical review letters} \textbf{86} 2361.

\bibitem{Lewiner;2347}
Lewiner C, Bastard G 1980 Indirect exchange interaction in extremely non-parabolic zero-gap semiconductors \emph{Journal of Physics C: Solid State Physics} \textbf{13} 2347.

\bibitem{Huang;859}
Huang B L, Chuu C P, Lin M F 2019 Asymmetry-enriched electronic and optical properties of bilayer graphene \emph{Scientific reports} 9 859.

\bibitem{Sarma;1936}
Sarma S D, Hwang E H 1996 Dynamical response of a one-dimensional quantum-wire electron system \emph{Physical Review B} \textbf{54} 1936.

\bibitem{Gusynin;157402}
Gusynin V P, Sharapov S G, Carbotte J P 2007 Anomalous absorption line in the magneto-optical response of graphene \emph{Physical review letters} \textbf{98} 157402.

\bibitem{Futia;1626}
Futia G, Schlup P, Winters D G, Bartels R A 2011 Spatially-chirped modulation imaging of absorbtion and fluorescent objects on single-element optical detector \emph{Optics express} \textbf{19} 1626.

\bibitem{Kimura;178}
Kimura M, Toshima K 1998 Vibration sensor using optical-fiber cantilever with bulb-lens \emph{Sensors and Actuators A: Physical} \textbf{66} 178.

\bibitem{Li;826}
Li H, Tadesse S A, Liu Q, Li M 2015 Nanophotonic cavity optomechanics with propagating acoustic waves at frequencies up to 12 GHz \emph{Optica} \textbf{2} 826.

\bibitem{13PRB34;2}
Shung K W K 1986
Lifetime effects in low-stage intercalated graphite systems
\emph{Phys. Rev. B} \textbf{34} 2.

\bibitem{13PRB74;085406}
Ho J H, Lu C L, Hwang C C, Chang C P, and Lin M F 2006
Coulomb excitations in AA- and AB-stacked bilayer graphites
\emph{Phys. Rev. B} \textbf{74} 085406.

\bibitem{13PLA352;446}
Ho J H, Chang C P, and Lin M F 2006
Electronic excitations of the multilayered graphite
\emph{Phys. Lett. A} \textbf{352} 446.

\bibitem{13REVSCIINS83;093108}
Podobedov V B, Miller C C, and Nadal M E 2012 Performance of the
NIST goniocolorimeter with a broad-band source and multichannel
charged coupled device based spectrometer
\emph{Rev. Sci. Instrum.} \textbf{83} 093108.

\bibitem{13APPOPT48;5713}
Martinez L F L, Garcia R C , Navarro R E B, and Martinez A L 2009
Microreflectance difference spectrometer based on a
charge coupled device camera: Surface distribution of
polishing-related linear defect density in GaAs (001)
\emph{Appl. Opt.} \textbf{48} 5713.


\bibitem{Novikov;245435}
Novikov D S 2007 Elastic scattering theory and transport in graphene \emph{Physical Review B} \textbf{76} 245435.

\bibitem{Mak;1341}
Mak K F, Ju L, Wang F, Heinz T F 2012 Optical spectroscopy of graphene: from the far infrared to the ultraviolet \emph{Solid State Communications} \textbf{152} 1341.


\bibitem{Yaokawa;10657}
Yaokawa R, Ohsuna T, Morishita T, Hayasaka Y, Spencer M J S and Nakano H
2016 Monolayer-to-bilayer transformation of silicenes and their structural analysis \emph{Nat. Comm.} \textbf{7} 10657.

\bibitem{13SciRep7;40600}
Shih P H, Chiu Y H, Wu J Y, Shyu F L, and Lin M F 2017
Coulomb excitations of monolayer germanene
\emph{Sci. Rep.} \textbf{7} 40600.

\bibitem{13PRB94;045410}
Chen S C, Wu C L, Wu J Y and Lin M F 2016 Magnetic quantization of sp3 bonding in monolayer gray tin \emph{Phys. Rev. B} \textbf{94} 045410.


\bibitem{13PRB98;115117}
Yu J, Katsnelson M I and Yuan S 2018 Tunable electronic and magneto-optical properties of monolayer arsenene: From GW0 approximation to large-scale tight-binding propagation simulations \emph{Phys. Rev. B} \textbf{98} 115117.

\bibitem{13NJP20;062001}
Chen S C, Wu J Y and Lin M F 2018 Feature-rich magneto-electronic properties of bismuthene \emph{New Jour. Phys. Fast track communication} \textbf{20} 062001.

\bibitem{13PRB97;2018}
Shih P H, Chiu C W, Wu J Y, Do T N and Lin M F 2018 Coulomb scattering rates of excited states in monolayer electron-doped germanene \emph{Phys. Rev. B} \textbf{97} 195302.














%%

\bibitem{Kostelnik;1574}
Kostelnik P, Seriani N, Kresse G, Mikkelsen A, Lundgren E, Blum V, Schmid M 2007 The Pd (100)-(5¡Ñ5) R27¢X¡VO surface oxide: A LEED, DFT and STM study \emph{Surface science} \textbf{601} 1574.

\bibitem{Hafner;6}
Hafner J 2007 Materials simulations using VASP¡Xa quantum perspective to materials science \emph{Computer physics communications} \textbf{177} 6.

\bibitem{Neaton;1298}
Neaton J B, Muller D A, Ashcroft N W 2000 Electronic properties of the Si/SiO 2 interface from first principles \emph{Physical review letters} \textbf{85} 1298.

\bibitem{Eelbo;136804}
Eelbo T, Wasniowska M, Thakur P, Gyamfi M, Sachs B, Wehling T O, Wiesendanger R 2013 Adatoms and clusters of 3d transition metals on graphene: Electronic and magnetic configurations \emph{Physical review letters} \textbf{110} 136804.

\bibitem{Barabash;155704}
Barabash S V, Ozolins V, Wolverton C 2008 First-principles theory of competing order types, phase separation, and phonon spectra in thermoelectric AgPb m SbTe m+ 2 alloys \emph{Physical review letters} \textbf{101} 155704.

\bibitem{Lin;207}
Lin S Y, Chang S L, Shyu F L, Lu J M, Lin M F 2015 Feature-rich electronic properties in graphene ripples \emph{Carbon} \textbf{86} 207.

\bibitem{Tran;10623}
Tran N T T, Lin S Y, Glukhova O E, Lin M F 2015 Configuration-induced rich electronic properties of bilayer graphene \emph{The Journal of Physical Chemistry C} \textbf{119} 10623.

\bibitem{Lin;26443}
Lin S Y, Chang S L, Tran N T T, Yang P H, Lin M F 2015 H¡VSi bonding-induced unusual electronic properties of silicene: a method to identify hydrogen concentration \emph{Physical Chemistry Chemical Physics} \textbf{17} 26443.

\bibitem{Lin;209}
Lin S Y, Lin Y T, Tran N T T, Su W P, Lin M F 2017 Feature-rich electronic properties of aluminum-adsorbed graphenes \emph{Carbon} \textbf{120} 209.

\bibitem{Huang;84}
Huang H C, Lin S Y, Wu C L, Lin M F 2016 Configuration-and concentration-dependent electronic properties of hydrogenated graphene \emph{Carbon} \textbf{103} 84.

\bibitem{Lin;1722}
Lin Y T, Lin S Y, Chiu Y H, Lin M F 2017 Alkali-created rich properties in grapheme nanoribbons: Chemical bondings \emph{Scientific reports} 7 1722.















\bibitem{Lu;1123}
Lu J P 1995 Novel magnetic properties of carbon nanotubes \emph{Physical review letters} \textbf{74} 1123.

\bibitem{Tsai;075411}
Tsai C C, Shyu F L, Chiu C W, Chang C P, Chen R B, Lin M F 2004 Magnetization of armchair carbon tori \emph{Physical Review B}, \textbf{70} 075411.

\bibitem{Jonsson;572}
Jonsson D, Norman P, Ruud K, Agren H, Helgaker T 1998 Electric and magnetic properties of fullerenes \emph{The Journal of chemical physics} \textbf{109} 572.

\bibitem{Moran;6746}
Moran D, Stahl F, Bettinger H F, Schaefer H F, Schleyer P V R 2003 Towards graphite: magnetic properties of large polybenzenoid hydrocarbons \emph{Journal of the American Chemical Society} \textbf{125} 6746.

\bibitem{Searles;017403}
Searles T A, Imanaka Y, Takamasu T, Ajiki H, Fagan J A, Hobbie E K, Kono J 2010 Large anisotropy in the magnetic susceptibility of metallic carbon nanotubes \emph{Physical review letters} \textbf{105} 017403.

\bibitem{Kibis;741}
Kibis O V 2002 Electronic phenomena in chiral carbon nanotubes in the presence of a magnetic field \emph{Physica E: Low-dimensional Systems and Nanostructures} \textbf{12} 741.

\bibitem{Murmu;5069}
Murmu T, McCarthy M A, Adhikari S 2012 Vibration response of double-walled carbon nanotubes subjected to an externally applied longitudinal magnetic field: a nonlocal elasticity approach \emph{Journal of Sound and Vibration} \textbf{331} 5069.

\bibitem{13RevModPhys81;109}
Castro Neto A H, Guinea F, Peres N M R, Novoselov K S and Geim A K
2009 The electronic properties of graphene \emph{Rev. Mod. Phys.}
\textbf{81} 109-162.

\bibitem{13Rep76;056503}
McCann E and Koshino M 2013 The electronic properties of bilayer
graphene \emph{Rep. Prog. Phys.} \textbf{76} 056503.

\bibitem{13PRB80;165409}
Koshino M and McCann E 2009 Trigonal warping and Berry's phase N pi
in ABC-stacked multilayer graphene \emph{Phy. Rev. B} \textbf{80}
165409.

\bibitem{13PRB77;115313}
Koshino M and Ando T 2008 Magneto-optical properties of multilayer
graphene \emph{Phy. Rev. B } \textbf{77} 115313.


\bibitem{Tabert;085434}
Tabert C J and Nicol E J 2013 Magneto-optical conductivity of silicene and other buckled honeycomb lattices \emph{Phys. Rev. B} \textbf{88} 085434.

\bibitem{Matthes;395305}
Matthes L, Pulci O, Bechstedt F 2013 Massive Dirac quasiparticles in the optical absorbance of graphene, silicene, germanene, and tinene \emph{Journal of Physics: Condensed Matter} \textbf{25} 395305.


\end{thebibliography}
\end{document}